\documentclass[12pt,letterpaper]{JHEP3}
\pdfoutput=1

\usepackage{amsmath,bbm,array,amsfonts,graphicx,rotating,wrapfig}

\newcommand{\be}{\begin{equation}}
\newcommand{\ee}{\end{equation}}
\newcommand{\beq}{\begin{equation}}
\newcommand{\beql}[1]{\begin{equation}\label{#1}}
\newcommand{\eeq}{\end{equation}}
\newcommand{\ba}{\begin{array}}
\newcommand{\ea}{\end{array}}
\newcommand{\bea}{\begin{eqnarray}}
\newcommand{\beal}[1]{\begin{eqnarray}\label{#1}}
\newcommand{\eea}{\end{eqnarray}}
\newcommand{\ben}{\begin{enumerate}}
\newcommand{\een}{\end{enumerate}}
\newcommand{\bean}{\begin{eqnarray*}}
\newcommand{\eean}{\end{eqnarray*}}
\newcommand{\eref}[1]{(\ref{#1})}
\newcommand{\sref}[1]{\S\ref{#1}}
\newcommand{\tref}[1]{Table~\ref{#1}}
\newcommand{\nn}{\nonumber}

\newcommand{\bvec}{\left(\ba{c}}
\newcommand{\evec}{\ea\right)}
\newcommand{\bvset}[1]{\left\{\ba{#1}}
\newcommand{\evset}{\ea\right\}}
\newcommand{\fref}[1]{Figure \ref{#1}}
\newcommand{\btab}[1]{\begin{tabular}{#1}}
\newcommand{\etab}{\end{tabular}}

\def \Black {\pdfsetcolor{0 0 0 1}}

\newcommand{\comment}[1]{}

\newtheorem{theorem}{\bf Theorem}

\newtheorem{observation}[theorem]{\bf Observation}

\newtheorem{proposition}[theorem]{Proposition}

\newenvironment{definition}[1][Definition]{\begin{trivlist}
\item[\hskip \labelsep {\bfseries #1}]}{\end{trivlist}}

\newcommand{\qed}{\nobreak \ifvmode \relax \else
      \ifdim\lastskip<1.5em \hskip-\lastskip
      \hskip1.5em plus0em minus0.5em \fi \nobreak
      \vrule height0.75em width0.5em depth0.25em\fi}

\title{Symmetries of Abelian Orbifolds}
\author{Amihay Hanany and Rak-Kyeong Seong\\
Theoretical Physics Group \\
The Blackett Laboratory \\
Imperial College London, Prince Consort Road\\
London,  SW7 2AZ,  UK \\
{\tt a.hanany, rak-kyeong.seong@imperial.ac.uk}}

\preprint{Imperial/TP/10/AH/05}

\abstract{Using the Polya Enumeration Theorem, we count with particular attention to $\mathbb{C}^3/\Gamma$ up to $\mathbb{C}^6/\Gamma$, abelian orbifolds in various dimensions which are invariant under cycles of the permutation group $S_D$. This produces a collection of multiplicative sequences, one for each cycle in the Cycle Index of the permutation group. A multiplicative sequence is controlled by its values on prime numbers and their pure powers. Therefore, we pay particular attention to orbifolds of the form $\mathbb{C}^D/\Gamma$ where the order of $\Gamma$ is $p^\alpha$. We propose a generalization of these sequences for any $D$ and any $p$.
}

\begin{document}

%\pagestyle{plain}
%\setcounter{page}{1}
%\newcounter{bean}
%\baselineskip16pt

%%%%%%%%%%%%%%%%%%%%%%%%%%%%%%%%%%%%%%%%%%%%%%%%%%%%%%%%%%%%%%%%%%%%%%%%%%%%%%%%%%%%%%%%%%%%%%%%%%%%%%%%%%%%%%%%%%%%%%%%%%%%

\section{Introduction \label{s1}}

Recent advances in enumerating and counting distinct abelian orbifolds \cite{HananyOrlando10,DaveyHananySeong10} have uncovered rich structures in the vast family of quiver gauge theories. In the past, quiver gauge theories \cite{DouglasMoore96,DouglasMoore97,DouglasGreeneMorrison97} as world volume theories of D$3$-branes probing toric non-compact Calabi-Yau (CY) singularities \cite{Klebanov:1998hh,Acharya:1998db} have been fruitfully studied \cite{HananyKen05,Hanany05,Yamazaki:2008bt,Davey:2009bp,Hanany:2005ss,Franco:2005sm,Kennaway:2007tq}. Brane tilings were instrumental in relating world volume gauge theories of D$3$-branes with probed toric non-compact Calabi-Yau geometries. Trailblazing examples of study were the abelian orbifolds of $\mathbb{C}^3$ \cite{Muto:1997pq,Kachru:1998ys,Lawrence:1998ja,Bershadsky:1998mb,Hanany:1998sd,Beasley:1999uz,Uranga:2000ck,Feng:2002kk}. A guiding principle has been the fact that an infinite sub-class of $(3+1)$-dimensional world volume gauge theories have moduli spaces which are abelian orbifolds of the form $\mathbb{C}^{3}/\Gamma$ with $\Gamma$ being an abelian subgroup of $SU(3)$. The moduli spaces are toric, and for abelian orbifolds of $\mathbb{C}^{3}$ the toric diagrams are always elegantly triangles. Accordingly, from the geometrical perspective, two distinct abelian orbifolds of $\mathbb{C}^{3}$ have toric triangles which are not related under a $GL(2,\mathbb{Z})$ transformation. A thought-provoking example is the abelian orbifold of the form $\mathbb{C}^3/\mathbb{Z}_{30}$ with action $(2,3,25)$ whose toric triangle cannot be $GL(2,\mathbb{Z})$ equivalent to an orbifold with an action of the unnecessarily restrictive but commonly used form $(1,a,-1-a)$. This and many other untouched orbifolds lead to the problem of classifying and counting distinct abelian orbifolds of $\mathbb{C}^3$ which has been solved in the pioneering work in \cite{HananyOrlando10} and \cite{DaveyHananySeong10}.\\

How about higher dimensional abelian orbifolds of $\mathbb{C}^D$? The most recent breakthroughs which led towards studies on Calabi-Yau four-folds as orbifold backgrounds have been the works on ABJM theory \cite{BaggerLambert07,BaggerLambert08a,BaggerLambert08b,Gustavsson07,Gustavsson08,ABJM08}. These prompted an upgrade of brane tilings to accommodate the world volume gauge theories of M$2$-branes which probe toric non-compact CY $4$-folds. The world volume gauge theories of probe M$2$ branes are $\mathcal{N}=2$ $(2+1)$-dimensional quiver Chern-Simons theories \cite{Martelli:2008si,Hanany:2008cd,Hanany:2008gx}. The theories' Chern-Simons levels are represented in a modified brane tiling \cite{Hanany:2008fj,Davey:2009sr,Davey:2009qx,Davey:2009et} which obviates the use of the initially proposed brane crystal constructions \cite{Lee:2006hw,Lee:2007kv}. The special connection to our work has been the observation that an infinite sub-class of $(2+1)$-dimensional $M2$-brane world volume gauge theories have moduli spaces which are abelian orbifolds of the form $\mathbb{C}^{4}/\Gamma$ with $\Gamma$ being an abelian subgroup of $SU(4)$. As for the CY$3$ case, the moduli spaces are toric, and the associated toric diagrams elegantly turn out to be always tetrahedra \cite{Hanany:2008fj,Taki:2009wf}. Again, from a geometrical perspective two distinct abelian orbifolds of $\mathbb{C}^4$ have toric tetrahedra which are not related under a $GL(3,\mathbb{Z})$ transformation. Accordingly, not surprisingly we encounter from this special example of Chern-Simons gauge theories the familiar problem of enumerating and counting distinct abelian orbifolds of $\mathbb{C}^4$ \cite{HananyOrlando10,DaveyHananySeong10}.\\

\begin{figure}[ht!]
\begin{center}
\includegraphics[totalheight=5cm]{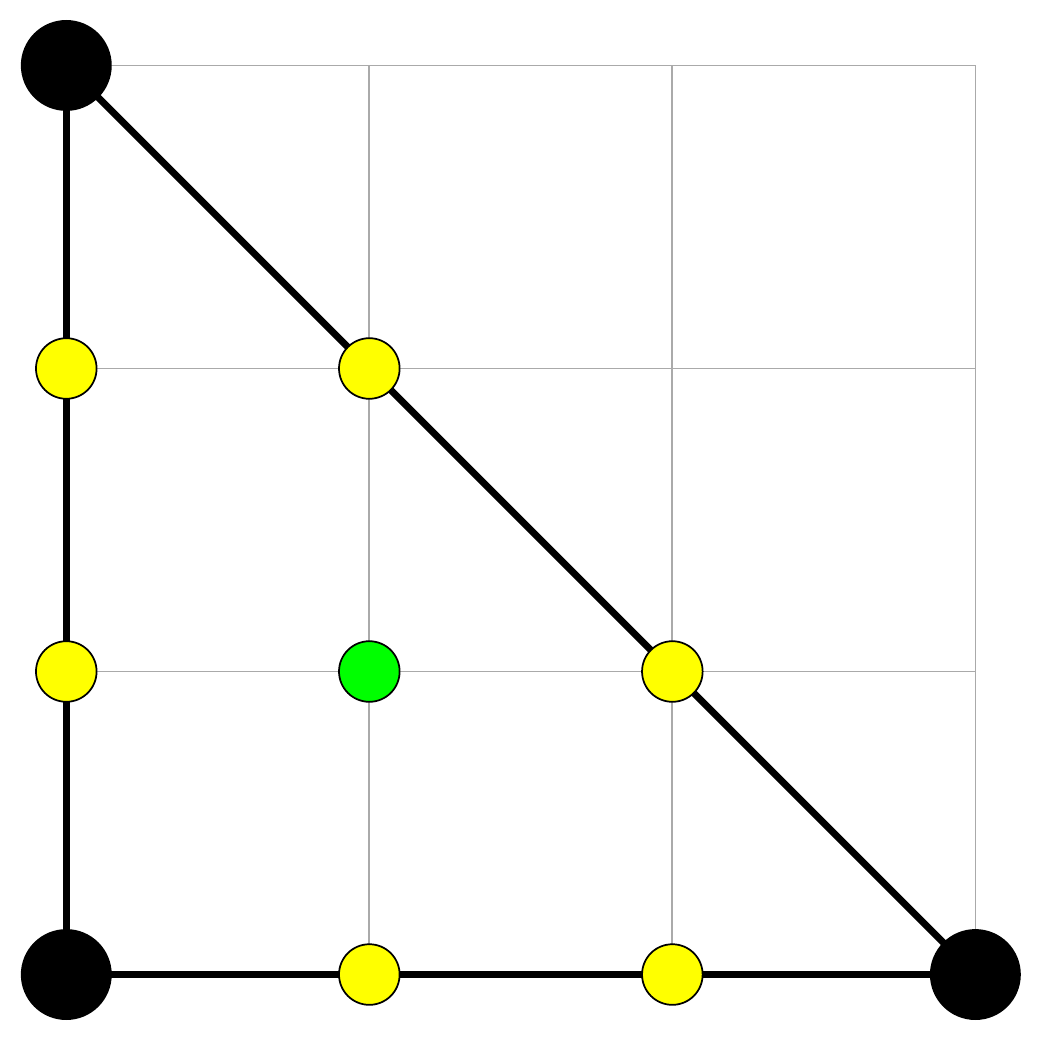}
\includegraphics[totalheight=5cm]{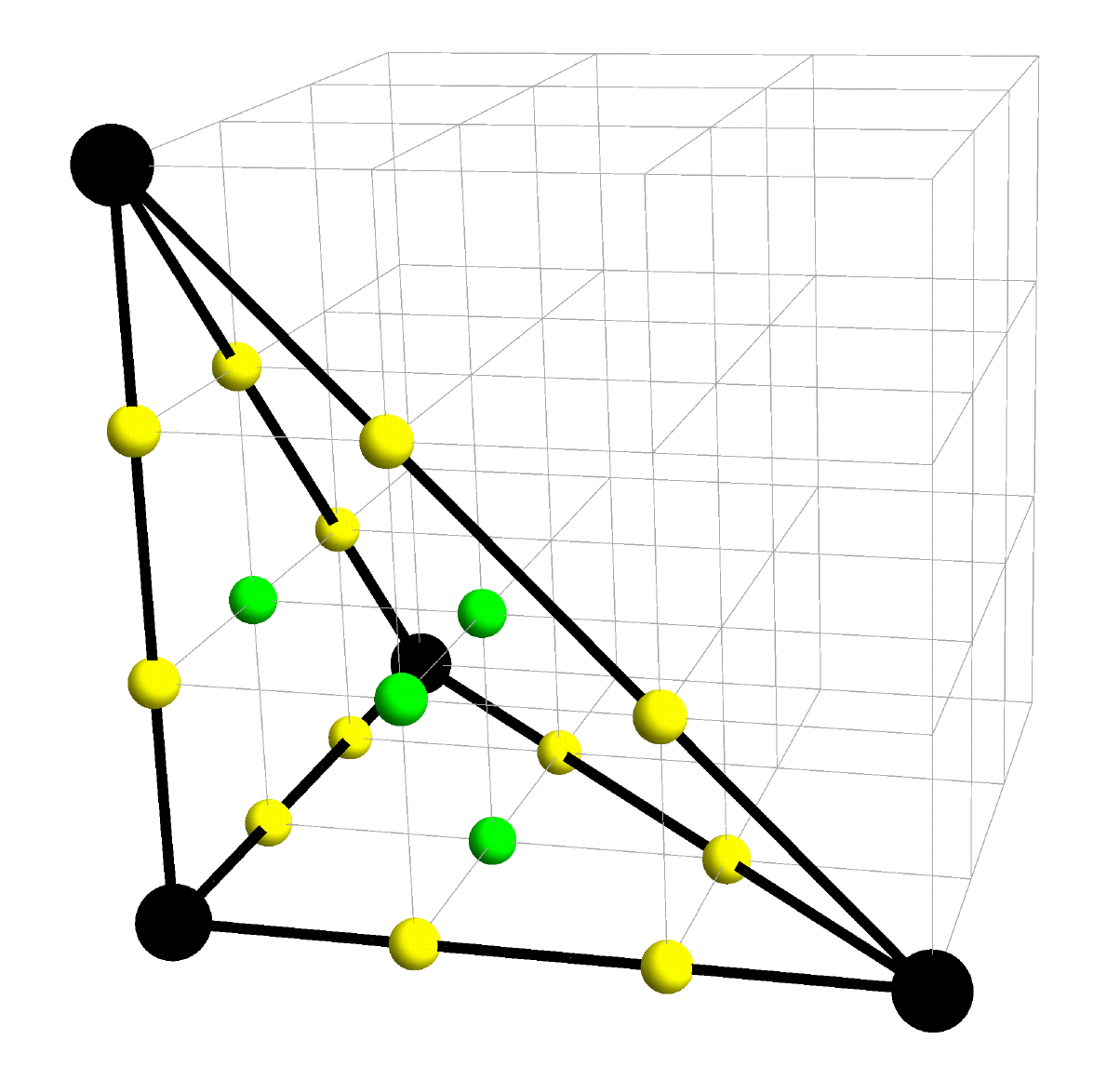}
\includegraphics[totalheight=8cm]{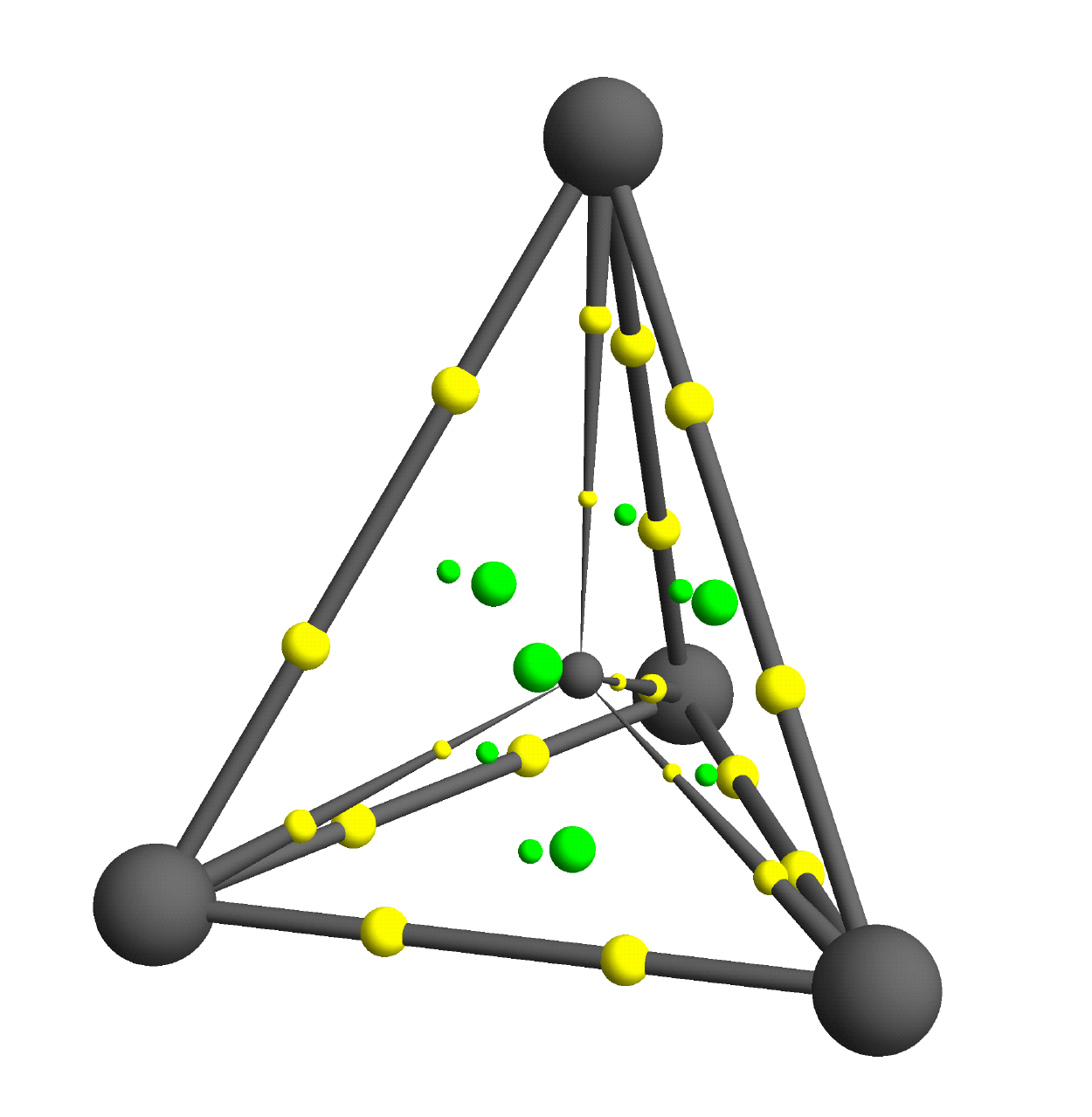}
\caption{The toric diagrams for the abelian orbifolds of the form $\mathbb{C}^{3}/\mathbb{Z}_{3}\times\mathbb{Z}_{3}$, $\mathbb{C}^{4}/\mathbb{Z}_{3}\times\mathbb{Z}_{3}\times\mathbb{Z}_{3}$ and $\mathbb{C}^{5}/\mathbb{Z}_{3}\times\mathbb{Z}_{3}\times\mathbb{Z}_{3}\times\mathbb{Z}_{3}$ respectively. The $4$-dimensional toric diagram of $\mathbb{C}^{5}/\mathbb{Z}_{3}\times\mathbb{Z}_{3}\times\mathbb{Z}_{3}\times\mathbb{Z}_{3}$ has been projected into $3$-space. $\mathbb{Z}^{D}$ lattice points on $1$-simplices and $2$-simplices are colored yellow and green respectively, whereas the defining vertex points are in black.}
  \label{projection}
 \end{center}
\end{figure}

By continuation, we expect that higher dimensional abelian orbifolds of the form $\mathbb{C}^{D}/\Gamma$ with $\Gamma$ being an abelian subgroup of $SU(D)$ have toric diagrams which are $(D-1)$-dimensional simplices embedded in $\mathbb{Z}^{D-1}$. An efficient method of testing $GL(D-1,\mathbb{Z})$ equivalence between toric simplices has been outlined in detail in \cite{DaveyHananySeong10}.\\

In the following we argue that discrete symmetries of an abelian orbifold of $\mathbb{C}^D$ can be observed directly through its toric diagram using the same method used to test $GL(D-1,\mathbb{Z})$ equivalence between toric simplices. Discrete symmetries have played an integral role in specifying the global symmetries of the gauge theory in $3+1$ dimensions in the past \cite{Beasley:2001zp,Feng:2002zw}, and so far, they have been identified only through the quiver or superpotential of the gauge theory. The method we present in this work to `measure' symmetries directly from the toric diagram of a given abelian orbifold of $\mathbb{C}^D$ is a novel approach whose unexpected by-product through Polya's Enumeration Theorem is the counting of distinct abelian orbifolds of $\mathbb{C}^D$ - something which we believe has never been done before.\\

We identify and count explicitly abelian orbifolds of $\mathbb{C}^3$ to $\mathbb{C}^6$ which are invariant under cycles of the permutation group $S_D$. This produces multiplicative sequences, each corresponding to a cycle in the Cycle Index of the permutation group $S_D$. Multiplicativity states that the sequence values at co-prime orders $n_1$ and $n_2$ give as a product the sequence value at order $n_1 n_2$. Accordingly, we put emphasis on orbifolds of the form $\mathbb{C}^D/\Gamma$ with the order of $\Gamma$ being a prime number. From this perspective, we propose a novel generalization of sequences which count distinct abelian orbifolds of  $\mathbb{C}^D$ and abelian orbifolds which are invariant under cycles of the permutation group $S_D$. Such a generalization enables us to probe and quantify the rich geometrical structure of abelian orbifolds of $\mathbb{C}^D$ in any dimension $D$. 
\\ 

The work is divided into the following sections:
\begin{itemize}
	\item Section \sref{s2} gives a short summary of how to identify distinct abelian orbifolds of $\mathbb{C}^{D}$ and toric diagrams which are invariant under cycles of the symmetric group $S_D$.% using toric diagrams. 
	\item Section \sref{s5} presents the results of counting for the orbifolds of $\mathbb{C}^{3}$, $\mathbb{C}^{4}$, $\mathbb{C}^{5}$ and $\mathbb{C}^{6}$, and reviews how these results can be encoded in terms of partition functions for the special cases of $\mathbb{C}^{3}/\Gamma_N$ and $\mathbb{C}^{4}/\Gamma_N$.
	\item Section \sref{sp1} presents the role of values on prime indices of sequences which count orbifolds that are invariant under cycles of $S_D$, and discusses how the values on prime indices affect the derivation of partition functions. We explicitly derive the partition function counting distinct $\mathbb{C}^5/\Gamma$.
	\item Section \sref{sp2} outlines generalizations for partition functions which count orbifolds that are invariant under certain cycles of $S_D$. In addition, a complete generalization is presented for sequences which count distinct abelian orbifolds of the form $\mathbb{C}^{D}/\Gamma$ and their symmetries where the order of $\Gamma$ is prime.
	\item Section \sref{sp3} describes all possible symmetry patterns of a single orbifold and counts the number of such patterns for a given order $N$ of
% order independent characteristics of orbifolds
$\mathbb{C}^{D}/\Gamma_N$.
\end{itemize}

\noindent\textbf{Notation and Nomenclature} \\

A list of the most common notation and nomenclature used in this work is presented below. The reader will be introduced to them in more detail in the main text.
\begin{itemize}
	\item A cycle $g$ of the permutation group $S_D$ is denoted by $g^\alpha$ to emphasize its correspondence to a conjugacy class $H_\alpha$ of $S_D$. A conjugacy class $H_\alpha \subset S_D$ is labeled by a cycle index variable $x^\alpha$. 
	
	\item Given a sequence $\mathsf{g}$ with elements $\mathsf{g}_n=\mathsf{g}(n)$ denoted by integer indices $n \in\mathbb{Z}^{+}$, we write a partition function of the sequence as $g(t)=\sum_n \mathsf{g}_n t^n$. 
	
	\item Given a sequence $\mathsf{g}$, the new sequence formed by picking elements $\mathsf{g}_p$ on prime indices $p$ is called a \textit{prime index sequence} of $\mathsf{g}$.
\end{itemize}

%%%%%%%%%%%%%%%%%%%%%%%%%%%%%%%%%%%%%%%%%%%%%%

\section{Background and Methods \label{s2a} \label{s2}}

\subsection{Distinguishing Orbifolds \label{sconject}}

As illustrated in \cite{DaveyHananySeong10}, there are different methods of distinguishing abelian orbifolds of $\mathbb{C}^{D}$. In general, two orbifolds of $\mathbb{C}^{D}$ are distinct if there is no $GL(D-1,\mathbb{Z})$ transformation which maps between the corresponding toric diagrams. We give here a short summary of the method which tests this condition efficiently.\\

\noindent\textbf{Toric Diagrams and Barycentric Coordinates}\\

Non-compact toric CY singularities are represented by toric diagrams. For abelian orbifolds of the form $\mathbb{C}^{2}/\Gamma_{N}$, the toric diagrams are lines in $\mathbb{Z}^{1}$ with length $N$. For abelian orbifolds of the form $\mathbb{C}^{3}/\Gamma_{N}$, the toric diagrams are triangles embedded in $\mathbb{Z}^{2}$ with area $N$. For abelian orbifolds of the form $\mathbb{C}^{4}/\Gamma_{N}$, the toric diagrams are tetrahedra embedded in $\mathbb{Z}^{3}$ with volume $N$. By continuation, abelian orbifolds of the form $\mathbb{C}^{D}/\Gamma_{N}$ have toric diagrams as $(D-1)$-simplices, henceforth denoted by $\sigma^{D-1}$, which are embedded in $\mathbb{Z}^{D-1}$ with hyper-volume $N$.\\

\begin{figure}[ht!]
\begin{center}
\includegraphics[totalheight=8cm]{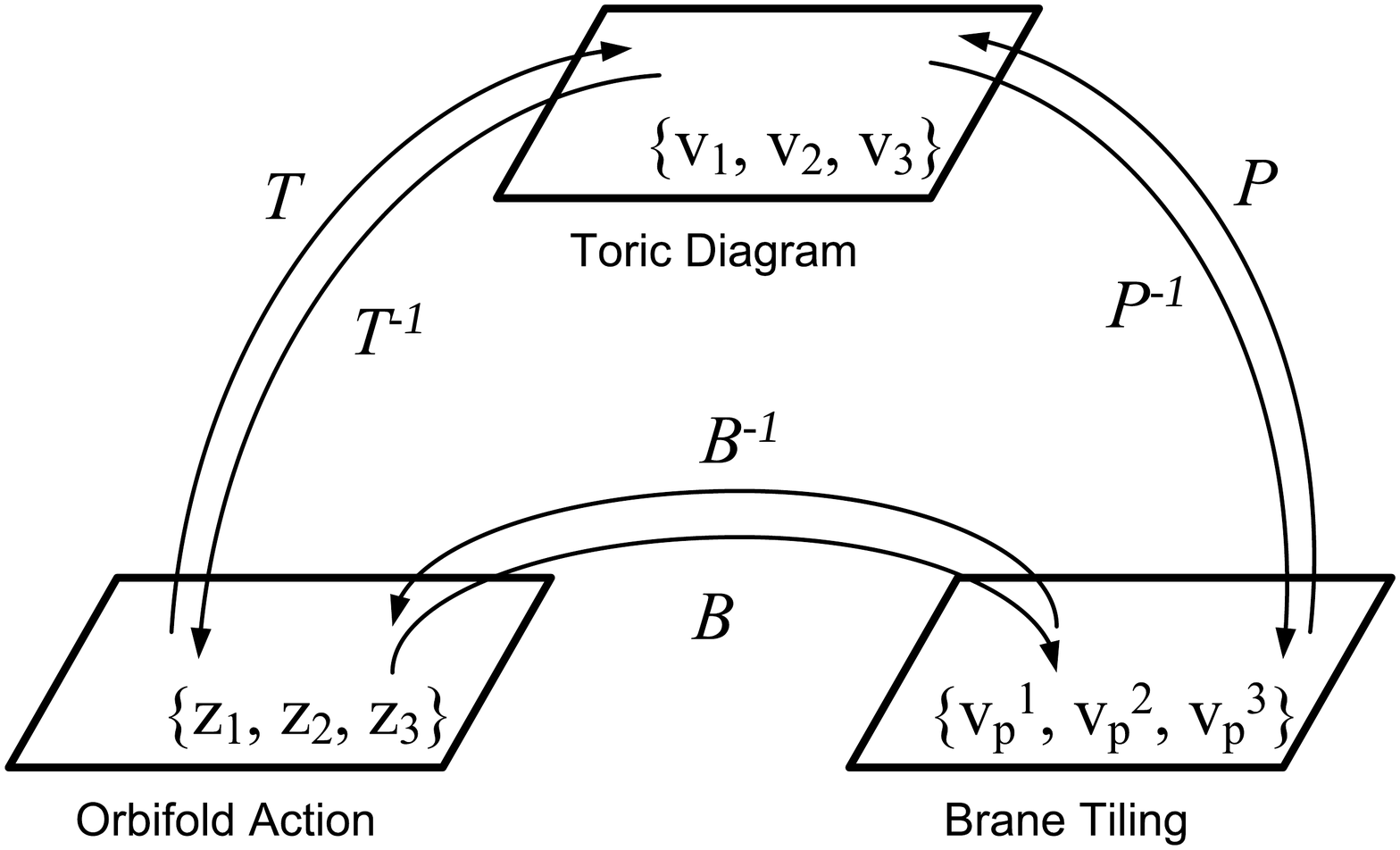}
\caption{The correspondence between barycentric coordinates of the toric triangle, coordinates of the hexagonal brane tiling and the complex coordinates of $\mathbb{C}^{3}$ as first illustrated in \cite{DaveyHananySeong10}.}
  \label{fcorres}
 \end{center}
\end{figure}

Every lattice point $w_k$ on and enclosed by the boundary of $\sigma^{D-1}$  ($w_k \in \sigma^{D-1}$) divides $\sigma^{D-1}$ into $D$ sub-simplices of dimension $D-1$ or less. These sub-simplices have $(D-1)$-dimensional hyper-volumes with values $\lambda_{k1},\lambda_{k2},\dots,\lambda_{kD}$. Accordingly, the lattice point $w_{k}\in\sigma^{D-1}$ can be given in terms of \textbf{barycentric coordinates} of the form
\beql{e7}
w_{k}~~=~~\frac{1}{N}(\lambda_{k1},\lambda_{k2},\dots,\lambda_{ki},\dots,\lambda_{kD})~~,
\eeq
where the barycentric coordinate axes are labeled by $i=1,\dots,D$ and $N$ is the $(D-1)$-dimensional hyper-volume of the simplex $\sigma^{D-1}$. \\

It has been proposed in \cite{DaveyHananySeong10} that the barycentric coordinates defined on toric simplices of $\mathbb{C}^D/\Gamma$ correspond to complex coordinates on $\mathbb{C}^D$ as well as for $D=3$ the zig-zag-paths on the hexagonal brane tiling of $\mathbb{C}^3$. The correspondence is illustrated in \fref{fcorres}.\\

\begin{figure}[ht!]
\begin{center}
\includegraphics[totalheight=5.2cm]{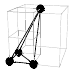}
\includegraphics[totalheight=5.2cm]{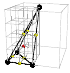}
\includegraphics[totalheight=5.2cm]{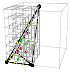}\\
$s_0=1$~~~~~~~~~~~~~~~~~~~~~~~~~~~
$s_1=s_3=2$~~~~~~~~~~~~~~~~~~~~~~~~~~~
$s_2=3$
\caption{Toric tetrahedra corresponding to $\mathbb{C}^{4}/\mathbb{Z}_{2}$ with orbifold action $A=((1,1,1,1),(0,0,0,0),(0,0,0,0))$ and scalings $s_{0}=1$, $s_{1}=s_{3}=2$ and $s_{2}=3$ respectively. Lattice points on edges ($I_{0}$), lattice points on faces ($I_{1}$) and internal lattice points ($I_{3}$) are colored yellow, green and red respectively.}
  \label{fscaling}
 \end{center}
\end{figure}

\noindent\textbf{The Topological Character and Scaling}\\

The topological character of a given toric simplex $\sigma^{D-1}$ is defined as the set of barycentric coordinates for all $w_{k}\in I(\sigma^{D-1})$. $I(\sigma^{D-1})$ defines the set of relevant lattice points of $\sigma^{D-1}$, and is defined as
\beql{em1}
I(\sigma^{D-1})~=~\bigcup_{d=0}^{D-1}{I_{d}(f_{s}(\sigma^{d-1}))}~~.
\eeq 
Here, $I_{d}(\sigma^{D-1})$ is the set of defining lattice points of all $d$-dimensional sub-simplices contained in $\sigma^{D-1}$. Accordingly, $I_{0}(\sigma^{D-1})$ is the set of $D$ corner points of $\sigma^{D-1}$ (\fref{fscaling}). $f_{s_{d}}(\sigma^{D-1})$ is a scaled simplex $\sigma^{D-1}$ such that $I_{d}(f_{s_{d}}(\sigma^{D-1}))\neq \emptyset$ with $s_d$ being the scaling coefficient. In \eref{em1} we use an overall scaling coefficient $s=\max{(s_{1},\dots,s_{D-1})}$.\\

\noindent\textbf{Example.} Let us take the example shown in \fref{fscaling} for the orbifold of the form $\mathbb{C}^4/\mathbb{Z}_{2}$. Here, $I_{0}$ is the set of the four corner points of the toric tetrahedron which are `visible' with scaling $s_0=1$. The internal (red) points and points on edges (yellow) forming the sets $I_{1}$ and $I_{3}$ respectively are visible only with scaling $s_1=s_3=2$. Finally, lattice points on faces of the tetrahedron (green) forming the set $I_3$ are visible only with an overall scaling $s_3=3$. In order to collect all topologically significant lattice points in the overall set $I$, we scale the toric tetrahedron of $\mathbb{C}^4/\mathbb{Z}_{2}$ to $\max{(s_0,s_1,s_2,s_3)}=3$.\\

Overall, the \textbf{topological character} of a toric simplex $\sigma^{D-1}$ is defined as
\beql{e9}
\tau~~=~~\left\{
\frac{1}{N}(\lambda_{k1},\lambda_{k2},\dots,\lambda_{ki},\dots,\lambda_{kD})~~\Big|~~
w_k\in I(\sigma^{D-1})
\right\}~~,
\eeq
where $w_k$ is the barycentric coordinate defined in \eref{e7} of a point in the set $I(\sigma^{D-1})$ defined in \eref{em1}.

\begin{observation}\label{cequiv}
Two toric simplices of $\mathbb{C}^D/\Gamma_N$ that are related under a $GL(D-1,\mathbb{Z})$ transformation, and hence are equivalent, have equal topological characters up to a permutation of the barycentric coordinate axes labeled by $i=1,\dots,D$.
\end{observation}

%%%%%%%%%%%%%%%%%%%%%%%%%%%%%%%%%%%%%%%%%%%%%%%%%%%%%%%%%%%%%%%%%%%%%%%%%%%%%%%%%%%%%%%%%%%%%%%%%%%%%%%%%%%%%%%%%%%%%%%%%%%%

\subsection{Hermite Normal Forms and Symmetries}

\begin{figure}[ht!]
\begin{center}
\includegraphics[totalheight=4cm]{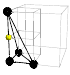}
\includegraphics[totalheight=4cm]{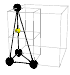}
\includegraphics[totalheight=4cm]{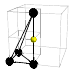}
\includegraphics[totalheight=4cm]{C4n2i4.pdf}
\includegraphics[totalheight=4cm]{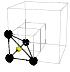}
\includegraphics[totalheight=4cm]{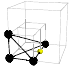}
\includegraphics[totalheight=4cm]{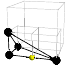}
\caption{The Hermite Normal Forms $D(2)$ for $\mathbb{C}^{4}/\Gamma_{2}$.}
  \label{fdomainex}
 \end{center}
\end{figure}

\noindent\textbf{Hermite Normal Forms}\\

The Hermite Normal Form (HNF) is an upper diagonal square matrix of size $D-1$ with non-negative integer entries. It takes the form

%HNF's are of the form
\beql{e95}
M~=~\left(\ba{cccccc}
m_{11} & m_{12} & \dots & m_{1j}     & \dots & m_{1(D-1)} \\
0 & m_{22} & \dots      & m_{2j}     & \dots & m_{2(D-1)} \\
0 & 0 &                 & m_{3j}     &       & m_{3(D-1)}  \\
\vdots & \vdots &       & \vdots     &       & \vdots  \\
0 & 0  &                & m_{(j-1)j} &       & m_{(j-1)(D-1)} \\
0 & 0 &                 & m_{jj}     &       & m_{j(D-1)} \\
0 & 0  &                & 0          &       & m_{(j+1)(D-1)}\\
\vdots & \vdots  &      & \vdots     &       & \vdots \\
0 & 0 & \dots           & 0          & \dots & m_{(D-1)(D-1)}
\ea\right)~~,
\eeq
where %$j=1,\dots,D-1$ such that 
$\det{M}=\prod_{j=1}^{D-1}{m_{jj}}=N$ and the off diagonal entries are restricted by the condition $0\leq m_{jk} < m_{jj}$ with  $m_{jk}\in\mathbb{N}_{0}$. For each such matrix one can construct a toric diagram with hyper-volume $N$ by multiplying the matrix on the Cartesian basis in $D$ dimensions, $\{(1,0,\ldots,0), (0,1,0,\ldots, 0), \ldots, (0,\ldots,0,1)\}$. The set of all toric diagrams will henceforth be called the set of HNF's. 
\\

All HNF's of order $N$ and given dimension $D$ form a set $D(N)$. Every cycle $g\in S_{D}$ is an automorphism of $D(N)$, 
\beql{e96ex}
gD(N) ~=~ D(N)~~.
\eeq

\begin{observation}\label{csubset}
Under all $g\in S_{D}$, $D(N)$ is partitioned into $\mathsf{g}^{D}(N)$ subsets where each subset $[\sigma^{D-1}]$ corresponds to a distinct abelian orbifold of the form $\mathbb{C}^{D}/\Gamma_{N}$.
\end{observation}

A consequence of the above observation is the following:

\begin{observation}
A subset $[\sigma^{D-1}]\in D(N)$ which corresponds to a distinct orbifold of the form $\mathbb{C}^{D}/\Gamma_N$ is mapped onto itself under all $g\in S_D$.
\end{observation}
\vspace{5mm}

\noindent\textbf{Example.} Let us consider an example with orbifolds of the form $\mathbb{C}^{4}/\Gamma_2$. The corresponding set of all possible HNF matrices $D(2)$ is given by
\beal{ehnfex}
\left\{\left(
\begin{array}{ccc}
 1 & 0 & 0 \\
 0 & 1 & 0 \\
 0 & 0 & 2
\end{array}
\right),\left(
\begin{array}{ccc}
 1 & 0 & 0 \\
 0 & 1 & 1 \\
 0 & 0 & 2
\end{array}
\right),\left(
\begin{array}{ccc}
 1 & 0 & 1 \\
 0 & 1 & 0 \\
 0 & 0 & 2
\end{array}
\right),\left(
\begin{array}{ccc}
 1 & 0 & 1 \\
 0 & 1 & 1 \\
 0 & 0 & 2
\end{array}
\right),\left(
\begin{array}{ccc}
 1 & 0 & 0 \\
 0 & 2 & 0 \\
 0 & 0 & 1
\end{array}
\right),\left(
\begin{array}{ccc}
 1 & 1 & 0 \\
 0 & 2 & 0 \\
 0 & 0 & 1
\end{array}
\right),\left(
\begin{array}{ccc}
 2 & 0 & 0 \\
 0 & 1 & 0 \\
 0 & 0 & 1
\end{array}
\right)\right\}~~.\nn\\
\eea
The corresponding toric tetrahedra are shown respectively in \fref{fdomainex}.\\ 

%%%%%%%%%%%%%%%%%%%%%%%%%%%%%%%%%%%%%%%%%%%%%%%%%%%%%%%%%%%%%%%%%%%%%%%%%%%%%%%%%%%%%%%%%%%%%%%%%%%%%%%%%%%%%%%%%%%%%%%%%%%%

\noindent\textbf{Orbifold Symmetries \label{s3}}\\

Let $C_g$ be a transformation on the topological character $\tau$ of a toric simplex $\sigma^{D-1}$ where $g\in S_{D}$. Given $g=\{\gamma^i\}$, where $\gamma^i=(m_{1}^{i},\dots,m_{n_i}^{i})\in g$ is a cyclic permutation of $n_i$ elements at positions $\{m_{1}^i,\dots,m_{n_i}^i\}$, we interpret the action of the transformation $C_{\gamma^i}$ as a cyclic permutation of $n_i$ barycentric coordinate axes at positions $\{m_{1}^i,\dots,m_{n_i}^i\}$. \\

If for a given transformation $C_g$ the topological character $\tau$ of $\sigma^{D-1}$ is invariant, then we call $C_g$ and the corresponding cycle $g\in S_D$ a \textbf{symmetry} of $\sigma^{D-1}$. \\

\begin{wrapfigure}{r}{3cm} % "l" or "r" for the side on the page. And the width parameter for the width of the image space.
\centering
		\includegraphics[totalheight=3cm]{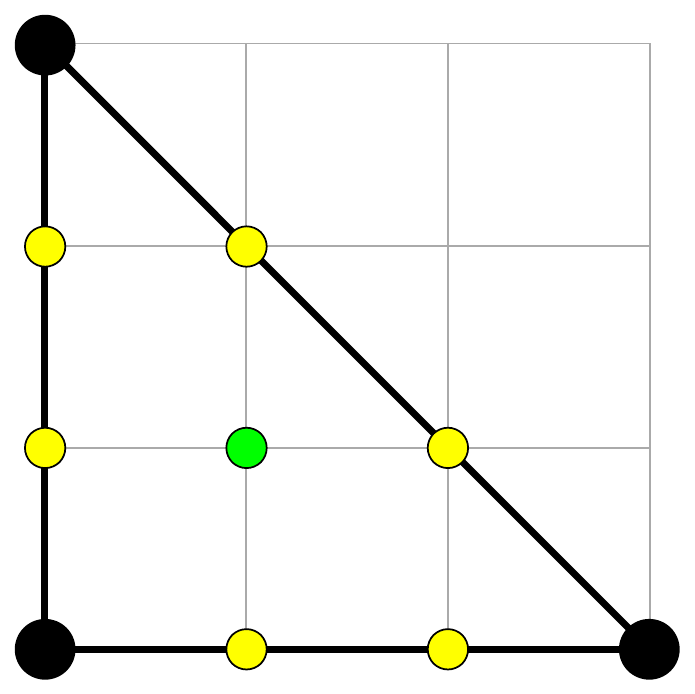}
\end{wrapfigure}

\noindent\textbf{Example.} Let us consider $4$ elements of the topological character of the orbifold of the form $\mathbb{C}^{3}/\mathbb{Z}_{3}\times\mathbb{Z}_{3}$ with the toric triangle shown on the right. The  $4$ elements correspond to the barycentric coordinates of the $3$ corner points and the green internal point, and are
\beal{etopch}
\tau=\left\{
(0,0,1),(0,1,0),(1,0,0),(1/3,1/3,1/3),\dots
\right\}~~.\nn
\eea
By transforming under $C_{(1,2,3)}$ which is a cyclic permutation of all $3$ barycentric coordinate axes, we see that the elements which correspond to the corner points are permuted whilst the element corresponding to the internal point is mapped onto itself. Accordingly, we note that under $C_{(1,2,3)}$, from considering just the first $4$ elements, $\tau$ is invariant under the cycle $(1,2,3)\in S_{3}$.
\\

More explicit examples of identifying orbifold symmetries using this method are shown in Appendix \ref{s3app}.\\

%%%%%%%%%%%%%%%%%%%%%%%%%%%%%%%%%%%%%%%%%%%%%%%%%%%%%%%%%%%%%%%%%%%%%%%%%%%%%%%%%%%%%%%%%%%%%%%%%%%%%%%%%%%%%%%%%%%%%%%%%%%%
\clearpage
\subsection{Counting Orbifold Symmetries \label{s4}}

%%%%%%%%%%%%%%%%%%%%%%%%%%%%%%%%%%%%%%%%%%%%%%%%%%%%%%%%%%%%%%%%%%%%%%%%%%%%%%%%
\noindent\textbf{The Cycle Index of $S_{D}$ \label{s4b}}\\

\begin{wrapfigure}[18]{r}{6cm} % "l" or "r" for the side on the page. And the width parameter for the width of the image space.
\centering
		\includegraphics[totalheight=6cm]{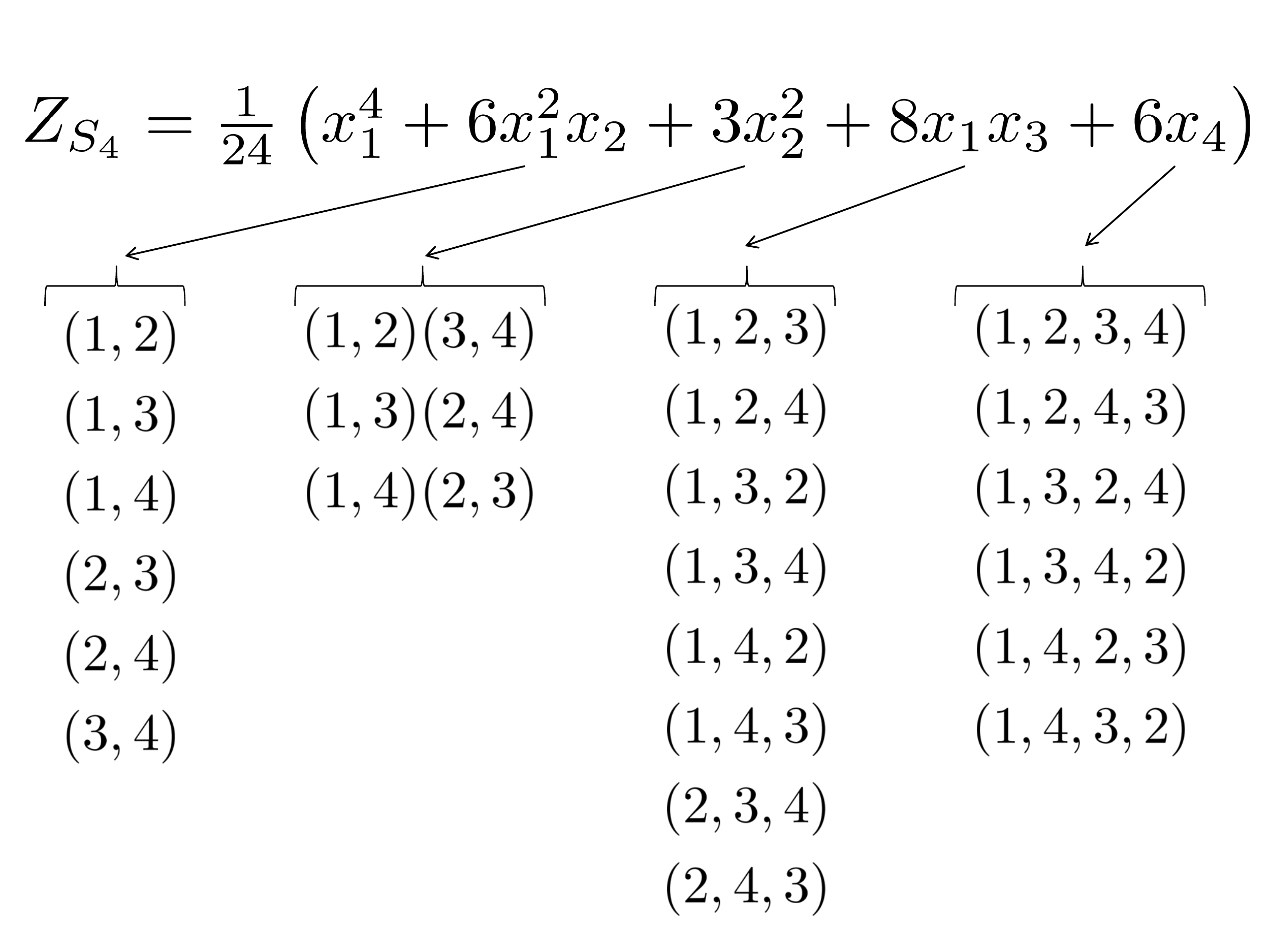}
\caption{The cycle index of $S_4$ and the $S_4$ cycles corresponding to terms of the cycle index.}
\label{fs4ex}
\end{wrapfigure}

Let a cycle $g\in S_D$ be denoted as $g=\{\gamma^i\}$ where $i=1,\dots,|g|=M$. Each sub-cycle $\gamma^i \in g$ permutes $n_i=|\gamma_{i}|$ elements at positions $\{m_{1}^{i},\dots,m_{n_i}^{i}\}$. Furthermore, let $\alpha=1,\dots,N_H$ be the index over conjugacy classes $H_\alpha$ of $S_D$. \\

Using this notation, the cycle index of $S_{D}$ is given by
\beql{e28}
Z_{S_{D}}~~=~~\frac{1}{|S_{D}|}
\sum_{\alpha=1}^{N_{H}}
\left(
|H_\alpha|
\prod_{i=1}^{M} x_{n_{i}(g_{\alpha})}
\right)
~~,
\eeq
The cycle index of $S_{D}$ can be found recursively using 
\beql{e32}
Z_{S_{D}}~=~\frac{1}{D}\sum_{r=1}^{D}{x_{r}~Z_{S_{D-r}}}~~,
\eeq
where $Z_{S_0}=1$. The first $9$ cycle indices are shown in \tref{fcycle}.\\

\begin{table}[ht!]
	\centering
		\begin{tabular}{c|c|l}
			\hline\hline
			$D$ & Orbifold & Cycle Index
			\\
			\hline
			$1$ & $\mathbb{C}$ & $Z_{S_1}~=~x_1$
			\\
			\hline
			$2$ & $\mathbb{C}^{2}/\Gamma_N$ & 
			$Z_{S_2}~=~\frac{1}{2}\left(x_{1}^{2}+x_{2}\right)$
			\\
			\hline
			$3$ & $\mathbb{C}^{3}/\Gamma_N$ & 
			$Z_{S_3}~=~\frac{1}{6}\left(x_{1}^{3}+3x_{1}x_{2}+2x_{3}\right)$
			\\
			\hline
			$4$ & $\mathbb{C}^{4}/\Gamma_N$ & 
			$Z_{S_4}~=~
			\frac{1}{24}\left(x_{1}^{4}+6x_{1}^{2}x_{2}+3x_{2}^{2}+8x_{1}x_{3}+6x_{4}\right)$
			\\
			\hline
			$5$ & $\mathbb{C}^{5}/\Gamma_N$ & 
			$Z_{S_5}~=~
			\frac{1}{120} \left(x_1^5+10 x_1^3 x_2+15 x_1 x_2^2+20 x_1^2 x_3+20 x_2 x_3+30 x_1 x_4+24 x_5\right)$
			\\
			\hline
			$6$ & $\mathbb{C}^{6}/\Gamma_N$ & 
			$\ba{rl}Z_{S_6}~=~ &
			\frac{1}{720} (x_1^6+15 x_1^4 x_2+45 x_1^2 x_2^2+15 x_2^3+40 x_1^3 x_3+120 x_1 x_2 x_3+40 x_3^2\\
			&+90 x_1^2 x_4+90 x_2 x_4+144 x_1 x_5+120 x_6)\ea$
			\\
			\hline
			$7$ & $\mathbb{C}^{7}/\Gamma_N$ & 
			$\ba{rl}Z_{S_7}~=~ &
			\frac{1}{5040} (x_1^7+21 x_1^5 x_2+105 x_1^3 x_2^2+105 x_1 x_2^3+70 x_1^4 x_3+420 x_1^2 x_2 x_3\\
			&+210 x_2^2 x_3+280 x_1 x_3^2+210 x_1^3 x_4+630 x_1 x_2 x_4+420 x_3 x_4\\
			&+504 x_1^2 x_5+504 x_2 x_5+840 x_1 x_6+720 x_7)
			\ea$
			\\
			\hline
			$8$ & $\mathbb{C}^{8}/\Gamma_N$ & 
			$\ba{rl}Z_{S_8}~=~ &
			\frac{1}{40320} (x_1^8+28 x_1^6 x_2+210 x_1^4 x_2^2+420 x_1^2 x_2^3+105 x_2^4+112 x_1^5 x_3\\
			&+1120 x_1^3 x_2 x_3+1680 x_1 x_2^2 x_3+1120 x_1^2 x_3^2+1120 x_2 x_3^2\\
			&+420 x_1^4 x_4+2520 x_1^2 x_2 x_4+1260 x_2^2 x_4+3360 x_1 x_3 x_4+1260 x_4^2\\
			&+1344 x_1^3 x_5+4032 x_1 x_2 x_5+2688 x_3 x_5+3360 x_1^2 x_6+3360 x_2 x_6\\
			&+5760 x_1 x_7+5040 x_8)
			\ea$
			\\
			\hline
			$9$ & $\mathbb{C}^{9}/\Gamma_N$ & 
			$\ba{rl}Z_{S_9}~=~ &
			\frac{1}{362880}(x_1^9+36 x_1^7 x_2+378 x_1^5 x_2^2+1260 x_1^3 x_2^3+945 x_1 x_2^4\\
			&+168 x_1^6 x_3+2520 x_1^4 x_2 x_3+7560 x_1^2 x_2^2 x_3+2520 x_2^3 x_3\\
			&+3360 x_1^3 x_3^2+10080 x_1 x_2 x_3^2+2240 x_3^3+756 x_1^5 x_4+7560 x_1^3 x_2 x_4\\
			&+11340 x_1 x_2^2 x_4+15120 x_1^2 x_3 x_4+15120 x_2 x_3 x_4+11340 x_1 x_4^2\\
			&+3024 x_1^4 x_5+18144 x_1^2 x_2 x_5+9072 x_2^2 x_5+24192 x_1 x_3 x_5\\
			&+18144 x_4 x_5+10080 x_1^3 x_6+30240 x_1 x_2 x_6+20160 x_3 x_6\\
			&+25920 x_1^2 x_7+25920 x_2 x_7+45360 x_1 x_8+40320 x_9)
			\ea$
			\\
			\hline\hline
		\end{tabular}
		\caption{The first nine cycle indices of $S_{D}$ and the corresponding abelian orbifolds.}
		\label{fcycle}
\end{table}

%%%%%%%%%%%%%%%%%%%%%%%%%%%%%%%%%%%%%%%%%%%%%%%%%%%%%%%%%%%%%%%%%%%%%%%%%%%%%%%%
\noindent\textbf{Polya's Enumeration Theorem \label{s4d}}\\

We recall that the set of HNF's $D(N)$ is invariant under all $g\in S_D$ and is partitioned into $\mathsf{g}^D(N)$ subsets under observation \sref{csubset}. Each subset corresponds to a distinct abelian orbifold of the form $\mathbb{C}^D/\Gamma_N$ and hence $\mathsf{g}^D(N)$ counts the number of distinct abelian orbifolds of the form $\mathbb{C}^D/\Gamma_N$ at order $N$.\\

A single HNF of $D(N)$ is invariant under $g\in S_D$ if $C_g$ is a symmetry of the corresponding toric simplex $\sigma^{D-1}$. Let $\mathsf{g}_{x^\alpha}(N)$ be the number of $g^\alpha$-symmetric HNF's in $D(N)$ where $g^\alpha \in H_{\alpha}$. $x^\alpha$ is a label of the $\alpha$-term in the cycle index of $S_D$, and the corresponding conjugacy class $H_\alpha$.\\

Under \textbf{Polya's Enumeration Theorem}, $Z_{S_{D}}=\mathsf{g}^{D}(N)$ if we insert for every monomial factor $x^\alpha$ in $Z_{S_{D}}$ the count $\mathsf{g}_{x^\alpha}(N)$ such that $x^\alpha=\mathsf{g}_{x^\alpha}(N)$. We recall that $\mathsf{g}^{D}(N)$ is the number of distinct toric simplices $\sigma^{D-1}$ of hyper-volume $N$ and equivalently the number of distinct abelian orbifolds of the form $\mathbb{C}^{D}/\Gamma_N$.\\

For the first four dimensions, the cycle indices are re-written as
\beal{exx1}
Z_{S_{1}}~~=~~x_{1}
~~&\Rightarrow&~~\nn\\
\mathsf{g}^{D=1}(N)~~=~~\mathsf{g}_{x_{1}}(N)&&
\nn\\
Z_{S_{2}}~~=~~\frac{1}{2}\left(x_{1}^{2}+x_{2}\right)
~~&\Rightarrow&~~\nn\\
\mathsf{g}^{D=2}(N)~~=~~\frac{1}{2}\left(\mathsf{g}_{x_{1}^{2}}(N)+\mathsf{g}_{x_{2}}(N)\right)&&
\nn\\
Z_{S_{3}}~~=~~
\frac{1}{6} \left(x_{1}^{3}+3 x_{1}x_{2}+2 x_{3}\right)
~~&\Rightarrow&~~\nn\\
\mathsf{g}^{D=3}(N)~~=~~
\frac{1}{6} \left(\mathsf{g}_{x_{1}^{3}}(N)+3 \mathsf{g}_{x_{1}x_{2}}(N)+2 \mathsf{g}_{x_{3}}(N)\right)&&
\nn\\
Z_{S_{4}}~~=~~
\frac{1}{24} \left(x_{1}^{4}+6 x_{1}^{2} x_{2}+3 x_{2}^{2}+8 x_{1} x_{3}+6 x_{4}\right)
~~&\Rightarrow&~~\nn\\
\mathsf{g}^{D=4}(N)~~=~~
\frac{1}{24} \left(\mathsf{g}_{x_{1}^{4}}(N)+6 \mathsf{g}_{x_{1}^{2} x_{2}}(N)+3 \mathsf{g}_{x_{2}^{2}}(N)+8 \mathsf{g}_{x_{1} x_{3}}(N)+6 \mathsf{g}_{x_{4}}(N)\right)&&~~.
\eea
\\

\noindent\textbf{The Counting Algorithm}\\

In summary, the following algorithm is used to count distinct orbifolds of the form $\mathbb{C}^D/\Gamma_N$ and HNF's symmetric under cycles of $S_D$:
\\
\begin{center}
\includegraphics[trim= 0mm 40mm 0mm 10mm, totalheight=7.5cm]{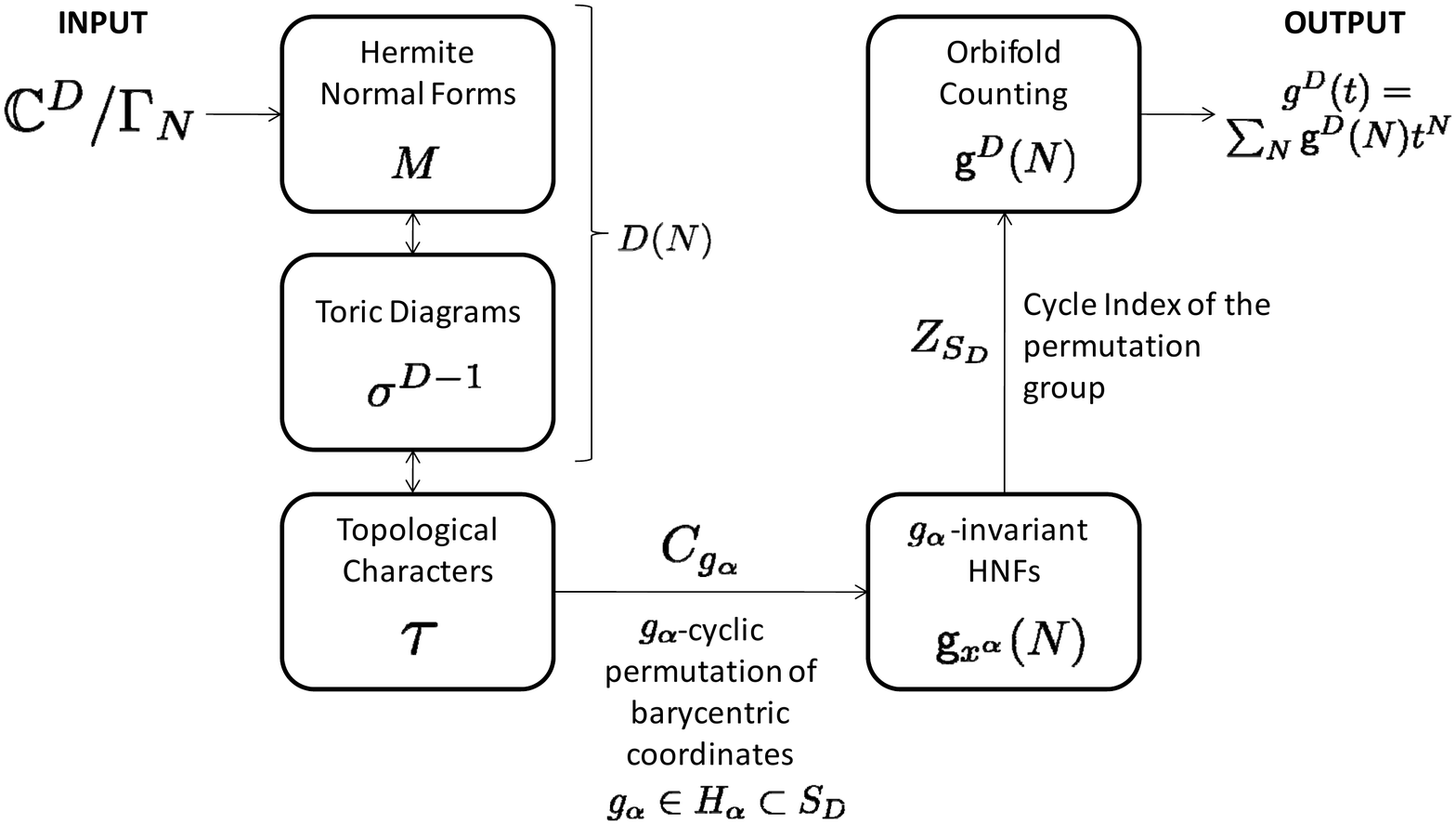}
\end{center}
The input of the algorithm is the dimension $D$ and the order $N$ of orbifolds of the form $\mathbb{C}^{D}/\Gamma_N$ where $\Gamma_N\subset SU(D)$. The output is the counting  $\mathsf{g}^{D}(N)$ of distinct abelian orbifolds of $\mathbb{C}^{D}$. A by-product is the counting $\mathsf{g}_{x^\alpha}(N)$ of HNF's which are invariant under the cycle $g_{\alpha}\in H_{\alpha}\subset S_D$ where $H_{\alpha}$ is a conjugacy class of $S_D$.\\

%%%%%%%%%%%%%%%%%%%%%%%%%%%%%%%%%%%%%%%%%%%%%%%%%%%%%%%%%%%%%%%%%%%%%
%%%%%%%%%%%%%%%%%%%%%%%%%%%%%%%%%%%%%%%%%%%%%%%%%%%%%%%%%%%%%%%%%%%%%

\section{The Symmetries of Abelian Orbifolds of $\mathbb{C}^{3}$, $\mathbb{C}^{4}$, $\mathbb{C}^{5}$ and $\mathbb{C}^{6}$ \label{s5}}

%%%%%%%%%%%%%%%%%%%%%%%%%%%%%%%%%%%%%%%%%%%%%%%%%%%%%%%%%%%%%%%%%%%%%%%%%
\subsection{Counting Symmetric Orbifolds}

Our explicit counting is presented in \tref{tc3sym} for $\mathbb{C}^{3}/\Gamma_{N}$, in \tref{tc4sym} for $\mathbb{C}^{4}/\Gamma_{N}$, in \tref{tc5sym} and \tref{tc5sym2} for $\mathbb{C}^{5}/\Gamma_{N}$ and in \tref{tc6sym} for $\mathbb{C}^{6}/\Gamma_{N}$.\\

The orbifold counting confirms the results presented in \cite{DaveyHananySeong10}. The sequences $\mathsf{g}_{x^\alpha}$ which count $g^\alpha$-symmetric HNF's of $\mathbb{C}^{3}/\Gamma_N$ and $\mathbb{C}^{4}/\Gamma_N$ also match the results in \cite{HananyOrlando10}. Accordingly, the counting method presented above gives a geometrical interpretation to the sequences in \cite{HananyOrlando10}.\\

%%%%%%%%%%%%%%%%%%%%%%%%%%%%%%%%%%%%%%%%%%%%%%%%%%%%%%%%%%%%%%%%%%%%%%%%%
\subsection{Partition Functions}

Let an infinite sequence $\mathsf{g}$ be expressed as a partition function $g(t)=\sum_{n=1}^{\infty}{\mathsf{g}(n)t^n}$. The partition functions $g^D(t)=\sum_{N=1}^{\infty}{\mathsf{g}^D(N) t^N}$ for sequences of $\mathbb{C}^{3}/\Gamma_N$ and $\mathbb{C}^{4}/\Gamma_N$ are presented in \cite{HananyOrlando10,DaveyHananySeong10}, and are summarized below.\\

\noindent\textbf{Partition Functions for $\mathbb{C}^{3}/\Gamma_N$.} The partition functions for the sequences that count $g^\alpha$-symmetric HNF's which correspond to abelian orbifolds of $\mathbb{C}^{3}$ can be presented in terms of rational functions $f(t)$. A partition function $g(t)$ is expressed as $g(t)=\sum_{k=1}^{\infty}f(t^k)$. The rational functions for the symmetries of $\mathbb{C}^{3}/\Gamma_N$ are
\beal{ex1}
&&
f_{x_{1}^{3}}(t)=
\frac{(1-t^6)}{(1-t)(1-t^2)(1-t^3)}-1
~~,~~
f_{x_{1}x_{2}}(t)=
\frac{(1+t^3)}{(1-t)(1+t^2)}-1 
~~,\nn\\
&&
f_{x_{3}}(t)=
\frac{(1-t^2)^2}{(1-t)(1-t^3)}-1 ~~,
\eea
such that the partition function for distinct $\mathbb{C}^{3}/\Gamma_N$ is
\beal{ex2}
g^{D=3}(t)~&=&~\sum_{k=1}^{\infty}{f^{D=3}(t^k)}\nn\\
~&=&~\frac{1}{6}\sum_{k=1}^{\infty}\left(f_{x_{1}^{3}}(t^k)+3f_{x_{1}x_{2}}(t^k)+2f_{x_3}(t^k)\right)~~.
\eea
The rational function for $g^{D=3}(t)$ is 
\beql{ex3}
f^{D=3}(t)~=~\frac{1}{(1-t)(1+t^2)(1-t^3)}-1~~.
\eeq
We note that the sequences which are generated in \eref{ex1} can be expressed as Dirichlet Series and in terms of Riemann zeta functions as shown in \cite{HananyOrlando10}.
\\

\noindent\textbf{Partition Functions for $\mathbb{C}^{4}/\Gamma_N$.} The rational functions for the symmetries of $\mathbb{C}^{4}/\Gamma_N$ are
\beal{ex4}
f_{x_{1}^{4}}(t)&=&
\sum_{n,m=1}^{\infty}{n m^2 t^{m n}}
~~,\nn\\
f_{x_{1}^{2}x_{2}}(t)&=&
\sum_{n,m=1}^{\infty}{m\left(t^{m n}-t^{2 m n}+4t^{4 m n}\right)}
~~,\nn\\
f_{x_{2}^{2}}(t)&=&
\sum_{n,m=1}^{\infty}{m\left(t^{m n}-t^{2 m n}+4 t^{4 m n}\right)}
~~,\nn\\
f_{x_{1}x_{3}}(t)&=&
\frac{1}{2}\left[\sum_{n,m=-\infty}^{\infty}{t^{n^2+4 m^{2}}}-1\right]
~~\nn\\
f_{x_{4}}(t)&=&
\frac{1}{2}\left[\sum_{n,m=-\infty}^{\infty}{t^{n^2+m n+7 m^2}}-1\right]~~.
\eea
These can also be expressed as Dirichlet Series and in terms of Riemann zeta functions. The partition function for distinct abelian orbifolds of $\mathbb{C}^{4}$ is
\beql{ex5}
g^{D=4}(t)~=~\frac{1}{24}\sum_{k=1}^{\infty}\left(
f_{x_{1}^{4}}(t^k)+6f_{x_{1}^{3}x_{2}}(t^k)+3f_{x_{2}^{2}}(t^k)+8f_{x_{1}x_{3}}(t^k)+6f_{x_{4}}(t^k)
\right)~~.
\eeq
\\

%%%%%%%%%%%%% C3 %%%%%%%%%%%%%%%%%%%%%%%%%%%%
\begin{table}[h!]
\begin{center}
\begin{tabular}{c|p{10mm}p{10mm}p{10mm}p{10mm}p{10mm}p{10mm}p{10mm}p{10mm}p{10mm}p{10mm}}
\hline
\hline
\multicolumn{11}{c}{$\mathbb{C}^{3}/\Gamma_{N}$}
\\
\hline
\hline
$N$ & 
1 & 2 & 3 & 4 & 5 & 6 & 7 & 8 & 9 & 10 	
\\
\hline
$\mathsf{g}_{x_{1}^{3}}$ & 
	1 &	3	&	4	&	7	&	6	&	12	&	8	&	15	&	13	& 18
\\
$\mathsf{g}_{x_{1}x_{2}}$ & 
1& 1& 2& 3& 2& 2& 2& 5& 3& 2
\\
$\mathsf{g}_{x_{3}}$ & 
1& 0& 1& 1& 0& 0& 2& 0& 1& 0
\\
\hline
$g^{D=3}$ & 
		1& 1& 2& 3& 2& 3& 3& 5& 4& 4
\\
\hline
\hline
$N$ & 
11 & 12 & 13 & 14 & 15 & 16 & 17 & 18 & 19 & 20 	
\\
\hline
$\mathsf{g}_{x_{1}^{3}}$ & 
12& 28& 14& 24& 24& 31& 18& 39& 20& 42
\\
$\mathsf{g}_{x_{1}x_{2}}$ & 
2& 6& 2& 2& 4& 7& 2& 3& 2& 6
\\
$\mathsf{g}_{x_{3}}$ & 
	0& 1& 2& 0& 0& 1& 0& 0& 2& 0
\\
\hline
$g^{D=3}$ & 
3& 8& 4& 5& 6& 9& 4& 8& 5& 10
\\
\hline
\hline
$N$ & 
21 & 22 & 23 & 24 & 25 & 26 & 27 & 28 & 29 & 30 	
\\
\hline
$\mathsf{g}_{x_{1}^{3}}$ & 
32& 36& 24& 60& 31& 42& 40& 56& 30& 72
\\
$\mathsf{g}_{x_{1}x_{2}}$ & 
4& 2& 2& 10& 3& 2& 4& 6& 2& 4
\\
$\mathsf{g}_{x_{3}}$ & 
2& 0& 0& 0& 1& 0& 1& 2& 0& 0
\\
\hline
$g^{D=3}$ & 
	8& 7& 5& 15& 7& 8& 9& 13& 6& 14
\\
\hline
\hline
$N$ & 
31 & 32 & 33 & 34 & 35 & 36 & 37 & 38 & 39 & 40 	
\\
\hline
$\mathsf{g}_{x_{1}^{3}}$ & 
32& 63& 48& 54& 48& 91& 38& 60& 56& 90
\\
$\mathsf{g}_{x_{1}x_{2}}$ & 
2& 9& 4& 2& 4& 9& 2& 2& 4& 10
\\
$\mathsf{g}_{x_{3}}$ & 
	2& 0& 0& 0& 0& 1& 2& 0& 2& 0
\\
\hline
$g^{D=3}$ & 
	7& 15& 10& 10& 10& 20& 8& 11&12& 20
\\
\hline
\hline
$N$ & 
41 & 42 & 43 & 44 & 45 & 46 & 47 & 48 & 49 & 50 	
\\
\hline
$\mathsf{g}_{x_{1}^{3}}$ & 
42& 96& 44& 84& 78& 72& 48& 124& 57& 93
\\
$\mathsf{g}_{x_{1}x_{2}}$ & 
2& 4& 2& 6& 6& 2& 2& 14& 3& 3,
\\
$g_{x_{3}}$ & 
	0& 0& 2& 0& 0& 0& 0& 1& 3& 0
\\
\hline
$g^{D=3}$ & 
	8& 18& 9& 17& 16& 13& 9& 28& 12& 17
\\
\hline
\hline
$N$ &
 51 & 52 & 53 & 54 & 55 & 56 & 57 & 58 & 59 & 60 	
\\
\hline
$\mathsf{g}_{x_{1}^{3}}$ & 
72& 98& 54& 120& 72& 120& 80& 90& 60& 168
\\
$\mathsf{g}_{x_{1}x_{2}}$ & 
4& 6& 2& 4& 4& 10& 4& 2& 2& 12
\\
$\mathsf{g}_{x_{3}}$ & 
	0& 2& 0& 0& 0& 0& 2& 0& 0& 0
\\
\hline
$g^{D=3}$ & 
	14& 20& 10& 22& 14& 25& 16& 16& 11& 34
\\
\hline
\end{tabular}
\caption{The symmetry count for $\mathbb{C}^{3}/\Gamma_{N}$ with cycle index $Z_{S_3}~=~\frac{1}{6}\left(x_{1}^{3}+3x_{1}x_{2}+2x_{3}\right)$.}
\label{tc3sym}
\end{center}
\end{table}

%%%%%%%%%%%%% C4 %%%%%%%%%%%%%%%%%%%%%%%%%%%%
\begin{table}[h!]
\begin{center}
\begin{tabular}{c|p{10mm}p{10mm}p{10mm}p{10mm}p{10mm}p{10mm}p{10mm}p{10mm}p{10mm}p{10mm}}
\hline
\hline
\multicolumn{11}{c}{$\mathbb{C}^{4}/\Gamma_{N}$}
\\
\hline
\hline
$N$ & 
1 & 2 & 3 & 4 & 5 & 6 & 7 & 8 & 9 & 10 	
\\
\hline
$\mathsf{g}_{x_{1}^{4}}$ & 
1 & 7 & 13 & 35 & 31 & 91 & 57 & 155 & 130 & 217 
\\
$\mathsf{g}_{x_{1}^{2}x_{2}}$ & 
1 & 3 & 5 & 11 & 7 & 15 & 9 & 31 & 18 & 21 
\\
$\mathsf{g}_{x_{2}^{2}}$ & 
1 & 3 & 5 & 11 & 7 & 15 & 9 & 31 & 18 & 21 
\\
$\mathsf{g}_{x_{1}x_{3}}$ & 
1 & 1 & 1 & 2 & 1 & 1 & 3 & 2 & 4 & 1 
\\
$\mathsf{g}_{x_{4}}$ & 
1 & 1 & 1 & 3 & 3 & 1 & 1 & 5 & 2 & 3 
\\
\hline
$g^{D=4}$ & 
1& 2& 3& 7& 5& 10& 7& 20& 14& 18
\\
\hline
\hline
$N$ & 
11 & 12 & 13 & 14 & 15 & 16 & 17 & 18 & 19 & 20
\\
\hline
$\mathsf{g}_{x_{1}^{4}}$ & 
133 & 455 & 183 & 399 & 403 & 651 & 307 & 910 & 381 & 1085
\\
$\mathsf{g}_{x_{1}^{2}x_{2}}$ & 
13 & 55 & 15 & 27 & 35 & 75 & 19 & 54 & 21 & 77
\\
$\mathsf{g}_{x_{2}^{2}}$ & 
13 & 55 & 15 & 27 & 35 & 75 & 19 & 54 & 21 & 77
\\
$\mathsf{g}_{x_{1}x_{3}}$ & 
1 & 2 & 3 & 3 & 1 & 3 & 1 & 4 & 3 & 2
\\
$\mathsf{g}_{x_{4}}$ & 
1 & 3 & 3 & 1 & 3 & 7 & 3 & 2 & 1 & 9
\\
\hline
$g^{D=4}$ & 
11& 41& 15& 28 & 31 & 58 &21 &60 &25 & 77
\\
\hline
\hline
$N$ & 
21 & 22 & 23 & 24 & 25 & 26 & 27 & 28 & 29 & 30
\\
\hline
$\mathsf{g}_{x_{1}^{4}}$ & 
741	&	931	&	553	&	2015	&	806	&	1281	&	1210	&	1995	&	871	& 2821
\\
$\mathsf{g}_{x_{1}^{2}x_{2}}$ & 
45	&	39	&	25	&	155	&	38	&	45	&	58	&	99	&	31	& 105
\\
$\mathsf{g}_{x_{2}^{2}}$ & 
45	&	39	&	25	&	155	&	38	&	45	&	58	&	99	&	31	& 105
\\
$\mathsf{g}_{x_{1}x_{3}}$ & 
3	&	1	&	1	&	2	&	2	&	3	&	7	&	6	&	1	& 1
\\
$\mathsf{g}_{x_{4}}$ & 
1	&	1	&	1	&	5	&	6	&	3	&	2	&	3	&	3	& 3
\\
\hline
$g^{D=4}$ & 
49	&	54	&	33	&	144	&	50	&	72	&	75	&	123	&	49	& 158
\\
\hline
\hline
$N$ & 
31 & 32 & 33 & 34 & 35 & 36 & 37 & 38 & 39 & 40
\\
\hline
$\mathsf{g}_{x_{1}^{4}}$ & 
993	&	2667	&	1729	&	2149	&	1767	&	4550	&	1407	&	2667	&	2379	& 4805
\\
$\mathsf{g}_{x_{1}^{2}x_{2}}$ & 
33	&	167	&	65	&	57	&	63	&	198	&	39	&	63	&	75	& 217
\\
$\mathsf{g}_{x_{2}^{2}}$ & 
33	&	167	&	65	&	57	&	63	&	198	&	39	&	63	&	75	& 217
\\
$\mathsf{g}_{x_{1}x_{3}}$ & 
3	&	3	&	1	&	1	&	3	&	8	&	3	&	3	&	3	& 2
\\
$\mathsf{g}_{x_{4}}$ & 
1	&	9	&	1	&	3	&	3	&	6	&	3	&	1	&	3	& 15
\\
\hline
$g^{D=4}$ & 
55	&	177	&	97	&	112	&	99	&	268	&	75	&	136	&	129	& 286
\\
\hline
\hline
$N$ & 
41 & 42 & 43 & 44 & 45 & 46 & 47 & 48 & 49 & 50
\\
\hline
$\mathsf{g}_{x_{1}^{4}}$ & 
1723	&	5187	&	1893	&	4655	&	4030	&	3871	&	2257	&	8463	&	2850	& 5642
\\
$\mathsf{g}_{x_{1}^{2}x_{2}}$ & 
43	&	135	&	45	&	143	&	126	&	75	&	49	&	375	&	66	& 114
\\
$\mathsf{g}_{x_{2}^{2}}$ & 
43	&	135	&	45	&	143	&	126	&	75	&	49	&	375	&	66	& 114
\\
$\mathsf{g}_{x_{1}x_{3}}$ & 
1	&	3	&	3	&	2	&	4	&	1	&	1	&	3	&	6	& 2
\\
$\mathsf{g}_{x_{4}}$ & 
3	&	1	&	1	&	3	&	6	&	1	&	1	&	7	&	2	& 6
\\
\hline
$g^{D=4}$ & 
89	&	268	&	97	&	249	&	218	&	190	&	113	&	496	&	146	& 280
\\
\hline
\hline
%$N$ & 
%51 & 52 & 53 & 54 & 55 & 56 & 57 & 58 & 59 & 60
%\\
%\hline
%$\mathsf{g}_{x_{1}^{4}}$ & 
%3991	&	6405	&	2863	&	8470	&	4123	&	8835	&	4953	&	6097	&	3541	& 14105
%\\
%$\mathsf{g}_{x_{1}^{2}x_{2}}$ & 
%95	&	165	&	55	&	174	&	91	&	279	&	105	&	93	&	61	& 385
%\\
%$\mathsf{g}_{x_{2}^{2}}$ & 
%95	&	165	&	55	&	174	&	91	&	279	&	105	&	93	&	61	& 385
%\\
%$\mathsf{g}_{x_{1}x_{3}}$ & 
%1	&	6	&	1	&	7	&	1	&	6	&	3	&	1	&	1	& 2
%\\
%$\mathsf{g}_{x_{4}}$ & 
%3	&	9	&	3	&	2	&	3	&	5	&	1	&	3	&	1	& 9
%\\
%\hline
%$g^{D=4}$ & 
%203	&	333	&	141	&	421	&	207	&	476	&	247	&	290	&	171	& 735
%\\
%\hline
%\hline
\end{tabular}
\caption{The symmetry count for $\mathbb{C}^{4}/\Gamma_{N}$ with cycle index $Z_{S_4}~=~	\frac{1}{24}\left(x_{1}^{4}+6x_{1}^{2}x_{2}+3x_{2}^{2}+8x_{1}x_{3}+6x_{4}\right)$.}
\label{tc4sym}
\end{center}
\end{table}

%%%%%%%%%%%%% C5 %%%%%%%%%%%%%%%%%%%%%%%%%%%%
\begin{table}[h!]
\begin{center}
\begin{tabular}{c|p{10mm}p{10mm}p{10mm}p{10mm}p{10mm}p{10mm}p{10mm}p{10mm}p{10mm}p{10mm}}
\hline
\hline
\multicolumn{11}{c}{$\mathbb{C}^{5}/\Gamma_{N}$}
\\
\hline
\hline
$N$ & 
1 & 2 & 3 & 4 & 5 & 6 & 7 & 8 & 9 & 10 
\\
\hline
$\mathsf{g}_{x_{1}^{5}}$ & 
1 & 15 & 40 & 155  	&156 &600 & 400	&1395 &1210 	& 2340
\\
$\mathsf{g}_{x_{1}^{3}x_{2}}$ & 
1 & 7 & 14 & 43  		&32	 &98 	& 58	&219 	&144	 	& 224
\\
$\mathsf{g}_{x_{1}x_{2}^{2}}$ & 
1 & 3 & 8 & 19 			&12	 &24 	& 16	&75 	&42 		& 36
\\
$\mathsf{g}_{x_{1}^{2}x_{3}}$ & 
1 & 3 & 4& 8				&6	 &12 	& 10	&18 	&22 		& 18
\\
$\mathsf{g}_{x_{2}x_{3}}$ & 
1 & 1 & 2& 4 				&2	 &2 	& 4		&6 		&6 			& 2
\\
$\mathsf{g}_{x_{1}x_{4}}$ & 
1 & 1 & 2& 3 				&4	 &2 	& 2		&7 		&4 			& 4
\\
$\mathsf{g}_{x_{5}}$ & 
1 & 0& 0& 0					&1	 &0 	& 0		&0 		&0 			& 0
\\
\hline
$g^{D=5}$ & 
1 & 2 & 4 & 10 			&8	 &19 	& 13	&45 	&33 		& 47
\\
\hline
\hline
$N$ & 
11 & 12 & 13 & 14 & 15 & 16 & 17 & 18 & 19 & 20 
\\
\hline
$\mathsf{g}_{x_{1}^{5}}$ & 
1464 	&6200 	&2380 	&6000 	&6240 	&11811 	& 5220 	&18150 	&7240 	&24180
\\
$\mathsf{g}_{x_{1}^{3}x_{2}}$ & 
134 	&602 		&184 		&406 		&448 		&995		& 308		&1008 	& 382		&1376 
\\
$\mathsf{g}_{x_{1}x_{2}^{2}}$ & 
24 		&152 		&28 		&48 		&96 		&251 		& 36		&126 		& 40		&228
\\
$\mathsf{g}_{x_{1}^{2}x_{3}}$ & 
12 		&32 		&16 		&30 		&24 		&39 		& 18 		&66 		& 22		&48
\\
$\mathsf{g}_{x_{2}x_{3}}$ & 
2 		&8 			&4 			&4 			&4 			&11 		& 2			&6	 		& 4			&8
\\
$\mathsf{g}_{x_{1}x_{4}}$ & 
2 		&6 			&4 			&2 			&8 			&19 		& 4			&4 			& 2		 	&12
\\
$\mathsf{g}_{x_{5}}$ & 
4 		&0 			&0 			&0 			&0 			&1 			& 0			&0 			&	0		 	&0
\\
\hline
$g^{D=5}$ & 
30 	&129 			&43		 	&96 		&108 		&226 		&78			&264 		&102 		&357
\\
\hline
\hline
$N$ & 
21 & 22 & 23 & 24 & 25 & 26 & 27 & 28 & 29 & 30 
\\
\hline
$\mathsf{g}_{x_{1}^{5}}$ & 
16000 & 21960	& 12720	& 55800		& 20306		& 35700		& 33880		& 62000		& 25260		& 93600
\\
$\mathsf{g}_{x_{1}^{3}x_{2}}$ & 
812 	& 938		& 554		& 3066		& 838			& 1288		& 1354		& 2494		& 872		& 3136
\\
$\mathsf{g}_{x_{1}x_{2}^{2}}$ & 
 	128	& 72		& 48		& 600		& 98			& 84		& 184		& 304		& 60		&288
\\
$\mathsf{g}_{x_{1}^{2}x_{3}}$ & 
 	40	& 36		& 24		& 72		& 32			& 48		& 85		& 80		& 30		&72
\\
$\mathsf{g}_{x_{2}x_{3}}$ & 
 	8		& 2		& 	2			& 12		& 4				& 4		& 13		& 16		& 2		&4 
\\
$\mathsf{g}_{x_{1}x_{4}}$ & 
 	4		& 2		& 		2		& 14		& 10			& 4		& 6		& 6		& 4		&8
\\
$\mathsf{g}_{x_{5}}$ & 
 	0		& 0		& 	0		& 0			& 1				& 0		& 0		& 0		& 0		&0
\\
\hline
$g^{D=5}$ & 
 	226	& 277		& 163	& 813		& 260				& 425		& 436		& 780		& 297		&1092
\\
\hline
\hline
$N$ & 
31 & 32 & 33 & 34 & 35 & 36 & 37 & 38 & 39 & 40 
\\
\hline
$\mathsf{g}_{x_{1}^{5}}$ & 
30784	&	97155 &	58560	&	78300	&	62400	&	187550	&	52060	&	108600	&	95200	& 217620
\\
$\mathsf{g}_{x_{1}^{3}x_{2}}$ & 
994		&	4251 &	1876	&	2156	&	1856	&	6192		&	1408	&	2674	&	2576	& 7008
\\
$\mathsf{g}_{x_{1}x_{2}^{2}}$ & 
64 		&	747
	&	192		&	108	&	192	&	798					&	76	&	120	&	224	& 900
\\
$\mathsf{g}_{x_{1}^{2}x_{3}}$ & 
34 		&	81 &	48		&	54	&	60	&	176					&	40	&	66	&	64	& 108
\\
$\mathsf{g}_{x_{2}x_{3}}$ & 
4 		&	15
	&	4			&	2	&	8	&	24							&	4	&	4	&	8	& 12
\\
$\mathsf{g}_{x_{1}x_{4}}$ & 
2 		&	31
	&	4			&	4	&	8	&	12								&	4	&	2	&	8	& 28
\\
$\mathsf{g}_{x_{5}}$ & 
4	 	&	0
	&	0			&	0	&	0	&	0								&	0	&	0	&	0	& 0
\\
\hline
$g^{D=5}$ & 
355	&	1281	&	678	&	856	&	712	&	2202		&	569	&	1155	& 1050		& 2537
\\
\hline
\hline
\end{tabular}
\caption{The symmetry count for $\mathbb{C}^{5}/\Gamma_{N}$ with cycle index $Z_{S_5}$ \textbf{(Part 1/2)}.
%~=~\frac{1}{120} \left(x_1^5+10 x_1^3 x_2+15 x_1 x_2^2+20 x_1^2 x_3+20 x_2 x_3+30 x_1 x_4+24 x_5\right)$.
}
\label{tc5sym}
\end{center}
\end{table}

%%%%%%%%%%%%% C5 %%%%%%%%%%%%%%%%%%%%%%%%%%%%
\begin{table}[h!]
\begin{center}
\begin{tabular}{c|p{10mm}p{10mm}p{10mm}p{10mm}p{10mm}p{10mm}p{10mm}p{10mm}p{10mm}p{10mm}}
\hline
\hline
\multicolumn{11}{c}{$\mathbb{C}^{5}/\Gamma_{N}$}
\\
\hline
\hline
$N$ & 
41 & 42 & 43 & 44 & 45 & 46 & 47 & 48 & 49 & 50 
\\
\hline
$\mathsf{g}_{x_{1}^{5}}$ & 
70644		&	240000	&	81400	&	226920	&	188760	&	190800	&	106080	&	472440	&	140050	& 304590
\\
$\mathsf{g}_{x_{1}^{3}x_{2}}$ & 
1724	&	5684	&	1894	&	5762	&	4608	&	3878	&	2258	&	13930	&	2908	& 5866
\\
$\mathsf{g}_{x_{1}x_{2}^{2}}$ & 
	84	&	384	&	88	&	456	&	504	&	144	&	96	&	2008	&	178	& 294
\\
$\mathsf{g}_{x_{1}^{2}x_{3}}$ & 
	42	&	120	&	46	&	96	&	132	&	72	&	48	&	156	&	76	& 96
\\
$\mathsf{g}_{x_{2}x_{3}}$ & 
	2	&	8	&	4	&	8	&	12	&	2	&	2	&	22	&	10	& 4
\\
$\mathsf{g}_{x_{1}x_{4}}$ & 
	4	&	4	&	2	&	6	&	16	&	2	&	2	&	38	&	4	& 10
\\
$\mathsf{g}_{x_{5}}$ & 
	4	&	0	&	0	&	0	&	0	&	0	&	0	&	0	&	0	& 0
\\
\hline
$g^{D=5}$ & 
	752	&	2544	&	856	&	2447	&	2048	&	1944	&	1093	&	5388	&	1447	& 3083
\\
\hline
\hline
$N$ & 
51 & 52 & 53 & 54 & 55 & 56 & 57 & 58 & 59 & 60 
\\
\hline
$\mathsf{g}_{x_{1}^{5}}$ & 
208800		&	368900	&	151740	&	508200	&	228384	&	558000	&	289600	&	378900	&	208920	& 967200
\\
$\mathsf{g}_{x_{1}^{3}x_{2}}$ & 
4312	&	7912	&	2864	&	9478	&	4288	&	12702	&	5348	&	6104	&	3542	& 19264
\\
$\mathsf{g}_{x_{1}x_{2}^{2}}$ & 
	288	&	532	&	108	&	552	&	288	&	1200	&	320	&	180	&	120	& 1824
\\
$\mathsf{g}_{x_{1}^{2}x_{3}}$ & 
	72	&	128	&	54	&	255	&	72	&	180	&	88	&	90 &60	& 192
\\
$\mathsf{g}_{x_{2}x_{3}}$ & 
	4	&	16	&	2	&	13	&	4	&	24	&	8	&	2	&	2	& 16
\\
$\mathsf{g}_{x_{1}x_{4}}$ & 
	8	&	12	&	4	&	6	&	8	&	14	&	4	&	4	&	2	& 24
\\
$\mathsf{g}_{x_{5}}$ & 
	0	&	0	&	0	&	0	&	4	&	0	&	0	&	0	&	0	& 0
\\
\hline
$g^{D=5}$ & 
	2150	&	3827	&	1527	&	5140	&	2312	&	5896	&	2916	&	3705	&	2062	& 9934
\\
\hline
\hline
$N$ & 
61 & 62 & 63 & 64 & 65 & 66 & 67 & 68 & 69 & 70 
\\
\hline
$\mathsf{g}_{x_{1}^{5}}$ & 
230764		&	461760	&	484000	&	788035	&	371280	&	878400	&	305320	&	809100	&	508800	& 936000
\\
$\mathsf{g}_{x_{1}^{3}x_{2}}$ & 
3784	&	6958	&	8352	&	17587	&	5888	&	13132	&	4558	&	13244	&	7756	& 12992
\\
$\mathsf{g}_{x_{1}x_{2}^{2}}$ & 
	124	&	192	&	672	&	2043	&	336	&	576	&	136	&	684	&	384	& 576
\\
$\mathsf{g}_{x_{1}^{2}x_{3}}$ & 
	64	&	102	&	220	&	166	&	96	&	144	&	70	&	144	&	96	& 180
\\
$\mathsf{g}_{x_{2}x_{3}}$ & 
	4	&	4	&	24	&	22	&	8	&	4	&	4	&	8	&	4	& 8
\\
$\mathsf{g}_{x_{1}x_{4}}$ & 
	4	&	2	&	8	&	51	&	16	&	4	&	2	&	12	&	4	& 8
\\
$\mathsf{g}_{x_{5}}$ & 
	4	&	0	&	0	&	0	&	0	&	0	&	0	&	0	&	0	& 0
\\
\hline
$g^{D=5}$ & 
	2267	&	4470	&	4856	&	8332	&	3684	&	8512	&	2954	&	7960	&	4952	& 8988
\\
\hline
\hline
$N$ & 
71 & 72 & 73 & 74 & 75 & 76 & 77 & 78 & 79 & 80 
\\
\hline
$\mathsf{g}_{x_{1}^{5}}$ & 
363024		&	\footnotesize{1687950}	&	394420	&	780900	&	812240	&	\footnotesize{1122200}	&	585600	&	\footnotesize{1428000}	&	499360	& \footnotesize{1842516}
\\
$\mathsf{g}_{x_{1}^{3}x_{2}}$ & 
	5114	&	31536	&	5404	&	9856	&	11732	&	16426	&	7772	&	18032	&	6322	& 31840
\\
$\mathsf{g}_{x_{1}x_{2}^{2}}$ & 
	144	&	3150	&	148	&	228	&	784	&	760	&	384	&	672	&	160	& 3012
\\
$\mathsf{g}_{x_{1}^{2}x_{3}}$ & 
	72	&	396	&	76	& 120 &	128	&	176	&	120	&	192	&	82	&	234	
\\
$\mathsf{g}_{x_{2}x_{3}}$ & 
	2	&	36	&	4	&	4	&	8	&	16	&	8	&	8	&	4	& 22
\\
$\mathsf{g}_{x_{1}x_{4}}$ & 
	2	&	28	&	4	&	4	&	20	&	6	&	4	&	8	&	2	& 76
\\
$\mathsf{g}_{x_{5}}$ & 
	4	&	0	&	0	& 0	&	0	&	0	&	0	&	0	&	0	& 1
\\
\hline
$g^{D=5}$ & 
	3483	&	17167	&	3770	&	7379	&	7872	&	10849	&	5598	&	13522	&	4723	& 18446
\\
\hline
\hline
\end{tabular}
\caption{The symmetry count for $\mathbb{C}^{5}/\Gamma_{N}$ with cycle index $Z_{S_5}$ \textbf{(Part 2/2)}.
%~=~\frac{1}{120} \left(x_1^5+10 x_1^3 x_2+15 x_1 x_2^2+20 x_1^2 x_3+20 x_2 x_3+30 x_1 x_4+24 x_5\right)$.
}
\label{tc5sym2}
\end{center}
\end{table}

%%%%%%%%%%%%% C6 %%%%%%%%%%%%%%%%%%%%%%%%%%%%
\begin{table}[h!]
\begin{center}
\begin{tabular}{c|p{10mm}p{10mm}p{10mm}p{10mm}p{10mm}p{10mm}p{10mm}p{10mm}p{10mm}p{10mm}}
\hline
\hline
\multicolumn{11}{c}{$\mathbb{C}^{6}/\Gamma_{N}$}
\\
\hline
\hline
$N$ & 
1 & 2 & 3 & 4 & 5 & 6 & 7 & 8 & 9 & 10 	
\\
\hline
$\mathsf{g}_{x_{1}^{6}}$ & 
	1	&	31		&	121			&	651			&		781		&	3751			&	2801			&	11811			&	11011			&24211
\\
$\mathsf{g}_{x_{1}^{4}x_{2}}$ & 
	1	&	15		&	41			&	171			&		157		&	615			&		401		&	1651			&	1251			&2355
\\
$\mathsf{g}_{x_{1}^{2}x_{2}^{2}}$ & 
	1	&	7			&	17			&	59			&		37		&	119			&		65		&	371			&	195			&259
\\
$\mathsf{g}_{x_{2}^{3}}$ & 
	1	&	7			&	17			&	59			&		37		&	119			&		65		&	371			&	195			&259
\\
$\mathsf{g}_{x_{1}^{3}x_{3}}$ & 
	1	&	7			&	13			&	36			&		31		&	91			&		59		&	162			&	157			&217
\\
$\mathsf{g}_{x_{1}x_{2}x_{3}}$ & 
	1	&	3			&	5			&	12			&		7		&			15	&				11&			34	&	27			&21
\\
$\mathsf{g}_{x_{1}^{2}x_{4}}$ & 
	1	&	3			&	5			&	11			&		9		&			15	&				9&			35	&	19			&27
\\
$\mathsf{g}_{x_{2}x_{4}}$ & 
	1	&	3			&	5			&	11			&		9		&			15	&				9&			35	&	19			&27
\\
$\mathsf{g}_{x_{3}^{2}}$ & 
	1	&	1			&	4			&	6			&			1	&				4&				17&				6& 22				&1
\\
$\mathsf{g}_{x_{1}x_{5}}$ & 
	1	&	1			&	1			&	1			&			1	&				1&				1&				1&	1			&1
\\
$\mathsf{g}_{x_{6}}$ & 
	1	&	1			&	2			&	2			&			1	&				2&				5&				2&	6			&1
\\
\hline
$g^{D=6}$ & 
	1	&	3			&	6			&	17			&		13		&		40		&			27	&		106		&		78		&127
\\
\hline
\hline
$N$ & 
11 & 12 & 13 & 14 & 15 & 16 & 17 & 18 & 19 & 20 	
\\
\hline
$\mathsf{g}_{x_{1}^{6}}$ & 
	16105	&	78771		&		30941	&	86831		&	94501		&	200787		&	88741		&	341341		&	137561		&508431
\\
$\mathsf{g}_{x_{1}^{4}x_{2}}$ & 
	1465	&	7011		&		2381	&	6015		&	6437		&	14547		&	5221		&		18765			&	7241		&26847
\\
$\mathsf{g}_{x_{1}^{2}x_{2}^{2}}$ & 
	145	&		1003	&		197	&	455		&	629						&	1987		&	325		&		1365				&	401		&2183
\\
$\mathsf{g}_{x_{2}^{3}}$ & 
	145	&		1003	&		197	&	455		&	629						&	1987		&	325		&	1365					&	401		&2183
\\
$\mathsf{g}_{x_{1}^{3}x_{3}}$ & 
	133	&		468	&		185	&	413		&	403							&	687		&	307		&		1099					&	383		&1116
\\
$\mathsf{g}_{x_{1}x_{2}x_{3}}$ & 
	13	&		60	&		17	&	33		&	35							&	87		&	19		&	81							&	23		&84
\\
$\mathsf{g}_{x_{1}^{2}x_{4}}$ & 
	13	&		55	&		17	&	27		&	45							&	115		&	21		&	57							&	21		&99
\\
$\mathsf{g}_{x_{2}x_{4}}$ & 
	13	&		55	&		17	&	27		&	45							&	115		&	21		&	57							&	21		&99
\\
$\mathsf{g}_{x_{3}^{2}}$ & 
	1	&			24&		29	&	17		&		4								&	27		&	1		&	22							&	41		&6
\\
$\mathsf{g}_{x_{1}x_{5}}$ & 
	5	&		1	&		1	&	1		&	1												&	2		&	1		&	1										&	1		&1
\\
$\mathsf{g}_{x_{6}}$ & 
	1	&		4	&		5	&	5		&	2												&	7		&	1		&	6										&	5		&2
\\
\hline
$g^{D=6}$ & 
79		&		391	&	129		&		321	&	358							&	832		&	285		&	1070						&	409		& 1549
\\
\hline
\hline
\end{tabular}
\caption{The symmetry count for $\mathbb{C}^{6}/\Gamma_{N}$ with cycle index $Z_{S_6}$.}
\label{tc6sym}
\end{center}
\end{table}

\clearpage

%%%%%%%%%%%%%%%%%%%%%%%%%%%%%%%%%%%%%%%%%%%%%%%%%%%%%%%%%%%%%%%%%%%%%%%%%%%%%%%%%
%%%%%%%%%%%%%%%%%%%%%%%%%%%%%%%%%%%%%%%%%%%%%%%%%%%%%%%%%%%%%%%%%%%%%%%%%%%%%%%%%
\section{Prime Index Sequences and Series Convolutions \label{sp1}}

%%%%%%%%%%%%%%%%%%%%%%%%%%%%%%%%%%%%%%%%%%%%%%%%%%%%%%%%%%%%%%%%%%%%%%%%%
\subsection{Series Convolutions}

Sequences that count $g^\alpha$-symmetric HNF's which correspond to abelian orbifolds of $\mathbb{C}^{D}$ can be expressed in terms of sequence convolutions. A sequence $\mathsf{g}=\{\mathsf{g}(1),\mathsf{g}(2),\mathsf{g}(3),\dots\}$ is related to its corresponding partition function by $g(t)=\sum_{n=1}^{\infty}{\mathsf{g}(n)t^n}$.
\\

\noindent\textbf{Partition Functions and Sequence Convolutions.} As outlined in \cite{HananyOrlando10} and \cite{He:2010mh}, given a sequence $\mathsf{q}=\mathsf{r} * \mathsf{s}$ generated by a convolution of the sequences $\mathsf{r}$ and $\mathsf{s}$, the partition function for the sequence $\mathsf{q}$, $q(t)$, is expressed as,
\beql{es1}
q(t)~=~\sum_{m,k=1}^{\infty}{\mathsf{r}(m) \mathsf{s}(k) t^{m k}}
~=~ \sum_{m=1}^{\infty}{\mathsf{r}(m)s(t^m)}
~=~ \sum_{m=1}^{\infty}{\mathsf{s}(m)r(t^m)} ~~,
\eeq
where $r(t)$ and $s(t)$ are the partition functions of the sequences $\mathsf{r}$ and $\mathsf{s}$ respectively. We invert \eref{es1} as follows
\beql{ess1b}
r(t) ~=~ \sum_{m=1}^{\infty}{q(t^k)\mathsf{s}(k)\mu(k)}~~,
\eeq
where $\mu(n)$ is the M\"{o}bius function. It is expected that the above inversion is valid for particular sequences $\mathsf{r}$ and $\mathsf{s}$ which are discussed and used below.
\\

\noindent\textbf{Multiplicative Sequences.} As first noted in \cite{HananyOrlando10}, the sequences $\mathsf{g}_{x^\alpha}$ in Tables \ref{tc3sym}-\ref{tc6sym} which count $g^\alpha$-symmetric HNF's are \textit{multiplicative}. Multiplicativity of $\mathsf{g}_{x^\alpha}$ says that given two integers $q_1$ and $q_2$ with $\gcd{(q_1,q_2)}=1$, we have
\beql{ee1}
\mathsf{g}_{x^\alpha}(q_1)\mathsf{g}_{x^\alpha}(q_2) = \mathsf{g}_{x^\alpha}(q_1 q_2)~~.
\eeq
This property can be seen from the counting of orbifold symmetries and is related to the convolution property in \eref{es1}.
\\

\noindent\textbf{Standard Sequences.} Convolution preserves multiplicativity, and therefore it is useful to discuss basic multiplicative sequences. 
\begin{itemize}
	\item The unit sequence:
	\beql{exs1}
	\mathsf{u}=\{1,1,1,\dots\} ~\Leftrightarrow~ u(t)=\sum_{n=1}^{\infty}{t^n}=t+t^2+t^3+\dots
	\eeq
	
	\item The natural number sequence:
	\beql{exs2}
	\mathsf{N}=\{1,2,3,\dots\} ~\Leftrightarrow~ N(t)=\sum_{n=1}^{\infty}{n t^n}=t+2t^2+3t^3+\dots
	\eeq
	
	\item Powers of the natural number sequence:
	\beql{exs2}
	\mathsf{N}^d=\{1^d,2^d,3^d,\dots\} ~\Leftrightarrow~ N^d(t)=\sum_{n=1}^{\infty}{n^d t^n}=t+2^d t^2+3^d t^3+\dots~~,
	\eeq
	where $\mathsf{N}^{0}=\mathsf{u}$.\\
	
	\item The Dirichlet character $\chi_{k,m}$ of modulo $k$ and index $m$ is defined under the conditions
	\beal{exs3}
	\chi_{k,m}(1)&=&1 \nn\\
	\chi_{k,m}(a)&=&\chi_{k,m}(a+k) \nn\\
	\chi_{k,m}(a)\chi_{k,m}(b) &=& \chi_{k,m}(a b) \nn\\
	\chi_{k,m}(a) &=& 0 ~~~~\mbox{if}~~\gcd{(k,a)\neq 1} ~~.
	\eea
	Under these conditions there are several solutions which are parameterized by $m$. The Dirichlet characters up to modulo $10$ used in this work are\\
	\begin{tabular}{ll}
	$\chi_{1,1} = \mathsf{u}$		&
	$\chi_{8,1} = \{1,0,1,0,1,0,1,0,\dots\}$
	\\
	$\chi_{2,1} = \{1,0,\dots\}$	&
	$\chi_{8,2} = \{1,0,1,0,-1,0,-1,0,\dots\}$
	\\
	$\chi_{3,1} = \{1,1,0,\dots\}$ &
	$\chi_{8,3} = \{1,0,-1,0,1,0,-1,0,\dots\}$
	\\
	$\chi_{3,2} = \{1,-1,0,\dots\}$ &
	$\chi_{8,4} = \{1,0,-1,0,-1,0,1,0,\dots\}$
	\\
	$\chi_{4,1} = \{1,0,1,0,\dots\}$ &
	$\chi_{9,1} = \{1,1,0,1,1,0,1,1,0,\dots\}$
	\\
	$\chi_{4,2} = \{1,0,-1,0,\dots\}$ &
	$\chi_{9,2} = \{1,\omega,0,\omega^2,-\omega^2,0,-\omega,-1,0,\dots\}$
	\\
	$\chi_{5,1} = \{1,1,1,1,0,\dots\}$ &
	$\chi_{9,3} = \{1,\omega^2,0,-\omega,-\omega,0,\omega^2,1,0,\dots\}$
	\\
	$\chi_{5,2} = \{1,i,-i,-1,0,\dots\}$ &
	$\chi_{9,4} = \{1,-1,0,1,-1,0,1,-1,0,\dots\}$
	\\
	$\chi_{5,3} = \{1,-1,-1,1,0,\dots\}$ &
	$\chi_{9,5} = \{1,-\omega,0,\omega^2,\omega^2,0,-\omega,1,0,\dots\}$
	\\
	$\chi_{5,4} = \{1,-i,i,-1,0,\dots\}$ &
	$\chi_{9,6} = \{1,-\omega^2,0,-\omega,\omega,0,\omega^2,-1,0,\dots\}$
	\\
	$\chi_{6,1} = \{1,0,0,0,1,0\dots\}$ &
	$\chi_{10,1} = \{1,0,1,0,0,0,1,0,1,0,\dots\}$
	\\
	$\chi_{6,2} = \{1,0,0,0,-1,0\dots\}$ &
	$\chi_{10,2} = \{1,0,i,0,0,0,-i,0,-1,0,\dots\}$
	\\
	$\chi_{7,1} = \{1,1,1,1,1,1,0\dots\}$ &
	$\chi_{10,3} = \{1,0,-1,0,0,0,-1,0,1,0,\dots\}$
	\\
	$\chi_{7,2} = \{1,-\omega,\omega^2,\omega^2,-\omega,1,0\dots\}$ &
	$\chi_{10,4} = \{1,0,-i,0,0,0,i,0,-1,0,\dots\}$
	\\
	$\chi_{7,3} = \{1,\omega^2,\omega,-\omega,-\omega^2,-1,0\dots\}$ &
	\\
	$\chi_{7,4} = \{1,1,-1,1,-1,-1,0\dots\}$ &
	\\
	$\chi_{7,5} = \{1,-\omega,-\omega^2,\omega^2,\omega,-1,0\dots\}$ &
	\\
	$\chi_{7,6} = \{1,\omega^2,-\omega,-\omega,\omega^2,1,0\dots\}$ &
	
	\end{tabular}\\
	where the first elements given above are the periods of the infinite sequences, and $\omega=\exp{\frac{i\pi}{3}}$.\\	
\end{itemize}

The number of distinct Dirichlet characters of period $k$ is given by the Euler totient function $\varphi(k)$. It is defined as the number of integers less than or equal to $k$ which are co-prime to $k$. For primes $p$, the totient function takes the values
\beql{exs4bb} 
\varphi(p)=p-1~~.
\eeq
Moreover, the direct sum of all distinct Dirichlet characters of period $k$ is given by
\beql{exs4bc}
\sum_{m=1}^{\varphi(k)}{\chi_{k,m}(n)}=\varphi(k)~\delta_{n,1 \bmod{k}}+\delta_{kn}~~.
\eeq

The totient function $\varphi$ is related to the natural number sequence $\mathsf{N}$ under
\beql{exs4bd}
\varphi * \mathsf{u} = \mathsf{N} \Leftrightarrow  \varphi = \mu * \mathsf{N} ~~.
\eeq
With $\mathsf{N}$ being a multiplicative sequence, both the Euler totient function $\varphi(n)$ and M\"{o}bius function $\mu(n)$ are multiplicative.
\\

A direct product of any of the above multiplicative sequences,
\beal{exs5}
\mathsf{A}\mathsf{B}=\{\mathsf{A}(1)\mathsf{B}(1), \mathsf{A}(2)\mathsf{B}(2), \mathsf{A}(3)\mathsf{B}(3), \dots \}
\Leftrightarrow
\mathsf{A}\mathsf{B}(n)=\mathsf{A}(n)\mathsf{B}(n)
~~,
\eea
is a multiplicative sequence as well. An example is the direct product of $\chi_{3,2}$ and $\mathsf{N}$ which gives
\beql{exs5b}
\mathsf{N}\chi_{3,2}=\{1,-2,0,4,-5,0,7,-8,0,\dots\}~~.
\eeq
Furthermore, the direct product of two Dirichlet characters is another Dirichlet character.
\\

%%%%%%%%%%%%%%%%%%%%%%%%%%%%%%%%%%%%%%%%%%%%%%%%%%%%%%%%%%%%%%%%%%%%%%%%%
\subsection{Functions on Primes for Prime Index Sequences \label{sp11}}

Multiplicative sequences are determined by their values at indices which are prime numbers or pure powers of prime. The values on prime indices of sequences in \tref{tc3sym} to \tref{tc6sym} for orbifolds of $\mathbb{C}^{3}$ to $\mathbb{C}^{6}$ are shown in \tref{primet} and \tref{primet2}. 
\\

It is of interest to find for a given sequence $\mathsf{g}_{x^\alpha}(p)$ in \tref{primet} and \tref{primet2} a function on primes $p$, $P_{\mathsf{g}_{x^\alpha}}(p)$, which takes the values $P_{\mathsf{g}_{x^\alpha}}(p)=\mathsf{g}_{x^\alpha}(p)$. 
\\

\begin{table}[ht!]
\begin{center}
\begin{tabular}{c|p{5mm}p{5mm}p{5mm}p{5mm}p{5mm}p{5mm}p{5mm}p{5mm}p{5mm}p{5mm}p{5mm}p{5mm}p{5mm}p{5mm}p{5mm}p{5mm}}
\hline
\hline
\multicolumn{17}{c}{$\mathbb{C}^{3}/\Gamma_{N}$}
\\
\hline
\hline
$N=p$ & 
2 & 3 & 5 & 7 & 11 & 13 & 17 & 19 
& 23 & 29 & 31 & 37 & 41 & 43 & 47 & 53
\\
\hline
$\mathsf{g}_{x_{1}^{3}}$ & 
3 & 4 & 6 & 8 & 12 & 14 & 18 & 20
& 24 & 30 & 32 & 38 & 42 & 44 & 48 & 54
\\
$\mathsf{g}_{x_{1}x_{2}}$ & 
1 & 2 & 2 & 2 & 2 & 2 & 2 & 2
& 2 & 2 & 2 & 2 & 2 & 2 & 2 & 2
\\
$\mathsf{g}_{x_{3}}$ & 
0 & 1 & 0 & 2 & 0 & 2 & 0 & 2
& 0 & 0 & 2 & 2 & 0 & 2 & 0 & 0
%\\
%\hline
%$g^{D=3}$ & 
\\
\hline
\hline
\multicolumn{17}{c}{$\mathbb{C}^{4}/\Gamma_{N}$}
\\
\hline
\hline
$N=p$ & 
2 & 3 & 5 & 7 & 11 & 13 & 17 & 19 
& 23 & 29 & 31 & 37 & 41 & 43 & 47 & 53
\\
\hline
$\mathsf{g}_{x_{1}^{4}}$ & 
7 & 13 & 31 & 57 & 133 & 183 & 307 & 381
& 553 & 871 & 993 & 1407 & 1723 & 1893 & 2257 & 2863
\\
$\mathsf{g}_{x_{1}^{2}x_{2}}$ & 
3 & 5 & 7 & 9 & 13 & 15 & 19 & 21
& 25 & 31 & 33 & 39 & 43 & 45 & 49 & 55
\\
$\mathsf{g}_{x_{2}^{2}}$ & 
3 & 5 & 7 & 9 & 13 & 15 & 19 & 21
& 25 & 31 & 33 & 39 & 43 & 45 & 49 & 55
\\
$\mathsf{g}_{x_{1}x_{3}}$ & 
1 & 1 & 1 & 3 & 1 & 3 & 1 & 3
& 1 & 1 & 3 & 3 & 1 & 3 & 1 & 1
\\
$\mathsf{g}_{x_{4}}$ & 
1 & 1 & 3 & 1 & 1 & 3 & 3 & 1
& 1 & 3 & 1 & 3 & 3 & 1 & 1 & 3
\\
%\hline
%$g^{D=4}$ & 
%\\
\hline
\hline
\multicolumn{17}{c}{$\mathbb{C}^{5}/\Gamma_{N}$}
\\
\hline
\hline
$N=p$ & 
2 & 3 & 5 & 7 & 11 & 13 & 17 & 19 
& 23 & 29 & 31 & 37 & 41 & 43 & 47 & 53
\\
\hline
$\mathsf{g}_{x_{1}^{5}}$ & 
15 & 40 & 156 & 400 & 1464 & 2380 & 5220 & 7240
& \footnotesize{12720} & \footnotesize{25260} & \footnotesize{30784} & \footnotesize{52060} & \footnotesize{70644} & \footnotesize{81400} & \tiny{106080} & \tiny{151740}
\\
$\mathsf{g}_{x_{1}^{3}x_{2}}$ & 
7 & 14 & 32 & 58 & 134 & 184 & 308 & 382 
& 554 & 872 & 994 & 1408 & 1724 & 1894 & 2258 & 2864
\\
$\mathsf{g}_{x_{1}x_{2}^{2}}$ & 
3 & 8 & 12 & 16 & 24 & 28 & 36 & 40
& 48 & 60 & 64 & 76 & 84 & 88 & 96 & 108
\\
$\mathsf{g}_{x_{1}^{2}x_{3}}$ & 
3 & 4 & 6 & 10 & 12 & 16 & 18 & 22
& 24 & 30 & 34 & 40 & 42 & 46 & 48 & 54
\\
$\mathsf{g}_{x_{2}x_{3}}$ & 
1 & 2 & 2 & 4 & 2 & 4 & 2 & 4
& 2 & 2 & 4 & 4 & 2 & 4 & 2 & 2
\\
$\mathsf{g}_{x_{1}x_{4}}$ & 
1 & 2 & 4 & 2 & 2 & 4 & 4 & 2
& 2 & 4 & 2 & 4 & 4 & 2 & 2 & 4
\\
$\mathsf{g}_{x_{5}}$ & 
0 & 0 & 1 & 0 & 4 & 0 & 0 & 0
& 0 & 0 & 4 & 0 & 4 & 0 & 0 & 0
\\
\hline
\hline
\end{tabular}
\caption{Sequences of $\mathbb{C}^{3}/\Gamma_{N}$, $\mathbb{C}^{4}/\Gamma_{N}$ and $\mathbb{C}^{5}/\Gamma_{N}$ for prime $N$.}
\label{primet}
\end{center}
\end{table}

\begin{observation}
For every sequence $\mathsf{g}_{x^\alpha}$ which counts HNF's symmetric under the cycle $g^\alpha \in H_\alpha \subset S_D$, there is a well defined function $P_{\mathsf{g}_{x^\alpha}}(p)$ over primes $p$ that takes the values $P_{\mathsf{g}_{x^\alpha}}(p)=\mathsf{g}_{x^\alpha}(p)$.
\end{observation}

\noindent The function on primes for the sequences of the abelian orbifolds of $\mathbb{C}^{3}$ are as follows:
\beal{ee5}
P_{\mathsf{g}_{x_{1}^{3}}}(p) &=& 1+p
\\
P_{\mathsf{g}_{x_{1}x_{2}}}(p) &=& 
\left\{ 
\begin{array}{l l}
1 & \quad \text{if $p = 2$}\\
2 & \quad \text{if $p \neq 2$}\\
\end{array} \right. 
\\
P_{\mathsf{g}_{x_{3}}}(p) &=& \label{epex1}
\left\{ 
\begin{array}{l l}
2 & \quad \text{if $p = 1 \mod 3$}\\
0 & \quad \text{if $p = 2 \mod 3$}\\
1 & \quad \text{if $p = 3$}
\end{array} \right. ~~.
\eea

\noindent For the case of abelian orbifolds of $\mathbb{C}^{4}$, the functions on primes are of the form
\beal{ee6}
P_{\mathsf{g}_{x_{1}^{4}}}(p) &=& 1+p+p^2
\\
P_{\mathsf{g}_{x_{1}^{2}x_2}}(p) = P_{\mathsf{g}_{x_{2}^{2}}}(p) &=& 
\left\{ 
\begin{array}{l l}
3 & \quad \text{if $p=2$}\\
p+2 & \quad \text{if $p\neq 2$}\\
\end{array} \right. 
\\
P_{\mathsf{g}_{x_{1}x_{3}}}(p) &=& 
\left\{ 
\begin{array}{l l}
3 & \quad \text{if $p = 1 \mod 3$}\\
1 & \quad \text{if $p = 2 \mod 3$}\\
1 & \quad \text{if $p = 3$}\\
%2 & \quad \text{otherwise}
\end{array} \right.
\\
P_{\mathsf{g}_{x_{4}}}(p) &=& 
\left\{ 
\begin{array}{l l}
3 & \quad \text{if $p = 1 \mod 4$}\\
1 & \quad \text{if $p = 2 \mod 4$}\\
1 & \quad \text{if $p = 3 \mod 4$}
%2 & \quad \text{otherwise}
\end{array} \right. ~~.
\eea

\noindent For the case of abelian orbifolds of $\mathbb{C}^5$, the functions on primes are of the form
\beal{ee7}
P_{\mathsf{g}_{x_{1}^{5}}}(p) &=& 1+p+p^2+p^3
\\
P_{\mathsf{g}_{x_{1}^{3}x_2}}(p) &=& 
\left\{ 
\begin{array}{l l}
7 & \quad \text{if $p=2$}\\
p^2+p+2 & \quad \text{if $p\neq 2$}\\
\end{array} \right. 
\\
P_{\mathsf{g}_{x_{1}x_{2}^{2}}}(p) &=& 
\left\{ 
\begin{array}{l l}
3 & \quad \text{if $p=2$}\\
2p+2 & \quad \text{if $p\neq 2$}
\end{array} \right.
\\
P_{\mathsf{g}_{x_{1}^{2}x_{3}}}(p) &=& 
\left\{ 
\begin{array}{l l}
p+3 & \quad \text{if $p = 1\mod 3$}\\
p+1 & \quad \text{if $p = 2\mod 3$}\\
4		& \quad \text{if $p=3$}
\end{array} \right.\\
%\eea
%\beal{ee7b}
P_{\mathsf{g}_{x_{2}x_{3}}}(p) &=& 
\left\{ 
\begin{array}{l l}
4 & \quad \text{if $p = 1 \mod 3$}\\
2 & \quad \text{if $p = 2 \mod 3$}\\
1 & \quad \text{if $p = 2$}\\
2 & \quad \text{if $p = 3$}
\end{array} \right.
\\
P_{\mathsf{g}_{x_{1}x_{4}}}(p) &=& 
\left\{ 
\begin{array}{l l}
4 & \quad \text{if $p = 1 \mod 4$}\\
1 & \quad \text{if $p = 2 \mod 4$}\\
2 & \quad \text{if $p = 3 \mod 4$}
%3 & \quad \text{otherwise}
\end{array} \right.
\\
P_{\mathsf{g}_{x_{5}}}(p) &=& 
\left\{ 
\begin{array}{l l}
4 & \quad \text{if $p=1 \mod 5$}\\
0 & \quad \text{if $p=2,3,4 \mod 5$}\\
1 & \quad \text{if $p=5$}
\end{array} \right.~~.
\eea

\begin{table}[ht!]
\begin{center}
\begin{tabular}{c|p{5mm}p{5mm}p{5mm}p{5mm}p{5mm}p{5mm}p{5mm}p{5mm}p{5mm}p{5mm}p{5mm}p{5mm}p{5mm}p{5mm}p{5mm}p{5mm}}
\hline
\hline
\multicolumn{17}{c}{$\mathbb{C}^{6}/\Gamma_{N}$}
\\
\hline
\hline
$N=p$ & 
2 & 3 & 5 & 7 & 11 & 13 & 17 & 19 
& 23 & 29 & 31 & 37 & 41 & 43 & 47 & 53
\\
\hline
$\mathsf{g}_{x_{1}^{6}}$ & 
	31&		121&		781&		2801&		\footnotesize{16105}&		\footnotesize{30941}& 	\footnotesize{88741}	& \scriptsize{137561}
	& \scriptsize{292561} & \scriptsize{732541} & \scriptsize{954305} & \tiny{1926221} & \tiny{2896405} & \tiny{3500201} & \tiny{4985761} &\tiny{8042221}
\\
$\mathsf{g}_{x_{1}^{4}x_{2}}$ & 
		15&		41&		157&		401&		1465&		2381& 	5221	&7241
		& \footnotesize{12721} & \footnotesize{25261} & \footnotesize{30785} & \footnotesize{52061} & \footnotesize{70645} & \footnotesize{81401} & \scriptsize{106081} & \scriptsize{151741}
\\
$\mathsf{g}_{x_{1}^{2}x_{2}^{2}}$ & 
		7&		17&		37&		65&		145&		197& 325		&401
		& 577 & 901 & 1025 & 1445 & 1765 & 1937 & 2305  & 2917
\\
$\mathsf{g}_{x_{2}^{3}}$ & 
		7&		17&		37&		65&		145&		197& 325		&401
		& 577 & 901 & 1025 & 1445 & 1765 & 1937 & 2305  & 2917
\\
$\mathsf{g}_{x_{1}^{3}x_{3}}$ & 
		7&		13&		31&		59&		133&		185& 	307	&383
		& 553 & 871 & 995 & 1409 & 1723 & 1895 & 2257 & 2863
\\
$\mathsf{g}_{x_{1}x_{2}x_{3}}$ & 
		3&		5&		7&		11&		13&		17& 	19	&23
		& 25 & 31 & 35 & 41 & 43 & 47 & 49 & 55
\\
$\mathsf{g}_{x_{1}^{2}x_{4}}$ & 
		3&		5&		9&		9&		13&		17& 21		&21
		& 25 &  33& 33 & 41 & 45 & 45 & 49 & 57
\\
$\mathsf{g}_{x_{2}x_{4}}$ & 
		3&		5&		9&		9&		13&		17& 21		&21
		& 25 &  33& 33 & 41 & 45 & 45 & 49 & 57
\\
$\mathsf{g}_{x_{3}^{2}}$ & 
		1&		4&		1&		17&		1&		29& 	1	&41
		& 1 & 1 & 65 & 77 & 1 & 89 & 1 & 1
\\
$\mathsf{g}_{x_{1}x_{5}}$ & 
		1&		1&		1&		1&		5&		1& 		1&1
		& 1 & 1 & 5 & 1 & 5 & 1 & 1 & 1
\\
$\mathsf{g}_{x_{6}}$ & 
		1&		2&		1&		5&		1&		5& 	1	&5
		& 1 & 1 & 5 & 5 & 1 & 5 & 1 & 1
\\
\hline
\hline
\end{tabular}
\caption{Sequences of $\mathbb{C}^{6}/\Gamma_{N}$ for prime $N$.}
\label{primet2}
\end{center}
\end{table}

For the case of abelian orbifolds of $\mathbb{C}^6$, the functions on primes are of the form
\beal{ee7b}
P_{\mathsf{g}_{x_{1}^{6}}}(p) &=& 1+p+p^2+p^3+p^4
\\
P_{\mathsf{g}_{x_{1}^{4}x_{2}}}(p) &=& 
\left\{ 
\begin{array}{l l}
15 & \quad \text{if $p=2$}\\
p^3+p^2+p+2 & \quad \text{if $p\neq 2$}\\
\end{array} \right.
\\
P_{\mathsf{g}_{x_{1}^{2}x_{2}^{2}}}(p) =
P_{\mathsf{g}_{x_{2}^{3}}}(p) &=&
\left\{ 
\begin{array}{l l}
7 & \quad \text{if $p=2$}\\
p^2+2p+2 & \quad \text{if $p\neq 2$}\\
\end{array} \right.
\\
P_{\mathsf{g}_{x_{1}^{3}x_{3}}}(p) &=&
\left\{ 
\begin{array}{l l}
p^2+p+3 & \quad \text{if $p = 1\mod 3$}\\
p^2+p+1 & \quad \text{if $p = 2\mod 3$}\\
13			& \quad \text{if $p=3$}
\end{array} \right.\\
P_{\mathsf{g}_{x_{1}x_{2}x_{3}}}(p) &=&
\left\{ 
\begin{array}{l l}
p+4 & \quad \text{if $p = 1\mod 3$}\\
p+2 & \quad \text{if $p = 2\mod 3$}\\
3			& \quad \text{if $p=2$} \\
5			& \quad \text{if $p=3$}
\end{array} \right.\\
P_{\mathsf{g}_{x_{1}^{2}x_{4}}}(p) =
P_{\mathsf{g}_{x_{2}x_{4}}}(p) &=& 
\left\{ 
\begin{array}{l l}
p+4 & \quad \text{if $p = 1 \mod 4$}\\
p+2 & \quad \text{if $p = 3 \mod 4$}\\
3 & \quad \text{if $p = 2$}
\end{array} \right.
\\
P_{\mathsf{g}_{x_{3}^{2}}}(p) &=& 
\left\{ 
\begin{array}{l l}
2p+3 & \quad \text{if $p = 1\mod 3$}\\
1 & \quad \text{if $p = 2\mod 3$}\\
%1			& \quad \text{if $p=2$} \\
4			& \quad \text{if $p=3$}
\end{array} \right. \label{espec}
\eea
\beal{ee7bb}
P_{\mathsf{g}_{x_{1}x_{5}}}(p) &=& 
\left\{ 
\begin{array}{l l}
5 & \quad \text{if $p=1 \mod 5$}\\
1 & \quad \text{if $p=2,3,4 \mod 5$}\\
1 & \quad \text{if $p=5$}
\end{array} \right.\\
P_{\mathsf{g}_{x_{6}}}(p) &=& 
\left\{ 
\begin{array}{l l}
5 & \quad \text{if $p=1 \mod 6$}\\
1 & \quad \text{if $p=2 \mod 6$}\\
2 & \quad \text{if $p=3 \mod 6$}\\
1 & \quad \text{if $p=5 \mod 6$}
\end{array} \right.
~~.
\eea

\begin{figure}[ht!]
\begin{center}
\includegraphics[totalheight=7cm]{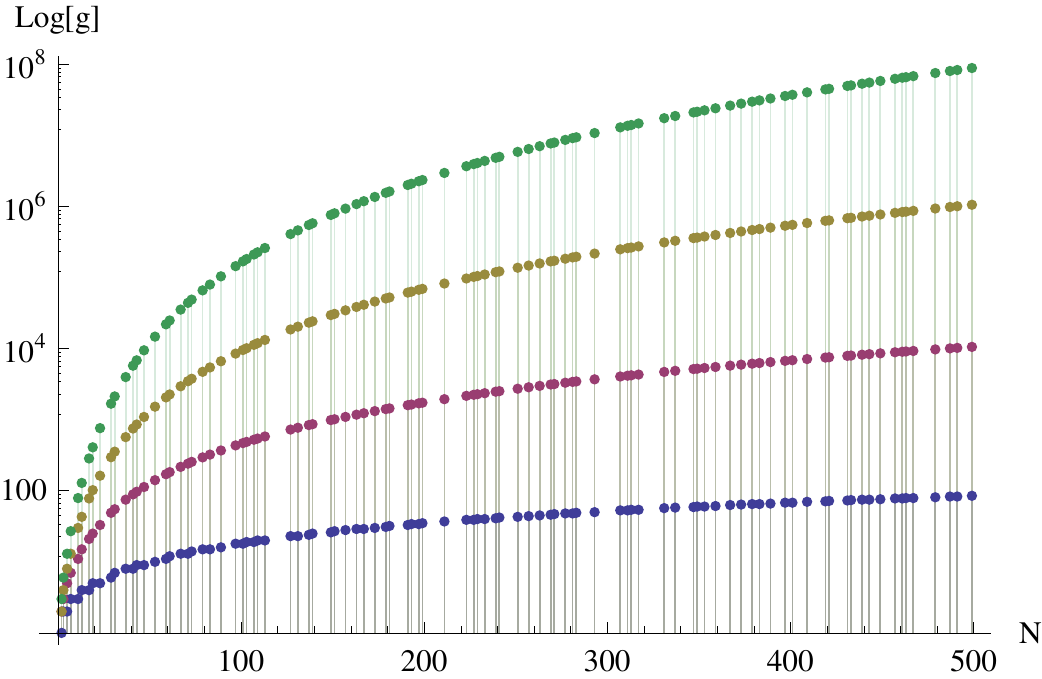}
\caption{The number of distinct orbifolds of $\mathbb{C}^{3}$ (blue), $\mathbb{C}^{4}$ (red), $\mathbb{C}^{5}$ (yellow) and $\mathbb{C}^{6}$ (green) respectively for prime $N$.}
  \label{primeplot}
 \end{center}
\end{figure}

%%%%%%%%%%%%%%%%%%%%%%%%%%%%%%%%%%%%%%%%%%%%%%%%%%%%%%%%%%%%%%%
\subsection{Series Convolutions from Functions on Primes \label{sconvol}}

The infinite sequences $\mathsf{u}=\{1,1,1\dots\}$ and $\mathsf{N}=\{1,2,3,\dots\}$ have functions on primes $P_{\mathsf{u}}(p)=1$ and $P_{\mathsf{N}}(p)=p$ respectively. If we now convolute the two infinite sequences to obtain $\mathsf{u}*\mathsf{N}=\{1,3,4,7,6,12,8,\dots\}$, the corresponding function on primes turns out to be $P_{\mathsf{u}*\mathsf{N}}(p)=P_{\mathsf{u}}(p)+P_{\mathsf{N}}(p)=1+p$.\\

\begin{observation}
Multiplicativity turns into additivity on prime indices. One can translate between a convolution and a function on primes with
\beql{ee2}
\mathsf{g} = *_{i=1}^{A}{\mathsf{N}^{d_i}} ~*_{j=1}^{B}{\mathsf{N}^{d_j}\chi_{k_j,m_j}} ~* \mathsf{C} ~\Leftrightarrow~ P_{\mathsf{g}}(p) = \sum_{i=1}^{A}{p^{d_i}} + \sum_{j=1}^{B}{p^{d_j}\chi_{k_j,m_j}}(p) + \mathsf{C}_{p}
\eeq
where $d_i$ is a non-negative integer. $\mathsf{C}$ can be any finite or infinite sequence with elements on prime indices denoted by $\mathsf{C}_{p}$.
\end{observation}
The aim is to keep $\mathsf{C}$ well-defined under the right combinations of $\mathsf{N}^d$ and $\chi_{k,m}$ in the convolution in \eref{ee2}.\\

\noindent\textbf{Example $x_3$.} An example is the sequence $\mathsf{g}_{x_3}$ that counts $x_3$-symmetric HNF's which correspond to the abelian orbifolds of $\mathbb{C}^{3}$. The sequence has a function of period $3$ on primes and is given in \eref{epex1}. The function on primes can be written in terms of the values on prime indices of basic multiplicative sequences as follows,
\beal{ee2d}
P_{\mathsf{g}_{x_3}}(p)
&=&
1+\chi_{3,2}(p) \nn\\
&=&
\chi_{3,1}(p)+\chi_{3,2}(p)
\eea

When considering the entire sequence with values on non-prime indices, the convolutions take the form
\beal{ee2e}
\mathsf{g}_{x_3}
&=&
\mathsf{u}*\chi_{3,2} \nn\\
&=&
\chi_{3,1}*\chi_{3,2}*(\sum_{a=0}^{\infty}{t^{3^a}})
~~,
\eea
where $\mathsf{C}=1$ and $\mathsf{C}=\sum_{a=0}^{\infty}{t^{3^a}}$ respectively. As desired, $\mathsf{C}$ is a well-defined partition function for both choices in \eref{ee2e}.\\

Under this scheme, sequences which count orbifolds that are invariant under cycles of $S_D$ can be re-written in terms of convolutions of the form \eref{ee2}. \tref{primetab} and \tref{primetab2} show choices of sequence convolutions for the orbifolds of $\mathbb{C}^2$ to $\mathbb{C}^5$. Convolutions for the sequences for the abelian orbifolds of $\mathbb{C}^3$ and $\mathbb{C}^4$ have been first presented in \cite{HananyOrlando10}. We present here the convolutions for the abelian orbifolds of $\mathbb{C}^5$.\\

In the section below, some generalizations are given for sequences on all indices. The reason why not all sequences on all indices can be generalized is that some sequences require finite term corrections on power of prime indices. This can be seen for sequences $\mathsf{g}_{x_1 x_{2}^2}$ and $\mathsf{g}_{x_1 x_{4}}$ in \tref{primetab} and \tref{primetab2}. However, a complete set of generalizations for the sequences on prime indices can be given. Using the cycle index of $S_D$, this set of generalizations lead to the counting of distinct abelian orbifolds of the form $\mathbb{C}^d/\Gamma$ with any prime order of $\Gamma$ and any dimension $D$.\\

\begin{table}[ht!]
\begin{center}
\begin{tabular}{r|l}
\hline
\hline
\multicolumn{2}{c}{$\mathbb{C}^{2}/\Gamma_{N}$}
\\
\hline
$x_{1}^{2}$ & $\mathsf{u}$ 
\\
$x_{2}$ & $\mathsf{u}$
\\
\\
\hline
\hline
\multicolumn{2}{c}{$\mathbb{C}^{3}/\Gamma_{N}$}
\\
\hline
$x_{1}^{3}$ & $\mathsf{u} * \mathsf{N}$ 
\\
$x_{1}x_{2}$ & $\mathsf{u} * \mathsf{u} * (t-t^2+2t^4)$%\{1,-1,0,2\}$ 
\\
$x_{3}$ & $\mathsf{u} * \chi_{3,2}$
\\
\\
\hline
\hline
\multicolumn{2}{c}{$\mathbb{C}^{4}/\Gamma_{N}$}
\\
\hline
$x_{1}^{4}$ & $\mathsf{u} * \mathsf{N} * \mathsf{N}^{2}$ 
\\
$x_{1}^{2}x_{2}$ & $\mathsf{u} * \mathsf{u} * \mathsf{N} * (t-t^2+4t^4)$%* \{1,-1,0,4\}$ 
\\
$x_{2}^{2}$ & $\mathsf{u} * \mathsf{u} * \mathsf{N} * (t-t^2+4t^4)$%\{1,-1,0,4\}$ 
\\
$x_{1}x_{3}$ & $\mathsf{u} * \mathsf{u} * \chi_{3,2} * (t-t^3+3t^9)$%\{1,0,-1,0,0,0,0,0,3\}$ 
\\
$x_{4}$ & $\mathsf{u} * \mathsf{u} * \chi_{4,2} * (t-t^2+2t^4)$%{1,-1,0,2\}$ 
\\
\\
\hline
\hline
\multicolumn{2}{c}{$\mathbb{C}^{5}/\Gamma_{N}$}
\\
\hline
$x_{1}^{5}$ & $\mathsf{u} * \mathsf{N} * \mathsf{N}^{2} * \mathsf{N}^{3}$ 
\\
$x_{1}^{3}x_{2}$ & $\mathsf{u} * \mathsf{u} * \mathsf{N} * \mathsf{N}^{2} * (t-t^2+8t^4)$%\{1,-1,0,8\}$ 
\\
$x_{1}x_{2}^{2}$ & $\mathsf{u} * \mathsf{u} * \mathsf{N} * \mathsf{N} * (t-3 t^2+14 t^4-12 t^8+16 t^{16})$
\\
$x_{1}^{2}x_{3}$ & $\mathsf{u} * \mathsf{u} * \mathsf{N} * \chi_{3,2} * (t-t^3+9t^9)$%\{1,0,-1,0,0,0,0,0,9\}$ 
\\
$x_{2}x_{3}$ & $\mathsf{u} * \mathsf{u} * \mathsf{u} * \chi_{3,2} * (t-t^2+2t^4) * (t-t^3+3t^9)$%\{1,-1,0,2\} * \{1,0,-1,0,0,0,0,0,3\}$
\\
$x_{1}x_{4}$ & $\mathsf{u} * \mathsf{u} * \mathsf{u} * \chi_{4,2} * (t-2t^{2}+3t^{4}+6t^{16}-8t^{32}+8t^{64})$
\\
$x_{5}$ & $\mathsf{u} * \chi_{5,2} * \chi_{5,3} * \chi_{5,4}$
\\
\\
\hline
\hline
\end{tabular}
\caption{Summary of the first choice of convolutions for orbifolds of $\mathbb{C}^{2}$, $\mathbb{C}^{3}$, $\mathbb{C}^{4}$ and $\mathbb{C}^{5}$.}
\label{primetab}
\end{center}
\end{table}

\begin{table}[ht!]
\begin{center}
\begin{tabular}{r|l}
\hline
\hline
\multicolumn{2}{c}{$\mathbb{C}^{2}/\Gamma_{N}$}
\\
\hline
$x_{1}^{2}$ & $\mathsf{u}$ 
\\
$x_{2}$ & $\chi_{2,1}*\left(\sum_{a=0}^{\infty}{t^{2^a}}\right)$
\\
\\
\hline
\hline
\multicolumn{2}{c}{$\mathbb{C}^{3}/\Gamma_{N}$}
\\
\hline
$x_{1}^{3}$ & $\mathsf{u}*\mathsf{N}$
\\
$x_{1}x_{2}$ & $\mathsf{u} * \chi_{2,1} * \left(t+2\sum_{a=0}^{\infty}{t^{2^{(a+2)}}}\right)$
\\
$x_{3}$ & $\chi_{3,1} * \chi_{3,2} *\left(\sum_{a=0}^{\infty}{t^{3^a}}\right)$
\\
\\
\hline
\hline
\multicolumn{2}{c}{$\mathbb{C}^{4}/\Gamma_{N}$}
\\
\hline
$x_{1}^{4}$ & $\mathsf{u} * \mathsf{N} * \mathsf{N}^{2}$ 
\\
$x_{1}^{2}x_{2}$ & $\mathsf{u} * \mathsf{N} * \chi_{2,1} * \left(t+4\sum_{a=0}^{\infty}{t^{2^{(a+2)}}}\right)$
\\
$x_{2}^{2}$ & $\mathsf{u} * \mathsf{N} * \chi_{2,1} * \left(t+4\sum_{a=0}^{\infty}{t^{2^{(a+2)}}}\right)$
\\
$x_{1}x_{3}$ & $\mathsf{u} * \chi_{3,1} * \chi_{3,2} * \left(t+3\sum_{a=0}^{\infty}{t^{3^{(a+2)}}}\right)$
\\
$x_{4}$ & $\mathsf{u} * \chi_{4,1} * \chi_{4,2} * \left(t+2\sum_{a=0}^{\infty}{t^{2^{(a+2)}}}\right)$
\\
\\
\hline
\hline
\multicolumn{2}{c}{$\mathbb{C}^{5}/\Gamma_{N}$}
\\
\hline
$x_{1}^{5}$ & $\mathsf{u} * \mathsf{N} * \mathsf{N}^{2} * \mathsf{N}^{3}$ 
\\
$x_{1}^{3}x_{2}$ & $\mathsf{u} * \mathsf{N} * \mathsf{N}^{2} * \chi_{2,1} * \left(t+8\sum_{a=0}^{\infty}{t^{2^{(a+2)}}}\right)$
\\
$x_{1}x_{2}^{2}$ & $\mathsf{u} * \mathsf{N} * \mathsf{N} * \chi_{2,1} * \left(t+16\sum_{a=0}^{\infty}{t^{2^{(a+2)}}~-2t^2-4t^4-16t^8}\right)$
\\
$x_{1}^{2}x_{3}$ & $\mathsf{u} * \mathsf{N} * \chi_{3,1} * \chi_{3,2} * \left(t+9\sum_{a=0}^{\infty}{t^{3^{(a+2)}}}\right)$
\\
$x_{2}x_{3}$ & $\mathsf{u} * \chi_{2,1} * \chi_{3,1} * \chi_{3,2} * \left(t+2\sum_{a=0}^{\infty}{t^{2^{(a+2)}}}\right)*\left(t+3\sum_{a=0}^{\infty}{t^{3^{(a+2)}}}\right)$
\\
$x_{1}x_{4}$ & $\mathsf{u}*\mathsf{u}*\chi_{4,1}*\chi_{4,2}*\left(t+8\sum_{a=0}^{\infty}{t^{2^{(a+2)}}}~-t^2-6t^4-6t^8-8t^{32}\right)$
\\
$x_{5}$ & $\chi_{5,1}*\chi_{5,2}*\chi_{5,3}*\chi_{5,4}*\left(\sum_{a=0}^{\infty}{t^{5^a}}\right)$
\\
\\
\hline
\hline
\end{tabular}
\caption{Summary of the second choice of convolutions for orbifolds of $\mathbb{C}^{2}$, $\mathbb{C}^{3}$, $\mathbb{C}^{4}$ and $\mathbb{C}^{5}$.}
\label{primetab2}
\end{center}
\end{table}

\clearpage

%%%%%%%%%%%%%%%%%%%%%%%%%%%%%%%%%%%%%%%%%%%%%%%%%%%%%%%%%%%%%%%%%%%%%%%%%
\section{Generalizations for Orbifold Symmetries of Abelian Orbifolds of $\mathbb{C}^D$\label{sp2}}

%%%%%%%%%%%%%%%%%%%%%%%%%%%%%%%%%%%%%%%%%%%%%%%%%%%%%%%%%%%%%%%%%%%%%%%%%
\subsection{Generalizations for Symmetry Sequences on all Indices \label{scomplete}}

Given sequences for abelian orbifolds of $\mathbb{C}^{2}$ to $\mathbb{C}^{6}$, and their corresponding sequence convolutions in \tref{primetab} and \tref{primetab2}, we attempt to derive sequences that count $g^\alpha$-symmetric HNF's which correspond to orbifolds of the form $\mathbb{C}^{D}/\Gamma_N$ on all integer orders $N$. We start with the simplest generalization, the identity sequence:\\

\begin{observation} The identity $x_{1}^{D}$ sequence which counts HNF's of a given orbifold order $N$ is expressed as
\beql{ee3}
\mathsf{g}_{x_{1}^{D}} = *_{d=0}^{D-2}{\mathsf{N}^d} ~\Leftrightarrow~ P_{\mathsf{g}_{x_{1}^{D}}}(p) = \sum_{d=0}^{D-2}{p^d} ~~,
\eeq
where $\mathsf{N}^{0}=\mathsf{u}$. The recursive form of the partition function of $\mathsf{g}_{x_{1}^{D}}$ is
\beql{es3}
g_{x_{1}^{D}}(t)~=~\sum_{k=1}^{\infty}{k^{D-2}g_{x_{1}^{D-1}}(t^{k})}~~,
\eeq
with $\mathsf{g}_{x_{1}^{D}}~=~\mathsf{g}_{x_{1}^{D-1}} * \mathsf{N}^{D-2}$ and $\mathsf{g}_{x_{1}}=\{1,0,0,\dots\}$. 
\end{observation}

\noindent A proof of the above observation is given in \cite{HananyOrlando10}.
\\

Let us proceed to sequences $\mathsf{g}_{x_q}$ with prime $q$ which count HNF's that are invariant under the cycle $(12\dots q)\in S_q$. The HNF's correspond to abelian orbifolds of $\mathbb{C}^{q}$.
\begin{observation}\label{prop1} The sequence $\mathsf{g}_{x_{q}}$ with prime $q$ has the convolutions
\beal{ep4}
\mathsf{g}_{x_{q}}&=&*_{m=1}^{q-1}{\chi_{q,m}}*\left(\sum_{a=0}^{\infty}{t^{q^a}}\right)\\
&=& \mathsf{u}*_{m=2}^{q-1}{\chi_{q,m}}~~,
\eea
where by convention, $\chi_{q,1}$ is the Dirichlet character with elements $\{0,1\}$ only. 
\end{observation}

\noindent 
\textbf{Example.} For the abelian orbifolds of $\mathbb{C}^{2}$, $\mathbb{C}^{3}$ and $\mathbb{C}^{5}$, we have the convolutions
\beal{e3e7}
\mathsf{g}_{x_2}&=&\chi_{2,1}*\left(\sum_{a=0}^{\infty}{t^{2^a}}\right)\nn\\
&=&\mathsf{u}
~~,\\
\mathsf{g}_{x_3}&=&\chi_{3,1}*\chi_{3,2}*\left(\sum_{a=0}^{\infty}{t^{3^a}}\right)\nn\\
&=&\mathsf{u}*\chi_{3,2}
~~,\\
\mathsf{g}_{x_5}&=&\chi_{5,1}*\chi_{5,2}*\chi_{5,3}*\chi_{5,4}*\left(\sum_{a=0}^{\infty}{t^{5^a}}\right)\nn\\
&=&\mathsf{u}*\chi_{5,2}*\chi_{5,3}*\chi_{5,4}
~~.
\eea  
These convolutions reproduce the sequences which have been obtained by explicit symmetry counting with the abelian orbifolds of $\mathbb{C}^3$ and $\mathbb{C}^5$. The orders for which the matching is checked explicitly are shown in \tref{tc3sym}, \tref{tc5sym} and \tref{tc5sym2}.
\\

The sequence $\mathsf{g}_{x_{1}^A x_q}$ with prime $q$ counts HNF's which are invariant under the cycle $(1)(2)\dots(A)(A+1\dots D)\in S_D$. The HNF's correspond to abelian orbifolds of $\mathbb{C}^D$. We observe that the sequence $\mathsf{g}_{x_{1}^A x_q}$ is related to $\mathsf{g}_{x_q}$ as follows:
\begin{observation}\label{prop2} Given the sequence $\mathsf{g}_{x_q}$ where $q$ is prime, the convolution for $\mathsf{g}_{x_{1}^{A}x_q}$ has the form
\beal{ep6}
\mathsf{g}_{x_{1}^{A}x_q}&=&*_{d=1}^{A}{\mathsf{N}^{d-1}}~*_{m=1}^{q-1}{\chi_{q,m}}~*\left(t+q^A \sum_{a=0}^{\infty}{t^{q^{(a+2)}}}\right)\\
&=&\mathsf{u}~*_{d=1}^{A}{\mathsf{N}^{d-1}}~*_{m=2}^{q-1}{\chi_{q,m}}~*(t-t^q+q^A t^{q^2})
~~,
\eea
where $\mathsf{N}^{0}=\mathsf{u}$ and $\chi_{q,1}$ is the Dirichlet character with elements $\{0,1\}$ only. 
\end{observation}

\noindent 
\textbf{Example.} For $q=2$, we have the sequences
\beal{e3e9}
\mathsf{g}_{x_{1}x_{2}}&=&\mathsf{u}*\chi_{2,1}*\left(t+2\sum_{a=0}^{\infty}{t^{2^{(a+2)}}}\right)\nn\\
&=& \mathsf{u}*\mathsf{u}*(t-t^2+2t^4)
~~,\\
\mathsf{g}_{x_{1}^2x_{2}}&=&\mathsf{u}*\mathsf{N}*\chi_{2,1}*\left(t+4\sum_{a=0}^{\infty}{t^{2^{(a+2)}}}\right)\nn\\
&=& \mathsf{u}*\mathsf{u}*\mathsf{N}*(t-t^2+4t^4)
~~,\\
\mathsf{g}_{x_{1}^3x_{2}}&=&\mathsf{u}*\mathsf{N}*\mathsf{N}^2*\chi_{2,1}*\left(t+8\sum_{a=0}^{\infty}{t^{2^{(a+2)}}}\right)\nn\\
&=& \mathsf{u}*\mathsf{u}*\mathsf{N}*\mathsf{N}^{2}*(t-t^2+8t^4)~~.
\eea
For $q=3$, we derive the sequences
\beal{e3e11}
\mathsf{g}_{x_{1}x_{3}}&=&\mathsf{u}*\chi_{3,1}*\chi_{3,2}*\left(t+3\sum_{a=0}^{\infty}{t^{3^{(a+2)}}}\right)\nn\\
&=& \mathsf{u}*\mathsf{u}*\chi_{3,2}*(t-t^3-3t^9)
~~,\\
\mathsf{g}_{x_{1}^2x_{3}}&=&\mathsf{u}*\mathsf{N}*\chi_{3,1}*\chi_{3,2}*\left(t+9\sum_{a=0}^{\infty}{t^{3^{(a+2)}}}\right)\nn\\
&=& \mathsf{u}*\mathsf{u}*\mathsf{N}*\chi_{3,2}*(t-t^3-9t^9)
~~.
\eea
The above sequences match the explicit counting for abelian orbifolds of $\mathbb{C}^3$, $\mathbb{C}^4$ and $\mathbb{C}^5$. The orders for which the matching was checked explicitly are shown in \tref{tc3sym}, \tref{tc4sym}, \tref{tc5sym} and \tref{tc5sym2}.
\\

\noindent 
For the abelian orbifolds of $\mathbb{C}^{6}$, we derive using the above observation the sequences 
\beal{e3e13}
\mathsf{g}_{x_{1}^4x_{2}}&=&\mathsf{u}*\mathsf{N}*\mathsf{N}^2*\mathsf{N}^3*\chi_{2,1}*\left(t+16\sum_{a=0}^{\infty}{t^{2^{(a+2)}}}\right)\nn\\
&=&\mathsf{u}*\mathsf{u}*\mathsf{N}*\mathsf{N}^2*\mathsf{N}^3*(t-t^2+16t^4)
~~,\\
\mathsf{g}_{x_{1}^3x_{3}}&=&\mathsf{u}*\mathsf{N}*\mathsf{N}^2*\chi_{3,1}*\chi_{3,2}*\left(t+27\sum_{a=0}^{\infty}{t^{3^{(a+2)}}}\right)\nn\\
&=&\mathsf{u}*\mathsf{u}*\mathsf{N}*\mathsf{N}^2*\chi_{3,2}*(t-t^3+27t^9)
~~,\\
\mathsf{g}_{x_{1}x_{5}}&=&\mathsf{u}*\chi_{5,1}*\chi_{5,2}*\chi_{5,3}*\chi_{5,4}*\left(t+5\sum_{a=0}^{\infty}{t^{5^{(a+2)}}}\right)\nn\\
&=&\mathsf{u}*\mathsf{u}*\chi_{5,2}*\chi_{5,3}*\chi_{5,4}*(t-t^5+5t^{25})
~~.
\eea
The convolution match the explicit counting for abelian orbifolds of $\mathbb{C}^6$ in \tref{tc6sym}. The matching was checked explicitly up to order $N=20$.
\\

The sequence $\mathsf{g}_{x_r x_s}$ with both prime $r$ and $s$ counts HNF's dual to abelian orbifolds of $\mathbb{C}^{r+s}$ which are invariant under the cycle $(1\dots r)(r+1\dots r+s)\in S_{r+s}$. We observe as follows that the sequence $\mathsf{g}_{x_r x_s}$ is related to the separate sequences $\mathsf{g}_{x_r}$ and $\mathsf{g}_{x_s}$.
\begin{observation}\label{prop3} Given the sequences $\mathsf{g}_{x_r}$ and $\mathsf{g}_{x_s}$ where $r\neq s$ and $r,s$ are both prime, the sequence $\mathsf{g}_{x_r x_s}$ has the convolution
\beal{ep8}
\mathsf{g}_{x_r x_s}
&=&
\mathsf{u}~*_{m=1}^{r-1}{\chi_{r,m}}~
*_{n=1}^{s-1}{\chi_{s,n}}~
*\left(t+r\sum_{a=0}^{\infty}{t^{r^{(a+2)}}}\right)
*\left(t+s\sum_{b=0}^{\infty}{t^{s^{(b+2)}}}\right)\\
&=&
\mathsf{u}*\mathsf{u}*\mathsf{u}~
*_{m=2}^{r-1}{\chi_{r,m}}~
*_{n=2}^{s-1}{\chi_{s,n}}~
*(t-t^r+r t^{r^2})
*(t-t^s+s t^{s^2})
~~.
\eea
\end{observation}

\noindent
\textbf{Example.} An example which confirms the above observation is the sequence 
\beal{e3e15}
\mathsf{g}_{x_2 x_3}
&=&
\mathsf{u}*\chi_{2,1}*\chi_{3,1}*\chi_{3,2}
*\left(t+2\sum_{a=0}^{\infty}{t^{2^{(a+2)}}}\right)
*\left(t+3\sum_{a=0}^{\infty}{t^{3^{(a+2)}}}\right)\nn\\
&=&
\mathsf{u}*\mathsf{u}*\mathsf{u}*\chi_{3,2}
*(t-t^2+2t^4)*(t-t^3+3t^9)
~~,
\eea
which matches the data presented in \tref{tc5sym} and \tref{tc5sym2}. The matching was checked explicitly up to the order given in the counting tables.
\\

The sequence which counts HNF's invariant under the cycle $(1)(2)\dots(A)(A+1\dots A+r)(A+r+1\dots A+r+s)\in S_{A+r+s}$ is denoted by $\mathsf{g}_{x_{1}^{A}x_r x_s}$. The HNF's are dual to the abelian orbifolds of $\mathbb{C}^{A+r+s}$. In the following we note that this sequence $\mathsf{g}_{x_{1}^{A}x_r x_s}$ is related to the sequence $\mathsf{g}_{x_r x_s}$.
\begin{observation}\label{prop4} Given the sequence $\mathsf{g}_{x_r x_s}$ where $r\neq s$ and $r,s$ are both prime, the sequence $\mathsf{g}_{x_{1}^{A}x_r x_s}$ has the convolution
\beal{ep10}
\mathsf{g}_{x_{1}^{A} x_{r} x_{s}}
&=&
\mathsf{u}~*_{d=1}^{A}{\mathsf{N}^{d}}~
*_{m=1}^{r-1}{\chi_{r,m}}~*_{n=1}^{s-1}{\chi_{s,n}}~\nn\\
&&
*\left(t+r^{(A+1)}\sum_{a=0}^{\infty}{t^{r^{(a+2)}}}\right)
*\left(t+s^{(A+1)}\sum_{b=0}^{\infty}{t^{s^{(b+2)}}}\right)\\
&=&
\mathsf{u}*\mathsf{u}*\mathsf{u}~*_{d=1}^{A}{\mathsf{N}^{d}}~
*_{m=2}^{r-1}{\chi_{r,m}}~*_{n=2}^{s-1}{\chi_{s,n}}~\nn\\
&&
*(t-t^r+r^{(A+1)} t^{r^2})
*(t-t^s+s^{(A+1)} t^{s^2})
~~.
\eea
\end{observation}

\noindent
\textbf{Example.} We have the sequence 
\beal{e3e17}
\mathsf{g}_{x_1 x_2 x_3}
&=&
\mathsf{u}*\mathsf{N}*\chi_{2,1}*\chi_{3,1}*\chi_{3,2}
*\left(t+4\sum_{a=0}^{\infty}{t^{2^{(a+2)}}}\right)
*\left(t+9\sum_{a=0}^{\infty}{t^{3^{(a+2)}}}\right)\nn\\
&=&
\mathsf{u}*\mathsf{u}*\mathsf{u}*\mathsf{N}*\chi_{3,2}
*(t-t^2+4t^4)
*(t-t^3+9t^9)
~~,
\eea
for the abelian orbifolds of $\mathbb{C}^6$. The explicit counting which is presented in \tref{tc6sym} matches the above sequence. The matching was checked explicitly up to order $N=20$.
\\

Let the sequence $\mathsf{g}_{x^\alpha}$ count HNF's which are invariant under the cycle $g^\alpha\in S_D$ where $H_\alpha \subset S_D$ is a conjugacy class. Given $g^\alpha$ does not contain a sub-cycle which has an odd-valued integer length, we observe the following property:
\begin{observation}\label{prop5} Given the sequence $\mathsf{g}_{x^\alpha}$ where $x_{1} \notin x^\alpha$ and $\forall x_{q}\in x^\alpha$ $q$ is an even integer, the following relation is satisfied
\beql{ep12}
\mathsf{g}_{x_{1}^2 x^\alpha}~=~\mathsf{g}_{x_{2} x^\alpha}~~.
\eeq
\end{observation}

\noindent
\textbf{Example.} A basic example is the relation 
\beal{e3e19}
\mathsf{g}_{x_{1}^{2}}~=~\mathsf{g}_{x_{2}}~=~\mathsf{u}~=~\chi_{2,1}*\left(\sum_{a=0}^{\infty}{t^{2^{a}}}\right)~~,
\eea
For the orbifolds of $\mathbb{C}^{4}$, we have through explicit counting which is presented in \tref{tc4sym}
\beal{e3e20}
\mathsf{g}_{x_{1}^{2}x_{2}}~=~\mathsf{g}_{x_{2}^{2}}
~=~\mathsf{u}*\mathsf{u}*\mathsf{N}*(t-t^2+4t^4)
~=~\mathsf{u}*\mathsf{N}*\chi_{2,1}*\left(t+4\sum_{a=0}^{\infty}{t^{2^{(a+2)}}}\right)~~.
\eea
For the orbifolds of $\mathbb{C}^{6}$, we derive
\beal{e3e21}
\mathsf{g}_{x_{1}^{2}x_{4}}~=~\mathsf{g}_{x_{2}x_{4}} ~~~,~~~\mathsf{g}_{x_{1}^{2}x_{2}^{2}}~=~\mathsf{g}_{x_{2}^{3}}~~,
\eea
which is confirmed by the counting in \tref{tc6sym}. The matching was checked explicitly up to order $N=20$.
\\

%%%%%%%%%%%%%%%%%%%%%%%%%%%%%%%%%%%%%%%%%%%%%%%%%%%%%%%%%%%%%%%%%%%%%%%%%
\subsection{Generalizations for Symmetry Sequences with only Prime Indices \label{sprimeprop}}

Let us restrict ourselves to elements on prime indices of sequences that count $g^\alpha$-symmetric HNF's which correspond to orbifolds of $\mathbb{C}^D$. The functions on primes which reproduce sequence elements on prime indices are fully generalizable to any orbifold dimension $D$. We observe in this section patterns of functions on primes and derive generalizations.\\

The first sequence which we consider is $\mathsf{g}_{x_a}$ where $a\in\mathbb{Z}^{+}$. This sequence counts HNF's which are invariant under the cycle $(12\dots a)\in S_a$. The HNF's are dual to abelian orbifolds of $\mathbb{C}^a$. On prime indices, the elements of the sequence are derived by the following function on primes:
\begin{proposition}\label{prop6}\label{propos6} Given the sequence $\mathsf{g}_{x_a}$ where $a\in\mathbb{Z}^{+}$, the corresponding function on primes is
\beal{ep1}
P_{\mathsf{g}_{x_a}}(p)
&=&
\sum_{m=1}^{\varphi(a)}{\chi_{a,m}(p)}
~+\sum_{\substack{k|a \\ 1<k<a}}{\sum_{m=1}^{\varphi(k)}{\chi_{k,m}(p) }}
~+\sum_{\substack{k|a \\ k=\text{prime}}}{\delta_{pk}} \\
&=&
\varphi(a)~\delta_{p,1\bmod{a}}
~+\sum_{\substack{k|a \\ 1<k<a}}{
\varphi(k)~\delta_{p,1\bmod{k}}
}
~+\sum_{\substack{k|a \\ k=\text{prime}}}{\delta_{pk}}~~,\label{ep1simp}
\eea
where $\varphi(k)$ is the Euler totient function which is the number of distinct Dirichlet characters of periodicity $k$. The simplification in \eref{ep1simp} comes from the property in \eref{exs4bc}.
\end{proposition}

\noindent
\textbf{Example.} From explicit counting we have
\beal{ep2}
P_{\mathsf{g}_{x_2}}(p)&=&
\chi_{2,1}(p) +\delta_{p2}\nn\\
&=&
\delta_{p,1\bmod{2}} +\delta_{p2}
\nn\\
P_{\mathsf{g}_{x_3}}(p)&=&
\chi_{3,1}(p)+\chi_{3,2}(p) +\delta_{p3} \nn\\
&=&
2\delta_{p,1\bmod{3}} +\delta_{p3}
\nn\\
P_{\mathsf{g}_{x_4}}(p)&=&
\chi_{2,1}(p)+\chi_{4,1}(p)+\chi_{4,2}(p)+\delta_{p2} \nn\\
&=&
\delta_{p,1\bmod{2}}+2\delta_{p,1\bmod{4}}+\delta_{p2}
\nn\\
P_{\mathsf{g}_{x_5}}(p)&=&
\chi_{5,1}(p)+\chi_{5,2}(p)+\chi_{5,3}(p)+\chi_{5,4}(p)+\delta_{p5} \nn\\
&=&
4\delta_{p,1\bmod{5}}+\delta_{p5}
\nn\\
P_{\mathsf{g}_{x_6}}(p)&=&
\chi_{2,1}(p)+\chi_{3,1}(p)+\chi_{3,2}(p)+\chi_{6,1}(p)+\chi_{6,2}(p)+\delta_{p,2}+\delta_{p,3}\nn\\
&=&
\delta_{p,1\bmod{2}}+2\delta_{p,1\bmod{3}}+2\delta_{p,1\bmod{6}}+\delta_{p2}+\delta_{p3}
~~.
\eea
The above functions reproduce the prime index elements of the sequences which have been obtained by explicit counting for the orbifolds of $\mathbb{C}^2$ to $\mathbb{C}^6$ (\tref{tc3sym} to \tref{tc6sym}).
\\

We recall that in Section \sref{s4b}, we mentioned that an element $g^\alpha \in S_D$ consists of $M$ disjoint cycles $\gamma^i$ of length $n_i=|\gamma^i|$. The general form of $x^\alpha$ which corresponds to a conjugacy class $H_\alpha \subset S_D$ and a term in the cycle index of $S_D$ is
\beql{ep3}
x^\alpha=\prod_{i=1}^{M}{x_{n_{i}}}~~.
\eeq
We call $M$ the partition number of the symmetry cycle. The dimension $D$ of the corresponding orbifold of $\mathbb{C}^{D}$ is given by $\sum_{k=1}^{M}{n_k}=D$. For example, the partition number of the following cycles are,
\beal{ep4}
M(x_{2}^{3})=3~~~,~~~M(x_{1}^2 x_2 x_3)=4 ~~~,~~~M(x_{2}^{2}x_{4})=3~~~.
\eea
Using the definition of the partition number, let us define an additional quantity which will be of use in our generalization.

\begin{definition} Given the cycle $g^\alpha$ of the conjugacy class $x^\alpha$ with corresponding partition number $M(x^\alpha)$, let the number of divisions by $m$ of the cycle $g^\alpha$ be defined as
\beql{ep7}
Q_{m}(x^\alpha)=\sum_{i=1}^{M(x^\alpha)}{\sum_{m|n_i}{1}}~~,
\eeq
where the dimension of the orbifold is given by $D=\sum_{k=1}^{M(x^\alpha)}{n_k}$. The number of divisions by $1$ is by definition the number of partitions of the cycle $g^\alpha$,
\beql{ep8}
Q_{1}(x^\alpha)=M(x^\alpha)~~.
\eeq
\end{definition}

Accordingly, we derive the number of division by $2$, $3$ and $4$ respectively for a cycle of the conjugacy class $x_{2}x_{4}$ as
\beql{ep9}
Q_{2}(x_2 x_4)=2~~~,~~~Q_{3}(x_2 x_4)=0~~~,~~~Q_{4}(x_2 x_4)=1~~~.
\eeq
Other examples are $x_{2}^{2}x_{3}$ and $x_{2}^{3}$ with
\beal{ep9}
&& Q_{2}(x_{2}^{2} x_{3})=2~~~,~~~Q_{3}(x_{2}^{2} x_{3})=1~~~,~~~Q_{4}(x_{2}^{2} x_{3})=0~~~, \nn\\
&& Q_{2}(x_{2}^{3})=3~~~,~~~Q_{3}(x_{2}^{3})=0~~~,~~~Q_{4}(x_{2}^{3})=0~~~.
\eea
\\

Let us consider now the most general sequence $\mathsf{g}_{x^\alpha}$ which counts HNF's that are invariant under the cycle $g^\alpha \in S_D$ where $g^\alpha$ is in the conjugacy class denoted by the cycle index variable $x^\alpha$. The elements of this sequence at prime indices are obtained from the function on primes $P_{x^\alpha}(p)$ which we generalize as follows:
\begin{proposition}\label{pmain} Given the cycle $g^\alpha$ with partition number $M(x^\alpha)>1$, the corresponding function on primes has the form
\beal{ep10}
P_{x^\alpha}(p)
&=&
\sum_{d=1}^{M(x^\alpha)-1}{p^{d-1}} ~+
\sum_{d=2}^{D}{
\sum_{q=1}^{Q_{d}(x^\alpha)}{
\sum_{m=1}^{\varphi(d)}{
p^{q-1}\chi_{d,m}(p)
}
} 
}
~+
\sum_{\substack{s|D  \\ s=prime \\ x_{1}\notin x^\alpha}}^{D}{
s^{Q_{s}(x^\alpha)-1} \delta_{ps}
}
\nn\\
\\
&=&
\sum_{d=1}^{M(x^\alpha)-1}{p^{d-1}} ~+
\sum_{d=2}^{D}{
\sum_{q=1}^{Q_{d}(x^\alpha)}{
p^{q-1}\varphi(d)~\delta_{p,1\bmod{d}}
} 
}
~+
\sum_{\substack{s|D  \\ s=prime \\ x_{1}\notin x^\alpha}}^{D}{
s^{Q_{s}(x^\alpha)-1} \delta_{ps}
}
~~.\nn\\
\eea
\end{proposition}

\noindent \textbf{Examples and Derivations.}
According to the above propositions, we are able to derive the functions on primes which correspond to any cycle $g^\alpha \in  S_D$. Tables \ref{gentab}, \ref{gentab2a}, \ref{gentab2b} and \ref{gentab3} present the derived functions on primes for the orbifolds of $\mathbb{C}^{2}$ to $\mathbb{C}^{9}$. The functions on primes reproduce the sequence elements on prime indices presented in \tref{tc3sym} to \tref{tc6sym} for the orbifolds of $\mathbb{C}^3$ to $\mathbb{C}^6$. The derived functions for the orbifolds of $\mathbb{C}^7$ to $\mathbb{C}^9$ have not been verified by an explicit counting.\\

We recall that these sequences count HNF's which are invariant under cycles of conjugacy classes of the permutation group $S_D$. The HNF's are dual to abelian orbifolds of $\mathbb{C}^D$ where \tref{primetab} and \tref{primetab2} present the results for dimensions $D=2,3,4,5$. Using the cycle index of the permutation group $S_D$, the sequences which count $g^\alpha$-invariant HNF's are combined to count distinct abelian orbifolds of $\mathbb{C}^D$.\\

\begin{table}[ht!]
\begin{center}

\begin{tabular}{r|l}

\hline
\hline
\multicolumn{2}{c}{$\mathbb{C}^{2}/\Gamma_{N}$}
\\
\hline
\hline
$x^{\alpha}$ & $P_{\mathsf{g_{x^{\alpha}}}}(p)$
\\
\hline
$x_{1}^{2}$ & 
$1$ 
\\
$x_{2}$ & 
$\delta_{p,1\bmod{2}}+\delta_{p2}$
\\
\\

\hline
\hline
\multicolumn{2}{c}{$\mathbb{C}^{3}/\Gamma_{N}$}
\\
\hline
\hline
$x^{\alpha}$ & $P_{\mathsf{g_{x^{\alpha}}}}(p)$
\\
\hline
$x_{1}^{3}$ & 
$1+p$
\\
$x_{1}x_{2}$ & 
$1+\delta_{p,1\bmod{2}}$
\\
$x_{3}$ & 
$2\delta_{p,1\bmod{3}}+\delta_{p3}$
\\
\\

\hline
\hline
\multicolumn{2}{c}{$\mathbb{C}^{4}/\Gamma_{N}$}
\\
\hline
\hline
$x^{\alpha}$ & $P_{\mathsf{g_{x^{\alpha}}}}(p)$
\\
\hline
$x_{1}^{4}$ & 
$1+p+p^2$
\\
$x_{1}^{2}x_{2}$ & 
$1+p+\delta_{p,1\bmod{2}}$
\\
$x_{2}^{2}$ & 
$1+(1+p)\delta_{p,1\bmod{2}}+2\delta_{p2}$
\\
$x_{1}x_{3}$ &
$1+2\delta_{p,1\bmod{3}}$
\\
$x_{4}$ & 
$\delta_{p,1\bmod{2}}+2\delta_{p,1\bmod{4}}+\delta_{p2}$
\\
\\

\hline
\hline
\multicolumn{2}{c}{$\mathbb{C}^{5}/\Gamma_{N}$}
\\
\hline
\hline
$x^{\alpha}$ & $P_{\mathsf{g_{x^{\alpha}}}}(p)$
\\
\hline
$x_{1}^{5}$ & 
$1+p+p^{2}+p^{3}$
\\
$x_{1}^{3}x_{2}$ & 
$1+p+p^{2}+\delta_{p,1\bmod{2}}$
\\
$x_{1}x_{2}^{2}$ & 
$1+p+\delta_{p,1\bmod{2}}+p\delta_{p,1\bmod{2}}$
\\
$x_{1}^{2}x_{3}$ & 
$1+p+2\delta_{p,1\bmod{3}}$
\\
$x_{2}x_{3}$ & 
$1+\delta_{p,1\bmod{2}}+2\delta_{p,1\bmod{3}}$
\\
$x_{1}x_{4}$ & 
$1+\delta_{p,1\bmod{2}}+2\delta_{p,1\bmod{4}}$
\\
$x_{5}$ & 
$4\delta_{p,1\bmod{5}}+\delta_{p5}$
\\
\\

\hline
\hline
\end{tabular}

\caption{Derived functions on primes for symmetries of orbifolds of the form $\mathbb{C}^{2}/\Gamma_{N}$, $\mathbb{C}^{3}/\Gamma_{N}$, $\mathbb{C}^{4}/\Gamma_{N}$ and $\mathbb{C}^{5}/\Gamma_{N}$ where $N$ is prime.}
\label{gentab}

\end{center}
\end{table}

\begin{table}[ht!]
\begin{center}

\begin{tabular}{r|l}

\hline
\hline
\multicolumn{2}{c}{$\mathbb{C}^{6}/\Gamma_{N}$}
\\
\hline
\hline
$x^{\alpha}$ & $P_{\mathsf{g_{x^{\alpha}}}}(p)$
\\
\hline
$x_{1}^{6}$ & 
$1+p+p^{2}+p^{3}+p^{4}$
\\
$x_{1}^{4}x_{2}$ & 
$1+p+p^{2}+p^{3}+\delta_{p,1\bmod{2}}$
\\
$x_{1}^{2}x_{2}^{2}$ & 
$1+p+p^{2}+(1+p)\delta_{p,1\bmod{2}}$
\\
$x_{2}^{3}$ & 
$1+p+(1+p+p^2)\delta_{p,1\bmod{2}}+4\delta_{p2}$
\\
$x_{1}^{3}x_{3}$ & 
$1+p+p^{2}+2\delta_{p,1\bmod{3}}$
\\
$x_{1}x_{2}x_{3}$ & 
$1+p+\delta_{p,1\bmod{2}}+2\delta_{p,1\bmod{3}}$
\\
$x_{1}^{2}x_{4}$ & 
$1+p+\delta_{p,1\bmod{2}}+2\delta_{p,1\bmod{4}}$
\\
$x_{2}x_{4}$ & 
$1+(1+p)\delta_{p,1\bmod{2}}+2\delta_{p,1\bmod{4}}+2\delta_{p2}$
\\
$x_{3}^{2}$ & 
$1+2(1+p)\delta_{p,1\bmod{3}}+3\delta_{p3}$
\\
$x_{1}x_{5}$ & 
$1+4\delta_{p,1\bmod{5}}$
\\
$x_{6}$ & 
$\delta_{p,1\bmod{2}}+2\delta_{p,1\bmod{2}}+2\delta_{p,1\bmod{6}}+\delta_{p2}+\delta_{p3}$
\\
\\

\hline
\hline
\multicolumn{2}{c}{$\mathbb{C}^{7}/\Gamma_{N}$}
\\
\hline
\hline
$x^{\alpha}$ & $P_{\mathsf{g_{x^{\alpha}}}}(p)$
\\
\hline
$x_{1}^{7}$ & 
$1+p+p^{2}+p^{3}+p^{4}+p^{5}$ 
\\
$x_{1}^{5}x_{2}$ & 
$1+p+p^{2}+p^{3}+p^{4}+\delta_{p,1\bmod{2}}$ 
\\
$x_{1}^{3}x_{2}^{2}$ & 
$1+p+p^{2}+p^{3}+(1+p)\delta_{p,1\bmod{2}}$ 
\\
$x_{1}x_{2}^{3}$ & 
$1+p+p^{2}+(1+p+p^2)\delta_{p,1\bmod{2}}$ 
\\
$x_{1}^{4}x_{3}$ & 
$1+p+p^{2}+p^{3}+2\delta_{p,1\bmod{3}}$ 
\\
$x_{1}^{2}x_{2}x_{3}$ & 
$1+p+p^{2}+\delta_{p,1\bmod{2}}+2\delta_{p,1\bmod{3}}$ 
\\
$x_{2}^{2}x_{3}$ & 
$1+p+(1+p)\delta_{p,1\bmod{2}}+2\delta_{p,1\bmod{3}}$ 
\\
$x_{1}x_{3}^{2}$ & 
$1+p+2(1+p)\delta_{p,1\bmod{3}}$ 
\\
$x_{1}^{3}x_{4}$ & 
$1+p+p^{2}+\delta_{p,1\bmod{2}}+2\delta_{p,1\bmod{4}}$ 
\\
$x_{1}x_{2}x_{4}$ & 
$1+p+(1+p)\delta_{p,1\bmod{2}}+2\delta_{p,1\bmod{4}}$ 
\\
$x_{3}x_{4}$ & 
$1+\delta_{p,1\bmod{2}}+2\delta_{p,1\bmod{3}}+2\delta_{p,1\bmod{4}}$ 
\\
$x_{1}^{2}x_{5}$ & 
$1+p+4\delta_{p,1\bmod{5}}$
\\
$x_{2}x_{5}$ & 
$1+\delta_{p,1\bmod{2}}+4\delta_{p,1\bmod{5}}$ 
\\
$x_{1}x_{6}$ & 
$1+\delta_{p,1\bmod{2}}+2\delta_{p,1\bmod{3}}+2\delta_{p,1\bmod{6}}$ 
\\
$x_{7}$ & 
$6\delta_{p,1\bmod{7}}+\delta_{p7}$ 
\\
\\

\hline
\hline
\end{tabular}

\caption{Derived functions on primes for symmetries of orbifolds of the form $\mathbb{C}^{6}/\Gamma_{N}$ and $\mathbb{C}^{7}/\Gamma_{N}$ where $N$ is prime.}
\label{gentab2a}

\end{center}
\end{table}

\begin{table}[ht!]
\begin{center}

\begin{tabular}{r|l}

\hline
\hline
\multicolumn{2}{c}{$\mathbb{C}^{8}/\Gamma_{N}$}
\\
\hline
\hline
$x^{\alpha}$ & $P_{\mathsf{g_{x^{\alpha}}}}(p)$
\\
\hline
$x_{1}^{8}$ & 
$1+p+p^{2}+p^{3}+p^{4}+p^{5}+p^{6}$
\\
$x_{1}^{6}x_{2}$ & 
$1+p+p^{2}+p^{3}+p^{4}+p^{5}+\delta_{p,1\bmod{2}}$
\\
$x_{1}^{4}x_{2}^{2}$ & 
$1+p+p^{2}+p^{3}+p^{4}+(1+p)\delta_{p,1\bmod{2}}$
\\
$x_{1}^{2}x_{2}^{3}$ & 
$1+p+p^{2}+p^{3}+(1+p+p^2)\delta_{p,1\bmod{2}}$
\\
$x_{2}^{4}$ & 
$1+p+p^{2}+(1+p+p^2+p^3)\delta_{p,1\bmod{2}}+8\delta_{p2}$
\\
$x_{1}^{5}x_{3}$ & 
$1+p+p^{2}+p^{3}+p^{4}+2\delta_{p,1\bmod{3}}$
\\
$x_{1}^{3}x_{2}x_{3}$ & 
$1+p+p^{2}+p^{3}+\delta_{p,1\bmod{2}}+2\delta_{p,1\bmod{3}}$
\\
$x_{1}x_{2}^{2}x_{3}$ & 
$1+p+p^{2}+(1+p)\delta_{p,1\bmod{2}}+2\delta_{p,1\bmod{3}}$
\\
$x_{1}^{2}x_{3}^{2}$ & 
$1+p+p^{2}+2(1+p)\delta_{p,1\bmod{3}}$
\\
$x_{2}x_{3}^{2}$ & 
$1+{p}+\delta_{p,1\bmod{2}}+2(1+p)\delta_{p,1\bmod{3}}$
\\
$x_{1}^{4}x_{4}$ & 
$1+p+p^{2}+p^{3}+\delta_{p,1\bmod{2}}+2\delta_{p,1\bmod{4}}$
\\
$x_{1}^{2}x_{2}x_{4}$ & 
$1+p+p^{2}+(1+p)\delta_{p,1\bmod{2}}+2\delta_{p,1\bmod{4}}$
\\
$x_{2}^{2}x_{4}$ & 
$1+p+(1+p)\delta_{p,1\bmod{2}}+2\delta_{p,1\bmod{2}}+4\delta_{p2}$
\\
$x_{1}x_{3}x_{4}$ & 
$1+p+2\delta_{p,1\bmod{3}}+2\delta_{p,1\bmod{4}}$
\\
$x_{4}^{2}$ & 
$1+(1+p)\delta_{p,1\bmod{2}}+2(1+p)\delta_{p,1\bmod{4}}$
\\
$x_{1}^{3}x_{5}$ & 
$1+p+p^{2}+4\delta_{p,1\bmod{5}}$
\\
$x_{1}x_{2}x_{5}$ & 
$1+p+4\delta_{p,1\bmod{5}}$
\\
$x_{3}x_{5}$ & 
$1+2\delta_{p,1\bmod{3}}+4\delta_{p,1\bmod{5}}$
\\
$x_{1}^{2}x_{6}$ & 
$1+p+\delta_{p,1\bmod{2}}+2\delta_{p,1\bmod{3}}+2\delta_{p,1\bmod{6}}$
\\
$x_{2}x_{6}$ & 
$1+(1+p)\delta_{p,1\bmod{2}}+2\delta_{p,1\bmod{3}}+2\delta_{p,1\bmod{6}}+2\delta_{p2}$
\\
$x_{1}x_{7}$ & 
$1+6\delta_{p,1\bmod{7}}$
\\
$x_{8}$ & 
$\delta_{p,1\bmod{2}}+2\delta_{p,1\bmod{4}}+4\delta_{p,1\bmod{8}}+\delta_{p2}$
\\
\\

\hline
\hline
\end{tabular}

\caption{Derived functions on primes for symmetries of orbifolds of the form $\mathbb{C}^{8}/\Gamma_{N}$ where $N$ is prime.}
\label{gentab2b}

\end{center}
\end{table}

\begin{table}[ht!]
\begin{center}

\begin{tabular}{r|l}

\hline
\hline
\multicolumn{2}{c}{$\mathbb{C}^{9}/\Gamma_{N}$}
\\
\hline
\hline
$x^{\alpha}$ & $P_{\mathsf{g_{x^{\alpha}}}}(p)$
\\
\hline
$x_{1}^{9}$ & 
$1+p+p^{2}+p^{3}+p^{4}+p^{5}+p^{6}+p^{7}$ 
\\
$x_{1}^{7}x_{2}$ & 
$1+p+p^{2}+p^{3}+p^{4}+p^{5}+p^{6}+\delta_{p,1\bmod{2}}$ 
\\
$x_{1}^{5}x_{2}^{2}$ & 
$1+p+p^{2}+p^{3}+p^{4}+p^{5}+(1+p)\delta_{p,1\bmod{2}}$ 
\\
$x_{1}^{3}x_{2}^{3}$ & 
$1+p+p^{2}+p^{3}+p^{4}+(1+p+p^2)\delta_{p,1\bmod{2}}$ 
\\
$x_{1}x_{2}^{4}$ & 
$1+p+p^{2}+p^{3}+(1+p+p^2+p^3)\delta_{p,1\bmod{2}}$ 
\\
$x_{1}^{6}x_{3}$ & 
$1+p+p^{2}+p^{3}+p^{4}+p^{5}+2\delta_{p,1\bmod{3}}$ 
\\
$x_{1}^{4}x_{2}x_{3}$ & 
$1+p+p^{2}+p^{3}+p^{4}+\delta_{p,1\bmod{2}}+2\delta_{p,1\bmod{3}}$ 
\\
$x_{1}^{2}x_{2}^{2}x_{3}$ & 
$1+p+p^{2}+p^{3}+(1+p)\delta_{p,1\bmod{2}}+2\delta_{p,1\bmod{3}}$ 
\\
$x_{2}^{3}x_{3}$ & 
$1+p+p^{2}+(1+p+p^2)\delta_{p,1\bmod{2}}+2\delta_{p,1\bmod{3}}$ 
\\
$x_{1}^{3}x_{3}^{2}$ & 
$1+p+p^{2}+p^{3}+2(1+p)\delta_{p,1\bmod{3}}$ 
\\
$x_{1}x_{2}x_{3}^{2}$ & 
$1+p+p^{2}+\delta_{p,1\bmod{2}}+2(1+p)\delta_{p,1\bmod{3}}$ 
\\
$x_{3}^{3}$ & 
$1+p+2(1+p+p^2)\delta_{p,1\bmod{3}}+9\delta_{p3}$ 
\\
$x_{1}^{5}x_{4}$ & 
$1+p+p^{2}+p^{3}+p^{4}+\delta_{p,1\bmod{2}}+2\delta_{p,1\bmod{4}}$ 
\\
$x_{1}^{3}x_{2}x_{4}$ & 
$1+p+p^{2}+p^{3}+(1+p)\delta_{p,1\bmod{2}}+2\delta_{p,1\bmod{4}}$ 
\\
$x_{1}x_{2}^{2}x_{4}$ & 
$1+p+p^{2}+(1+p+p^2)\delta_{p,1\bmod{2}}+2\delta_{p,1\bmod{4}}$ 
\\
$x_{1}^{2}x_{3}x_{4}$ & 
$1+p+p^{2}+\delta_{p,1\bmod{2}}+2\delta_{p,1\bmod{3}}+2\delta_{p,1\bmod{4}}$ 
\\
$x_{2}x_{3}x_{4}$ & 
$1+p+(1+p)\delta_{p,1\bmod{2}}+2\delta_{p,1\bmod{3}}+2\delta_{p,1\bmod{4}}$ 
\\
$x_{1}x_{4}^{2}$ & 
$1+p+(1+p)\delta_{p,1\bmod{2}}+2(1+p)\delta_{p,1\bmod{4}}$ 
\\
$x_{1}^{4}x_{5}$ & 
$1+p+p^{2}+p^{3}+4\delta_{p,1\bmod{5}}$ 
\\
$x_{1}^{2}x_{2}x_{5}$ & 
$1+p+p^{2}+\delta_{p,1\bmod{2}}+4\delta_{p,1\bmod{5}}$ 
\\
$x_{2}^{2}x_{5}$ & 
$1+p+(1+p)\delta_{p,1\bmod{2}}+4\delta_{p,1\bmod{5}}$ 
\\
$x_{1}x_{3}x_{5}$ & 
$1+p+2\delta_{p,1\bmod{3}}+4\delta_{p,1\bmod{5}}$ 
\\
$x_{4}x_{5}$ & 
$1+\delta_{p,1\bmod{2}}+2\delta_{p,1\bmod{4}}+4\delta_{p,1\bmod{5}}$ 
\\
$x_{1}^{3}x_{6}$ & 
$1+p+p^{2}+\delta_{p,1\bmod{2}}+2\delta_{p,1\bmod{3}}+2\delta_{p,1\bmod{6}}$ 
\\
$x_{1}x_{2}x_{6}$ & 
$1+p+(1+p)\delta_{p,1\bmod{2}}+2\delta_{p,1\bmod{3}}+2\delta_{p,1\bmod{6}}$ 
\\
$x_{3}x_{6}$ & 
$1+\delta_{p,1\bmod{2}}+2(1+p)\delta_{p,1\bmod{3}}+2\delta_{p,1\bmod{6}}+3\delta_{p3}$ 
\\
$x_{1}^{2}x_{7}$ & 
$1+p+6\delta_{p,1\bmod{7}}$ 
\\
$x_{2}x_{7}$ & 
$1+\delta_{p,1\bmod{2}}+6\delta_{p,1\bmod{7}}$
\\
$x_{1}x_{8}$ & 
$1+\delta_{p,1\bmod{2}}+2\delta_{p,1\bmod{4}}+4\delta_{p,1\bmod{8}}$ 
\\
$x_{9}$ & 
$2\delta_{p,1\bmod{3}}+6\delta_{p,1\bmod{9}}+\delta_{p3}$ 
\\
\\

\hline
\hline
\end{tabular}

\caption{Derived functions on primes for symmetries of orbifolds of the form $\mathbb{C}^{9}/\Gamma_{N}$ where $N$ is prime.}
\label{gentab3}

\end{center}
\end{table}

\clearpage

%%%%%%%%%%%%%%%%%%%%%%%%%%%%%%%%%%%%%%%%%%%%%%%%%%%%%%%%%%%%%%%%%%%%%%%%%
\subsection{Predictions for Prime Index Sequences for Orbifolds of $\mathbb{C}^{D}$ \label{apppriseq}}

\begin{figure}[ht!]
\begin{center}
\includegraphics[totalheight=9cm]{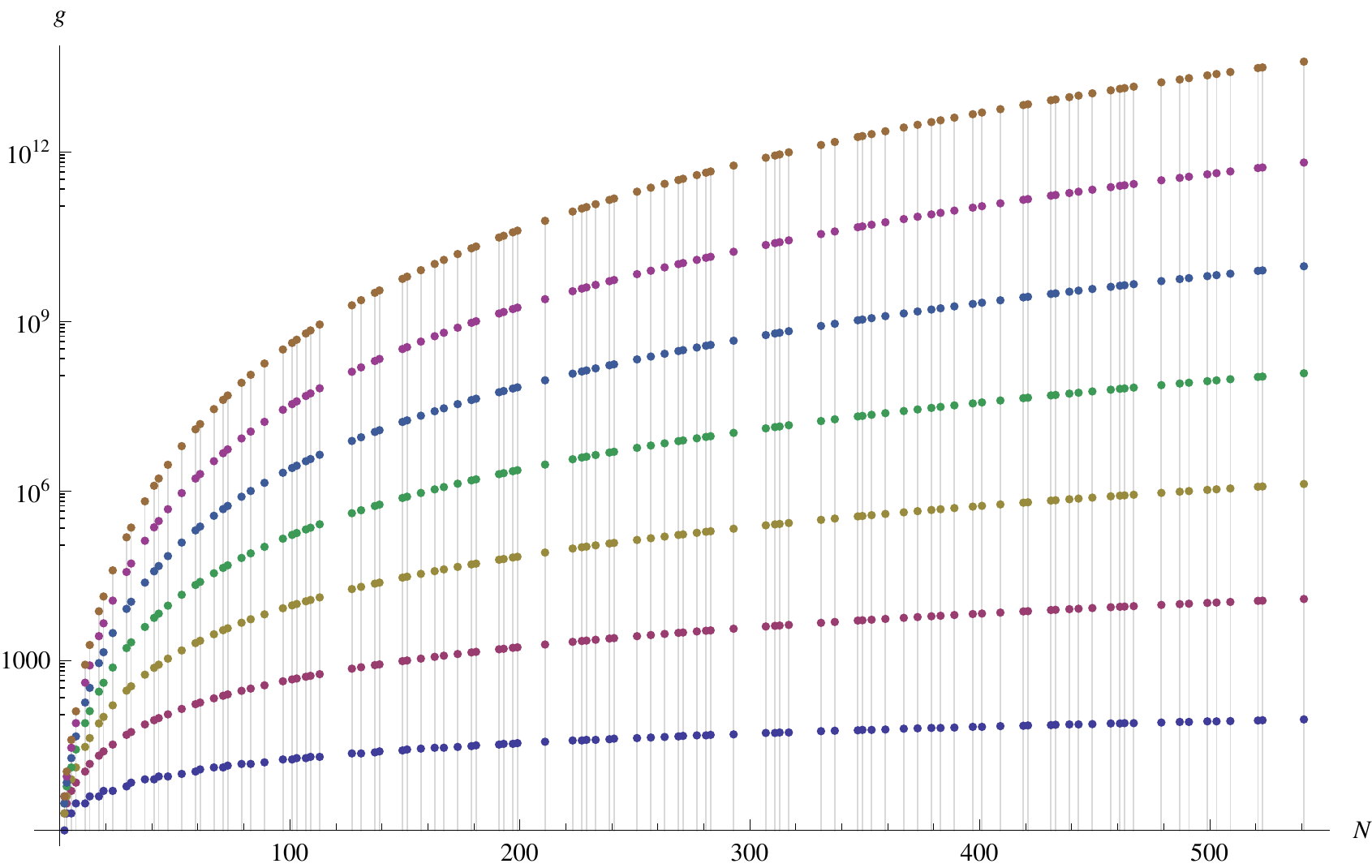}
\caption{The orbifold counting for $\mathbb{C}^{3}/\Gamma_{N}$ to $\mathbb{C}^{9}/\Gamma_{N}$ with prime $N$. The ordering of the sequences reflects the dimension of the orbifolds, with logarithmic differences between consecutive sequences approaching $log(p/D)$ at $p\rightarrow\infty$.}
  \label{primeplot2}
 \end{center}
\end{figure}

\textbf{Sequence Predictions for higher dimensional orbifolds.}
Using the observations in Section \sref{sprimeprop} and the cycle indices in \tref{fcycle}, we are able to derive the prime index sequences which count distinct orbifolds of the form $\mathbb{C}^{D}/\Gamma_{p}=\mathbb{C}^{D}/\mathbb{Z}_{p}$. The counting for distinct abelian orbifolds of the form $\mathbb{C}^{7}/\Gamma_{p}$, $\mathbb{C}^{8}/\Gamma_{p}$ and $\mathbb{C}^{9}/\Gamma_{p}$ are presented in \tref{predc7}, \tref{predc8} and \tref{predc9} respectively. The sequences which count $g^\alpha$-symmetric HNF's match the derivations in Section \sref{scomplete}. \textit{Explicit counting which matches with the predictions is marked by a $*$ in \tref{predc7}, \tref{predc8} and \tref{predc9}}.
\\

\textbf{The large $N$ limit.}
\fref{primeplot2} shows a logarithmic plot of the prime index sequences which count distinct orbifolds of the form $\mathbb{C}^{3}/\Gamma_p$ to $\mathbb{C}^{9}/\Gamma_p$. In the limit $p\rightarrow \infty$, the logarithmic difference between consecutive sequences becomes 
\beql{loglimit}
\lim_{p\rightarrow\infty}{\log{\left(\frac{\mathsf{g}^D(p)}{\mathsf{g}^{D-1}(p)}\right)}} = \log{\left(\frac{p}{D}\right)}~~.
\eeq
This confirms the asymptotic behavior analysis from \cite{HananyOrlando10}.

%%%%%%%%%%%%% C7 %%%%%%%%%%%%%%%%%%%%%%%%%%%%
\begin{table}[h!]
\begin{center}
\begin{tabular}{c|p{10mm}p{10mm}p{10mm}p{10mm}p{10mm}p{10mm}p{10mm}p{10mm}p{10mm}p{10mm}}
\hline
\hline
\multicolumn{11}{c}{$\mathbb{C}^{7}/\Gamma_{N}$}
\\
\hline
\hline
$N$ & 
2* & 3* & 5* & 7* & 11 & 13 & 17 & 19 & 23 & 29
\\
\hline

$x_{1}^{7}$ & 
63* & 364* & 3906* & 19608* & 177156 & 402234 & 
 \footnotesize{1508598} & \footnotesize{2613660} & \footnotesize{6728904} & \scriptsize{21243690}
\\

$x_{1}^{5}x_{2}$ & 
31* & 122* & 782* & 2802* & 16106 & 30942 & 
 88742 & 137562 & 292562 & 732542 
\\

$x_{1}^{3}x_{2}^{2}$ & 
15* & 44* & 162* & 408* & 1476 & 2394 & 
 5238 & 7260 & 12744 & 25290 
\\

$x_{1}x_{2}^{3}$ & 
7* & 26* & 62* & 114* & 266 & 366 & 614 &
 762 & 1106 & 1742 
\\

$x_{1}^{4}x_{3}$ & 
15* & 40* & 156* & 402* & 1464 & 2382 & 
 5220 & 7242 & 12720 & 25260 
\\

$x_{1}^{2}x_{2}x_{3}$ & 
7* & 14* & 32* & 60* & 134 & 186 & 308 & 
 384 & 554 & 872 
\\

$x_{2}^{2}x_{3}$ & 
3* & 8* & 12* & 18* & 24 & 30 & 36 & 
 42 & 48 & 60 
\\

$x_{1}x_{3}^{2}$ & 
3* & 4* & 6* & 24* & 12 & 42 & 18 & 
 60 & 24 & 30 
\\

$x_{1}^{3}x_{4}$ & 
7* & 14* & 34* & 58* & 134 & 186 & 310 &
 382 & 554 & 874 
\\

$x_{1}x_{2}x_{4}$ & 
3* & 8* & 14* & 16* & 24 & 30 & 38 & 
 40 & 48 & 62 
\\

$x_{3}x_{4}$ & 
1* & 2* & 4* & 4* & 2 & 6 & 4 & 4 & 
 2 & 4 
\\

$x_{1}^{2}x_{5}$ & 
3* & 4* & 6* & 8* & 16 & 14 & 18 & 
 20 & 24 &30 
\\

$x_{2}x_{5}$ & 
1* & 2* & 2* & 2* & 6 & 2 & 2 & 2 & 
 2 & 2 
\\

$x_{1}x_{6}$ & 
1* & 2* & 2* & 6* & 2 & 6 & 2 & 6 & 
 2 & 2 
\\

$x_{7}$ & 
0* & 0* & 0* & 1* & 0 & 0 & 0 & 0 & 0 & 6
\\

\hline

$g^{D=7}$ & 
3* &7* & 19* & 46* & 183 & 333 & 912 & 
 1421 & 3101 & 8307 
\\

\hline
\hline
$N$ & 
31 & 37 & 41 & 43 & 47 & 53 & 59 & 61 & 67 & 71
\\
\hline

$x_{1}^{7}$ & 
\scriptsize{29583456} & \scriptsize{71270178} & \tiny{118752606} & \tiny{150508644} & 
 \tiny{234330768} & \tiny{426237714} & \tiny{727250580} & \tiny{858672906} & 
 \tiny{1370581548} & \tiny{1830004056}
\\

$x_{1}^{5}x_{2}$ & 
954306 & \footnotesize{1926222} & \footnotesize{2896406} & \footnotesize{3500202} & 
 \footnotesize{4985762} & \footnotesize{8042222} & \scriptsize{12326282} & \scriptsize{14076606} & 
 \scriptsize{20456442} & \scriptsize{25774706}
\\

$x_{1}^{3}x_{2}^{2}$ & 
30816 & 52098 & 70686 & 81444 & 106128 & 
 151794 & 208980 & 230826 & 305388 & 363096 
\\

$x_{1}x_{2}^{3}$ &  
1986 & 2814 & 3446 & 3786 & 4514 & 
 5726 & 7082 & 7566 & 9114 & 10226 
\\

$x_{1}^{4}x_{3}$ & 
30786 & 52062 & 70644 & 81402 & 106080 & 
 151740 & 208920 & 230766 & 305322 & 363024 
\\

$x_{1}^{2}x_{2}x_{3}$ & 
996 & 1410 & 1724 & 1896 & 2258 & 
 2864 & 3542 & 3786 & 4560 & 5114 
\\

$x_{2}^{2}x_{3}$ & 
66 & 78 & 84 & 90 & 96 & 108 & 
 120 & 126 & 138 & 144 
\\

$x_{1}x_{3}^{2}$ & 
96 & 114 & 42 & 132 & 48 & 54 & 
 60 & 186 & 204 & 72 
\\

$x_{1}^{3}x_{4}$ & 
994 & 1410 & 1726 & 1894 & 2258 & 
 2866 & 3542 & 3786 & 4558 & 5114 
\\

$x_{1}x_{2}x_{4}$ & 
64 & 78 & 86 & 88 & 96 & 110 & 
 120 & 126 & 136 & 144 
\\

$x_{3}x_{4}$ & 
4 & 6 & 4 & 4 & 2 & 4 & 2 & 
 6 & 4 & 2 
\\

$x_{1}^{2}x_{5}$ & 
36 & 38 & 46 & 44 & 48 & 54 & 60 & 
 66 & 68 & 76 
\\

$x_{2}x_{5}$ & 
6 & 2 & 6 & 2 & 2 & 2 & 2 & 
 6 & 2 & 6 
\\

$x_{1}x_{6}$ & 
6 & 6 & 2 & 6 & 2 & 2 & 2 & 
 6 & 6 & 2 
\\

$x_{7}$ & 
0 & 0 & 0 & 6 & 0 & 0 & 0 & 0 & 0 & 6
\\

\hline

$g^{D=7}$ & 
11103 & 24235 & 38394 & 47619 & 71353 & 
 123855 & 203531 & 237709 & 368581 & 483987 \\

\hline
\hline

\end{tabular}
\caption{The derived symmetry count for the orbifolds of the form $\mathbb{C}^{7}/\Gamma_{N}$ with prime $N$. \textit{The values on indices marked by a * have been verified by explicit counting}.}
\label{predc7}
\end{center}
\end{table}

%%%%%%%%%%%%% C8 %%%%%%%%%%%%%%%%%%%%%%%%%%%%
\begin{table}[h!]
\begin{center}
\begin{tabular}{c|p{10mm}p{10mm}p{10mm}p{10mm}p{10mm}p{10mm}p{10mm}p{10mm}p{10mm}p{10mm}}
\hline
\hline
\multicolumn{11}{c}{$\mathbb{C}^{8}/\Gamma_{N}$}
\\
\hline
\hline
$N$ & 
2 & 3 & 5 & 7 & 11 & 13 & 17 & 19 & 23 & 29
\\
\hline

$x_{1}^{8}$ & 
127 & 1093 & 19531 & 137257 & \footnotesize{1948717} & 
 \footnotesize{5229043} & \scriptsize{25646167} & \scriptsize{49659541} & \scriptsize{154764793} &
 \scriptsize{616067011}
\\

$x_{1}^{6}x_{2}$ & 
63 & 365 & 3907 & 19609 & 177157 & 402235 &
 \footnotesize{1508599} & \footnotesize{2613661} & \footnotesize{6728905} & \footnotesize{21243691} 
\\

$x_{1}^{4}x_{2}^{2}$ & 
31 & 125 & 787 & 2809 & 16117 & 30955 & 
 88759 & 137581 & 292585 & 732571 
\\

$x_{1}^{2}x_{2}^{3}$ & 
15 & 53 & 187 & 457 & 1597 & 2563 &
 5527 & 7621 & 13273 & 26131
\\

$x_{2}^{4}$ & 
15 & 53 & 187 & 457 & 1597 & 2563 & 
 5527 & 7621 & 13273 & 26131 
\\

$x_{1}^{5}x_{3}$ & 
31 & 121 & 781 & 2803 & 16105 & 30943 &
 88741 & 137563 & 292561 & 732541 
\\

$x_{1}^{3}x_{2}x_{3}$ & 
15 & 41 & 157 & 403 & 1465 & 2383 & 
 5221 & 7243 & 12721 & 25261
\\

$x_{1}x_{2}^{2}x_{3}$ & 
7 &17 & 37 & 67 & 145& 199 & 325 & 
 403 & 577 & 901 
\\

$x_{1}^{2}x_{3}^{2}$ & 
7 & 13 & 31 & 73 & 133 & 211 & 307 & 
 421 & 553 & 871 
 \\

$x_{2}x_{3}^{2}$ & 
3 & 5 & 7 & 25 & 13 & 43 & 19 & 
 61 & 25 & 31 
 \\

$x_{1}^{4}x_{4}$ & 
15 & 41 & 159 & 401 & 1465 & 2383 & 
 5223 & 7241 & 12721 & 25263 
 \\

$x_{1}^{2}x_{2}x_{4}$ & 
7 & 17 & 39 & 65 & 145 & 199 & 327 & 
 401 &577 & 903 
 \\

$x_{2}^{2}x_{4}$ & 
7 & 17 & 39 & 65 & 145& 199 & 327 & 
 401 & 577 & 903 
 \\

$x_{1}x_{3}x_{4}$ & 
3 & 5 & 9 & 11 & 13 & 19 & 21 & 
 23 & 25 & 33 
 \\

$x_{4}^{2}$ & 
3 & 5 & 19 & 9 & 13 & 43 & 55 & 
 21 & 25 & 91 
 \\

$x_{1}^{3}x_{5}$ & 
7 & 13 & 31 & 57 & 137 & 183 & 307 & 
 381 & 553 & 871 
 \\

$x_{1}x_{2}x_{5}$ & 
3 & 5 & 7 & 9 & 17 & 15& 19 & 
 21 & 25 & 31 
 \\

$x_{3}x_{5}$ & 
1 & 1 & 1& 3 & 5 & 3 & 1& 
 3 & 1& 1
 \\

$x_{1}^{2}x_{6}$ & 
3 & 5 & 7 & 13& 13 & 19 & 19 & 
 25 & 25 & 31 
 \\

$x_{2}x_{6}$ & 
3 & 5 & 7 & 13 & 13 & 19 & 19 &
 25 & 25 & 31 
\\

$x_{1}x_{7}$ & 
1&1 & 1& 1& 1& 1& 1& 1& 1& 7 
\\

$x_{8}$ & 
1&1& 3& 1&1& 3& 7& 1&1& 
 3 
 \\

\hline

$g^{D=8}$ &  
4 & 9 & 29 & 79 & 411 & 829 & 2737 & 
 4611 & 11629 & 37379 
\\

\hline
\hline
\end{tabular}
\caption{The derived symmetry count for the orbifolds of the form $\mathbb{C}^{8}/\Gamma_{N}$ with prime $N$.}
\label{predc8}
\end{center}
\end{table}

%%%%%%%%%%%%% C9 %%%%%%%%%%%%%%%%%%%%%%%%%%%%
\begin{table}[h!]
\begin{center}
\begin{tabular}{c|p{10mm}p{10mm}p{10mm}p{10mm}p{10mm}p{10mm}p{10mm}p{10mm}p{10mm}p{10mm}}
\hline
\hline
\multicolumn{11}{c}{$\mathbb{C}^{9}/\Gamma_{N}$}
\\
\hline
\hline
$N$ & 
2 & 3 & 5 & 7 & 11 & 13 & 17 & 19 & 23 & 29
\\
\hline

$x_{1}^{9}$ & 
255 & 3280 & 97656 & 960800 & \scriptsize{21435888} & 
 \scriptsize{67977560} & \tiny{435984840} & \tiny{943531280} & \tiny{3559590240} & 
 \tiny{17865943320}
\\

$x_{1}^{7}x_{2}$ & 
127 & 1094 & 19532 & 137258 & \footnotesize{1948718} & 
 \footnotesize{5229044} & \scriptsize{25646168} & \scriptsize{49659542} & \tiny{154764794} & \tiny{616067012}
\\

$x_{1}^{5}x_{2}^{2}$ & 
63 & 368 & 3912 & 19616 & 177168 & 402248 & 
 \footnotesize{1508616} & \footnotesize{2613680} & \footnotesize{6728928} & \scriptsize{21243720}
\\

$x_{1}^{3}x_{2}^{3}$ & 
31 & 134 & 812 & 2858 & 16238 & 31124 & 
 89048 & 137942 & 293114 & 733412
\\

$x_{1}x_{2}^{4}$ & 
15 & 80 & 312 & 800 & 2928 & 4760 & 
 10440 & 14480 & 25440 & 50520 
\\

$x_{1}^{6}x_{3}$ & 
63 & 364 & 3906 & 19610 & 177156 & 402236 & 
 \footnotesize{1508598} & \footnotesize{2613662} & \footnotesize{6728904} & \scriptsize{21243690}
\\

$x_{1}^{4}x_{2}x_{3}$ & 
31 & 122 & 782 & 2804 & 16106 & 30944 & 
 88742 & 137564 & 292562 & 732542 
\\

$x_{1}^{2}x_{2}^{2}x_{3}$ & 
15 & 44 & 162 & 410 & 1476 & 2396 & 
 5238 & 7262 & 12744 & 25290 
\\

$x_{2}^{3}x_{3}$ & 
7 & 26 & 62 & 116 & 266 & 368 & 614 & 
 764 & 1106 & 1742 
\\

$x_{1}^{3}x_{3}^{2}$ & 
15 & 40 & 156 & 416 & 1464 & 2408 & 
 5220 & 7280 & 12720 & 25260 
\\

$x_{1}x_{2}x_{3}^{2}$ & 
7 & 14 & 32 & 74 & 134 & 212 & 308 & 
 422 & 554 & 872 
\\

$x_{3}^{3}$ & 
3 & 13 & 6 & 122 & 12 & 380 & 18 & 
 782 & 24 & 30 
\\

$x_{1}^{5}x_{4}$ & 
31 & 122 & 784 & 2802 & 16106 & 30944 & 
 88744 & 137562 & 292562 & 732544 
\\

$x_{1}^{3}x_{2}x_{4}$ & 
15 & 44 & 164 & 408 & 1476 & 2396 & 
 5240 & 7260 & 12744 & 25292 
\\

$x_{1}x_{2}^{2}x_{4}$ & 
7 & 26 & 64 & 114 & 266 & 368 & 616 & 
 762 & 1106 & 1744 
\\

$x_{1}^{2}x_{3}x_{4}$ & 
7 & 14 & 34 & 60 & 134 & 188 & 310 & 
 384 & 554 & 874 
\\

$x_{2}x_{3}x_{4}$ & 
3 & 8 & 14 & 18 & 24 & 32 & 38 & 
 42 & 48 & 62 
\\

$x_{1}x_{4}^{2}$ & 
3 & 8 & 24 & 16 & 24 & 56 & 72 & 
 40 & 48 & 120 
\\

$x_{1}^{4}x_{5}$ & 
15 & 40 & 156 & 400 & 1468 & 2380 & 
 5220 & 7240 & 12720 & 25260 
\\

$x_{1}^{2}x_{2}x_{5}$ & 
7 & 14 & 32 & 58 & 138 & 184 & 308 & 
 382 & 554 & 872 
\\

$x_{2}^{2}x_{5}$ & 
3 & 8 & 12 & 16 & 28 & 28 & 36 & 
 40 & 48 & 60 
\\

$x_{1}x_{3}x_{5}$ & 
3 & 4 & 6 & 10 & 16 & 16 & 18 & 
 22 & 24 & 30 
\\

$x_{4}x_{5}$ & 
1& 2 & 4 & 2 & 6 & 4 & 4 & 2 & 
 2 & 4 
\\

$x_{1}^{3}x_{6}$ & 
7 & 14 & 32 & 62 & 134 & 188 & 308 & 
 386 & 554 & 872 
\\

$x_{1}x_{2}x_{6}$ & 
3 & 8 & 12 & 20 & 24 & 32 & 36 & 
 44 & 48 & 60 
\\

$x_{3}x_{6}$ & 
1& 5 & 2 & 20 & 2 & 32 & 2 & 44 & 
 2 & 2 
\\

$x_{1}^{2}x_{7}$ & 
3 & 4 & 6 & 8 & 12 & 14 & 18 & 
 20 & 24 & 36 
\\

$x_{2}x_{7}$ & 
1& 2 & 2 & 2 & 2 & 2 & 2 & 2 & 
 2 & 8 
\\

$x_{1}x_{8}$ & 
1& 2 & 4 & 2 & 2 & 4 & 8 & 2 & 
 2 & 4 
\\

$x_{9}$ & 
0 & 1 & 0 & 2 & 0 & 2 & 0 & 8 & 0 & 0
\\

\hline

$g^{D=9}$ &  
4 & 11 & 40 & 128 & 853 & 1909 & 
 7544 & 13754 & 39904 & 153319 
\\

\hline
\hline
\end{tabular}
\caption{The derived symmetry count for the orbifolds of the form $\mathbb{C}^{9}/\Gamma_{N}$ with prime $N$.}
\label{predc9}
\end{center}
\end{table}

\clearpage

%%%%%%%%%%%%%%%%%%%%%%%%%%%%%%%%%%%%%%%%%%%%%%%%%%%%%%%%%%%%%%%%%%%%%%%%%
\section{Combinatorics of Symmetries \label{sp3}}

Above, $\mathsf{g}_{x^\alpha}(N)$ counts the number of HNF's in the set $D(N)$ which are symmetric under cycles $g^\alpha \in H_\alpha \subset S_D$. The set of all HNF's at order $N$, $D(N)$, is partitioned into $\mathsf{g}^{D}(N)$ subsets $[\sigma^{D-1}]$ where each subset corresponds to a distinct abelian orbifold of the form $\mathbb{C}^{D}/\Gamma_N$. The value of $\mathsf{g}^{D}(N)$ is obtained by averaging the values of $\mathsf{g}_{x^\alpha}(N)$ according to the cycle index of $S_D$.\\

As part of an alternative way of counting, let $\mathsf{g}_{x^\alpha}(N)$ count the number of $g^\alpha$-symmetric HNF's in a subset $[\sigma^{D-1}]\subset D(N)$. Accordingly, a distinct orbifold of the form $\mathbb{C}^{D}/\Gamma_N$ which corresponds to $[\sigma^{D-1}]$ is assigned a $N_H$-plet of integers, 
\beql{ech}
[\mathsf{g}_{x^1}(N),\dots,\mathsf{g}_{x^\alpha}(N),\dots,\mathsf{g}_{x^{N_{H}}}(N)]~~,
\eeq
where $N_H$ is the number of conjugacy classes of $S_D$. Let \eref{ech} be called the \textbf{Symmetry Characteristic} of the orbifold of the form $\mathbb{C}^D/\Gamma_N$ and the corresponding subset $[\sigma^{D-1}]\subset D(N)$. By definition, since a symmetry characteristic is assigned to a single abelian orbifold of the form $\mathbb{C}^D/\Gamma_N$, the values of the $N_H$-plet $[\mathsf{g}_{x^1}(N),\dots,\mathsf{g}_{x^\alpha}(N),\dots,\mathsf{g}_{x^{N_{H}}}(N)]$ averaged under the cycle index of $S_D$ gives always $\mathsf{g}^{D}(N)=1$.\\

Two distinct abelian orbifolds of the form $\mathbb{C}^D/\Gamma_N$ which correspond to two distinct subsets $[\sigma^{D-1}]\subset D(N)$ may have the same symmetry characteristic $[\mathsf{g}_{x^1}(N),\dots,\mathsf{g}_{x^\alpha}(N),\dots,\mathsf{g}_{x^{N_{H}}}]$. Accordingly, we are able to count how many subsets in $D(N)$ at order $N$ have a symmetry characteristic $[\mathsf{g}_{x^1}(N),\dots,\mathsf{g}_{x^\alpha}(N),\dots,\mathsf{g}_{x^{N_{H}}}]$. The explicit counting for distinct abelian orbifolds of $\mathbb{C}^3$, $\mathbb{C}^4$ and $\mathbb{C}^5$ is shown in Tables \ref{tc3symdist}-\ref{tc5symdist}.\\

%%%%%%%%%%%%% C3 %%%%%%%%%%%%%%%%%%%%%%%%%%%%
\begin{table}[ht!]
\begin{center}
\begin{tabular}{c|p{5mm}p{5mm}p{5mm}p{5mm}p{5mm}p{5mm}p{5mm}p{5mm}p{5mm}p{5mm}p{5mm}p{5mm}p{5mm}p{5mm}p{5mm}}
\hline
\hline
\multicolumn{16}{c}{$\mathbb{C}^{3}/\Gamma_{N}$}
\\
\hline
\hline
$N$ & 
1 & 2 & 3 & 4 & 5 & 6 & 7 & 8 & 9 & 10 & 11 & 12 & 13 & 14 & 15
\\
\hline
$[1,1,1]$ & 
1& 0& 1& 1& 0& 0& 0& 0& 1& 0& 0& 1& 0& 0& 0
\\
$[3,1,0]$ & 
0& 1& 1& 2& 2& 2& 2& 5& 2& 2& 2& 5& 2& 2& 4
\\
$[6,0,0]$ & 
0& 0& 0& 0& 0& 1& 0& 0& 1& 2& 1& 2& 1& 3& 2
\\
$[2,0,2]$ & 
0& 0& 0& 0& 0& 0& 1& 0& 0& 0& 0& 0& 1& 0& 0
\\
\hline
$g^{D=3}$ & 
1 & 1 & 2& 3& 2& 3& 3& 5& 4 & 4& 3& 8& 4& 5 & 6
\\
\hline
\hline
$N$ & 
16 & 17 & 18 & 19 & 20 & 21 & 22 & 23 & 24 & 25 & 26 & 27 & 28 & 29 & 30
\\
\hline
$[1,1,1]$ & 
1& 0& 0& 0& 0& 0& 0& 0& 0& 1& 0& 1& 0& 0& 0
\\
$[3,1,0]$ & 
6& 2& 3& 2& 6& 4& 2& 2& 10& 2& 2& 3& 6& 2& 4
\\
$[6,0,0]$ & 
2& 2& 5& 2& 4& 3&5& 3& 5& 4& 6& 5& 6& 4& 10
\\
$[2,0,2]$ & 
0& 0& 0& 1& 0& 1& 0& 0& 0& 0& 0& 0& 1& 0& 0
\\
\hline
$g^{D=3}$ & 
9 & 4 & 8 & 5 & 10 & 8& 7& 5& 15& 7& 8& 9& 13& 6 & 14
\\
\hline
\hline\end{tabular}
\caption{The number of distinct toric triangles in $D(N)$ with symmetry characteristic $[g_{x_{1}^{3}},g_{x_1 x_2},g_{x_3}]$.}
\label{tc3symdist}
\end{center}
\end{table}

Since there is only a finite number of subsets of $D(N)$ which correspond to distinct abelian orbifolds of the form $\mathbb{C}^D/\Gamma_N$, there is only a finite number of possibly assigned symmetry characteristics. With the property that the values $\mathsf{g}_{x^\alpha}$ in the $N_H$-plet of the symmetric character are required to give the $S_D$ cycle index value $\mathsf{g}^{D}(N)=1$, we are able to predict candidates of assigned symmetry characteristics. Let the following example illustrate this idea.\\

\begin{figure}[ht!]
\begin{center}
\includegraphics[totalheight=4cm]{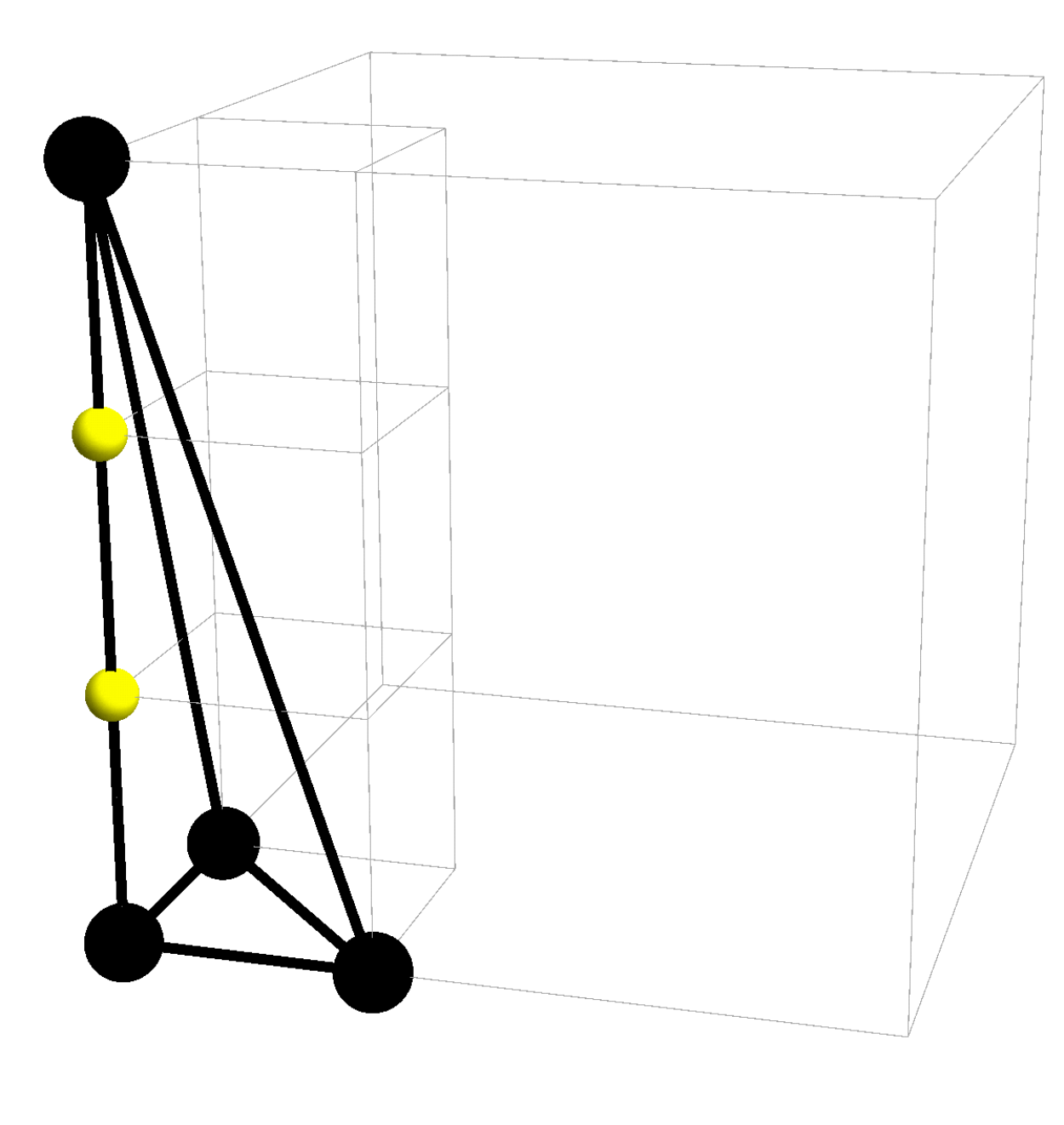}
\includegraphics[totalheight=4cm]{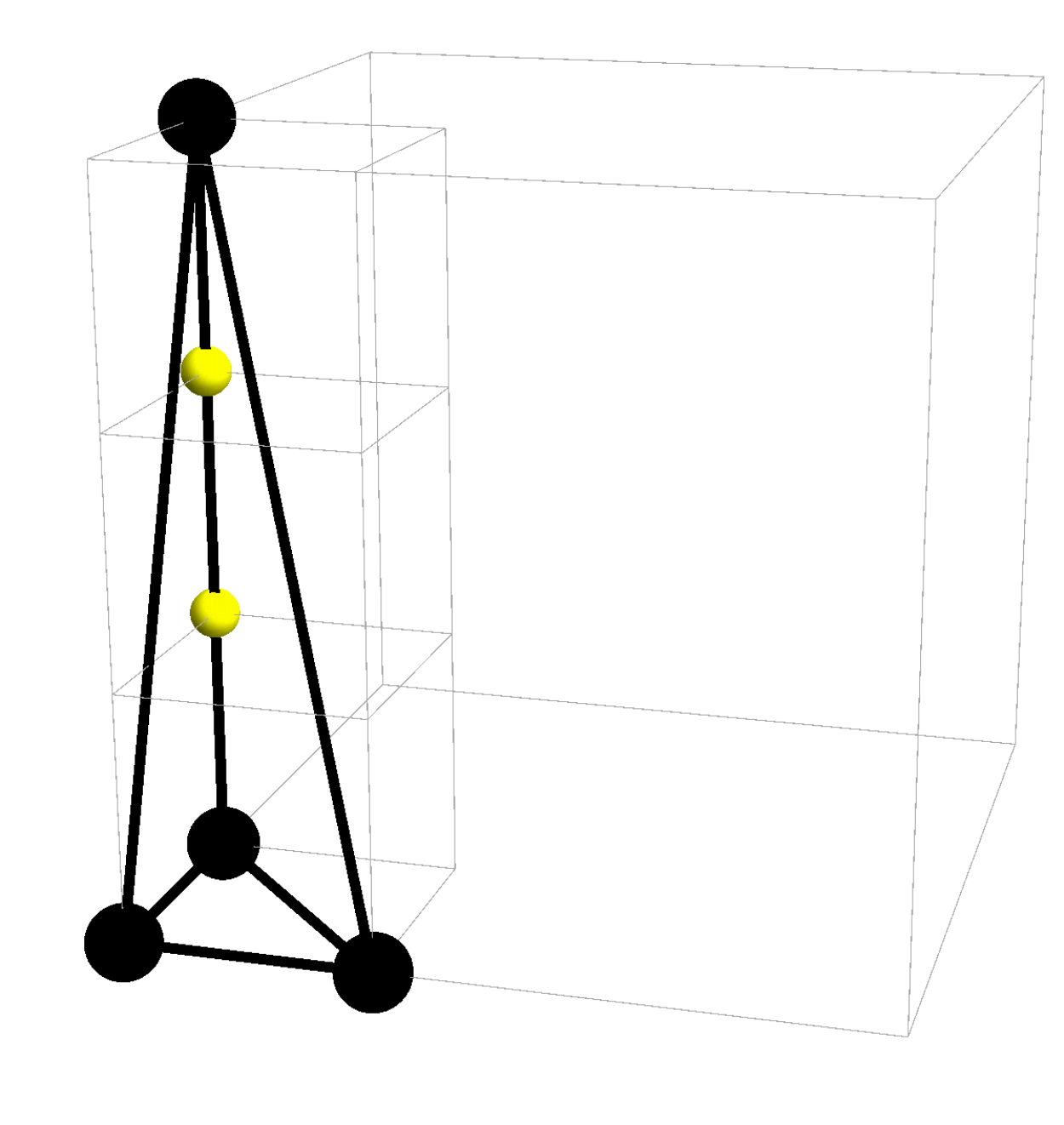}
\includegraphics[totalheight=4cm]{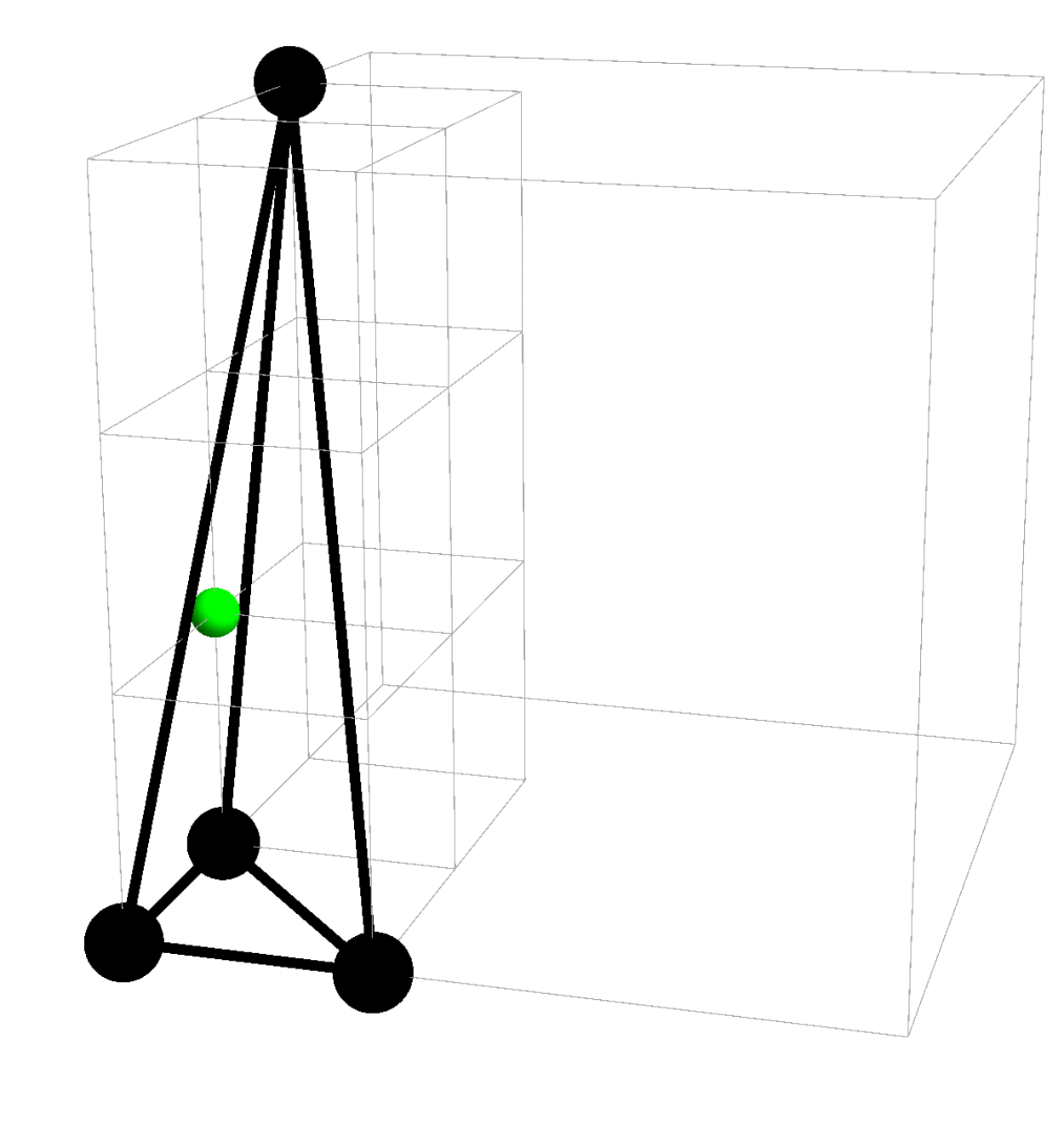}
\includegraphics[totalheight=4cm]{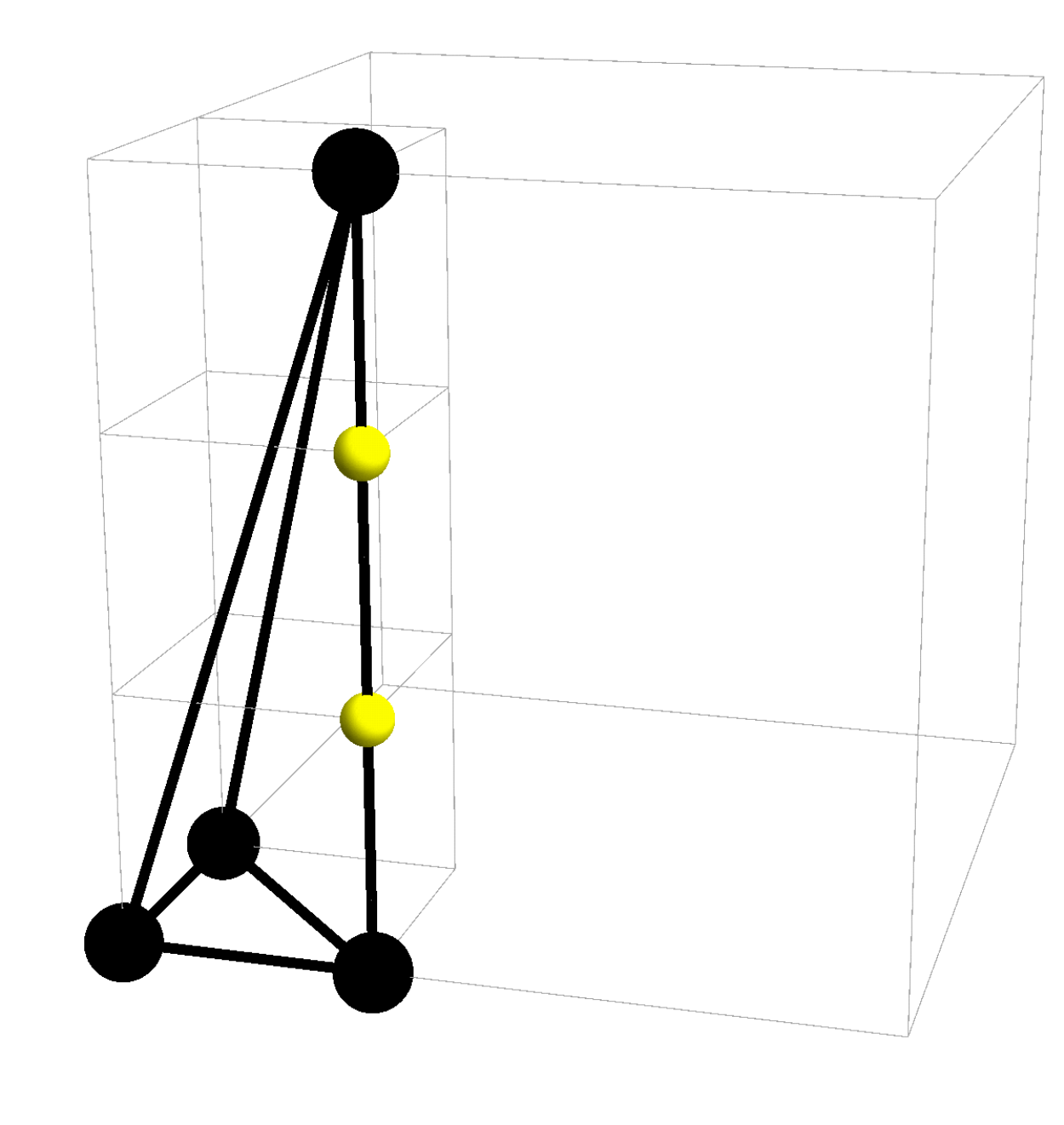}
\includegraphics[totalheight=4cm]{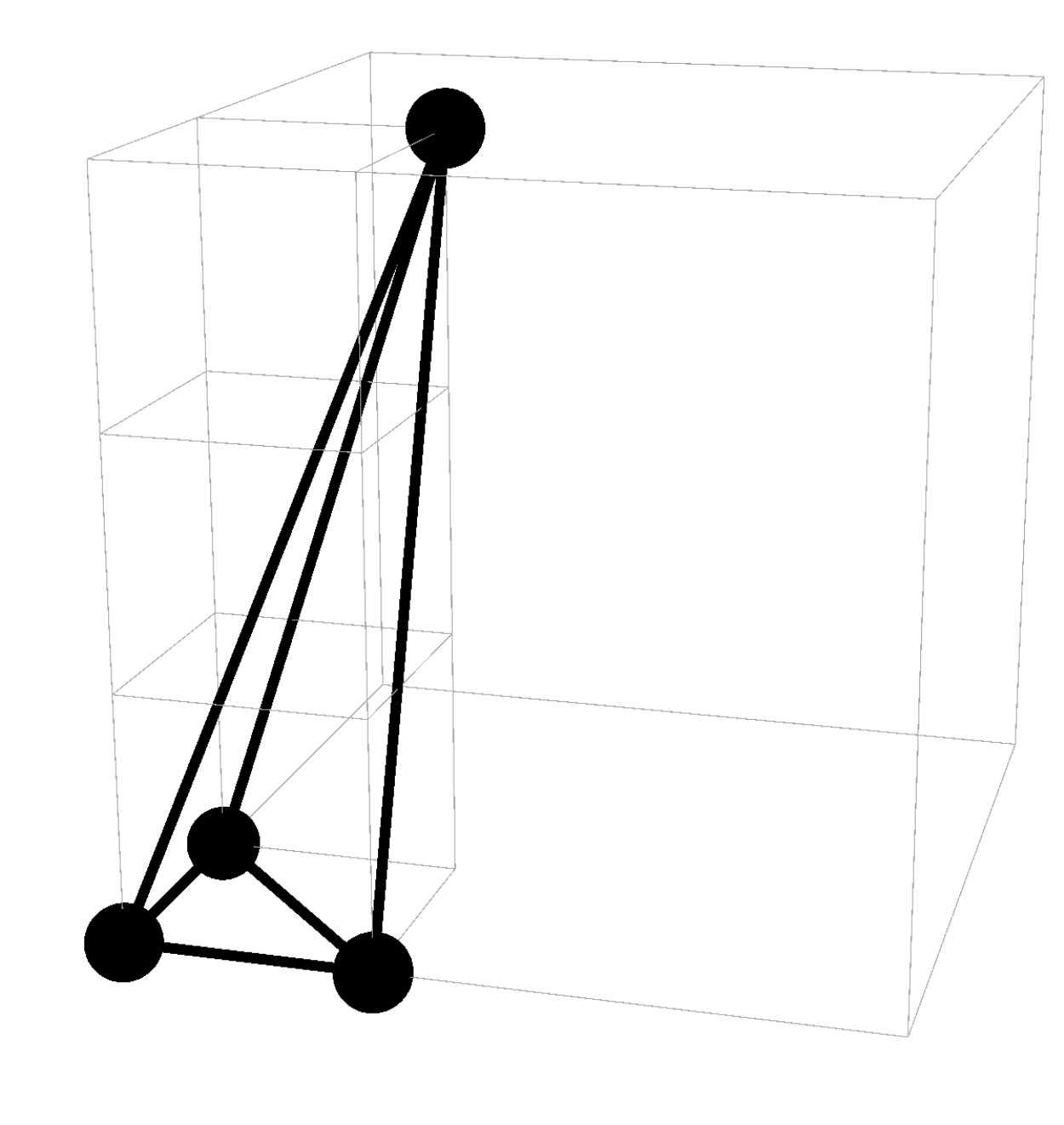}
\includegraphics[totalheight=4cm]{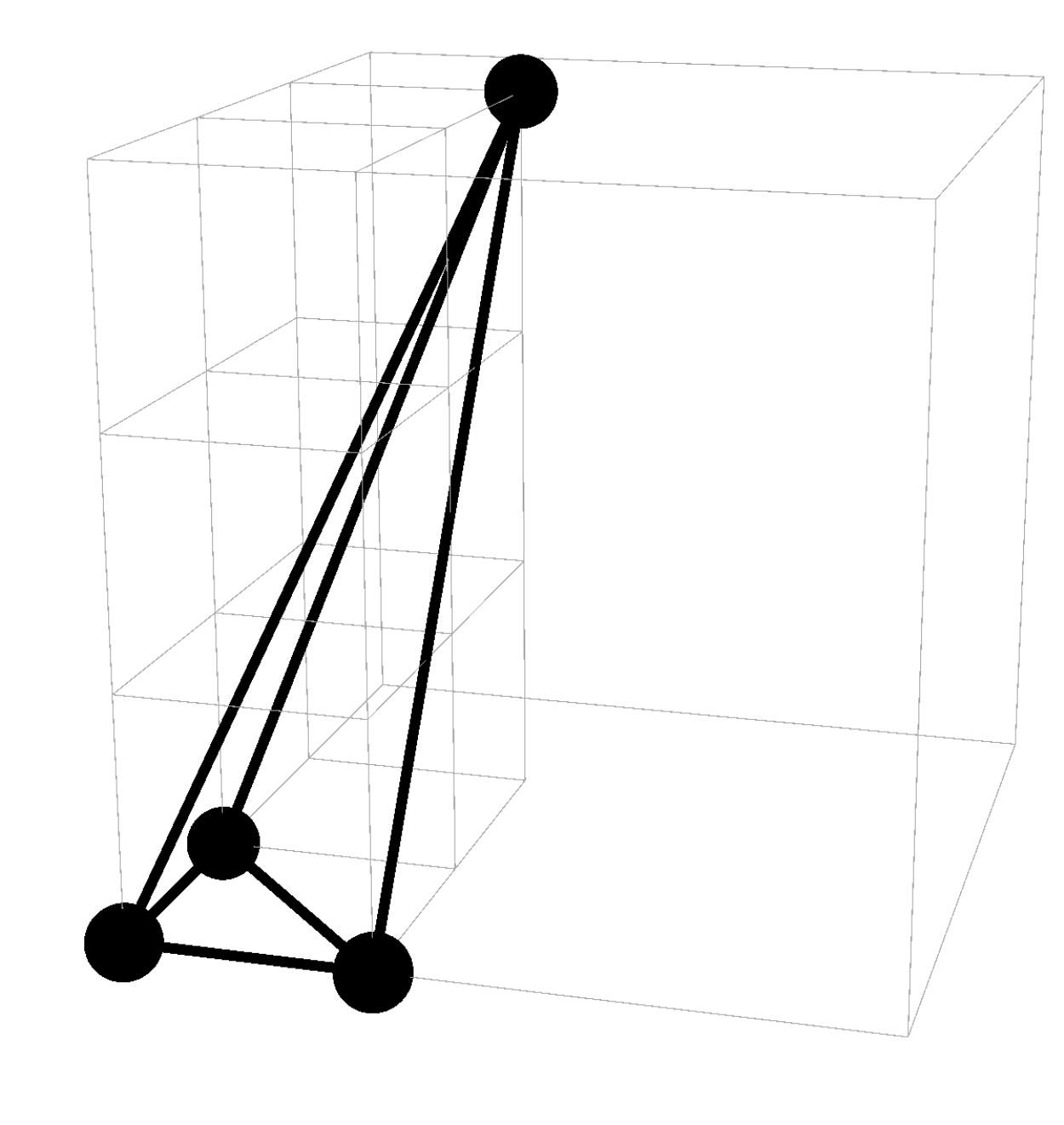}
\includegraphics[totalheight=4cm]{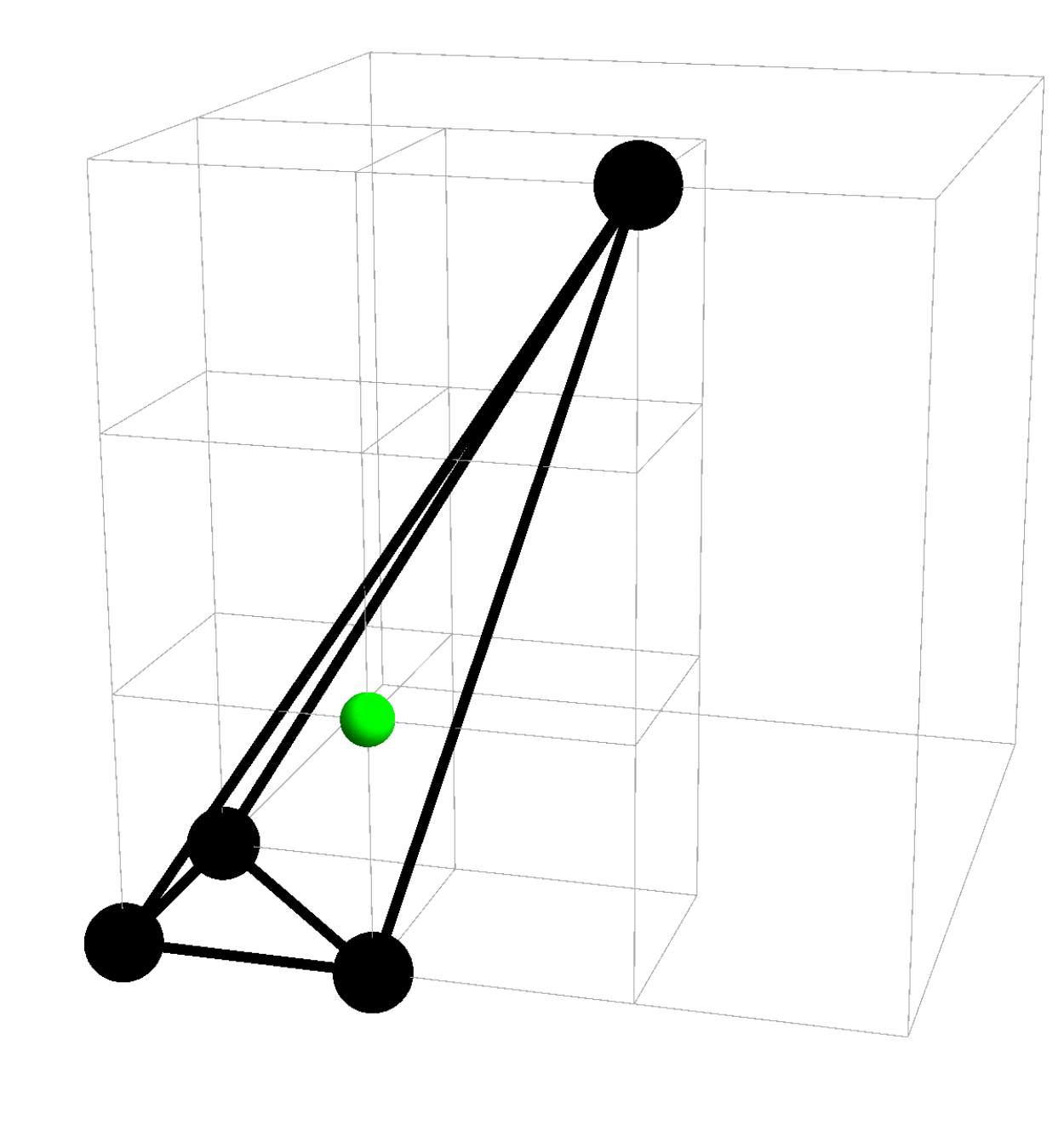}
\includegraphics[totalheight=4cm]{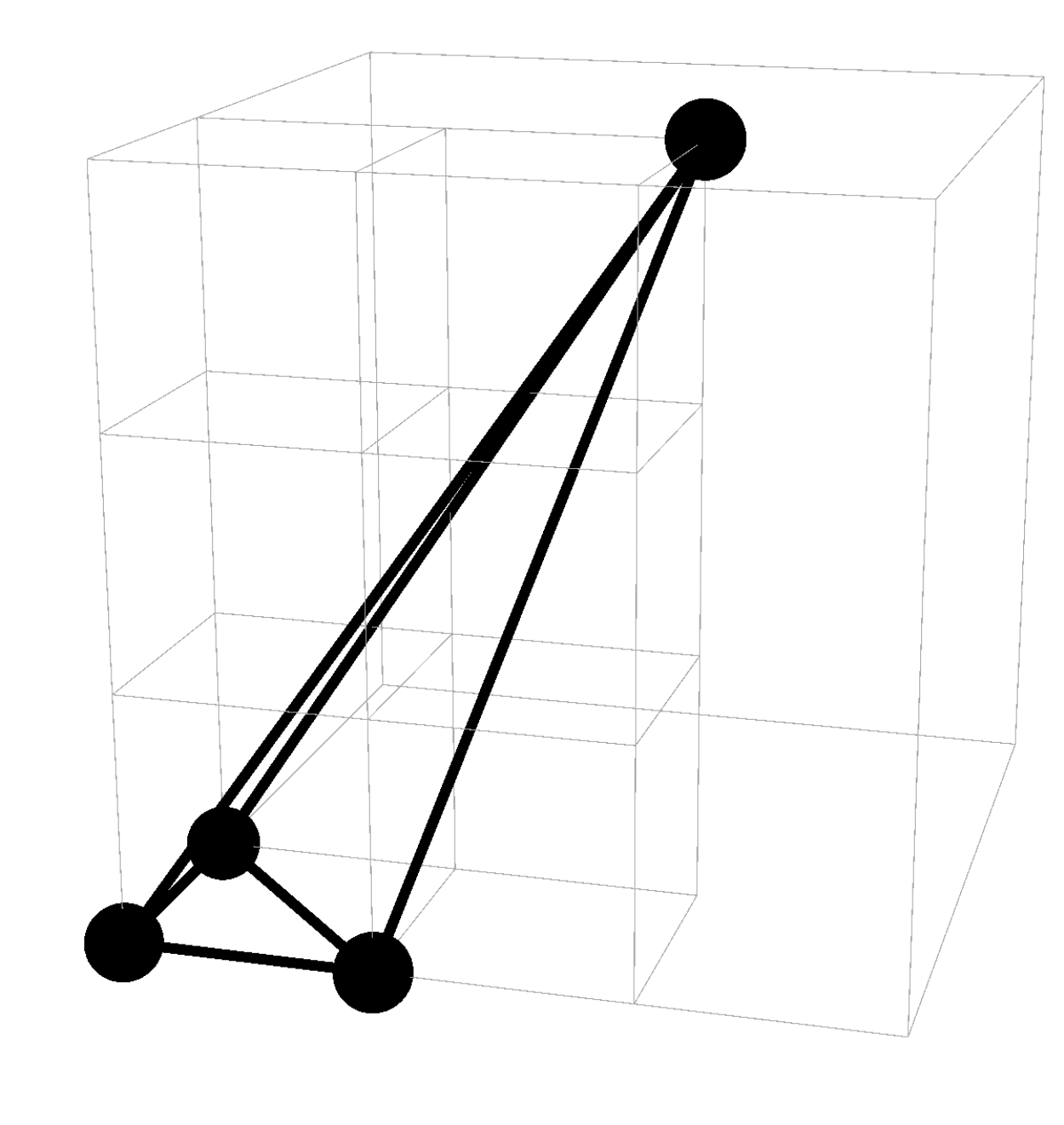}
\includegraphics[totalheight=4cm]{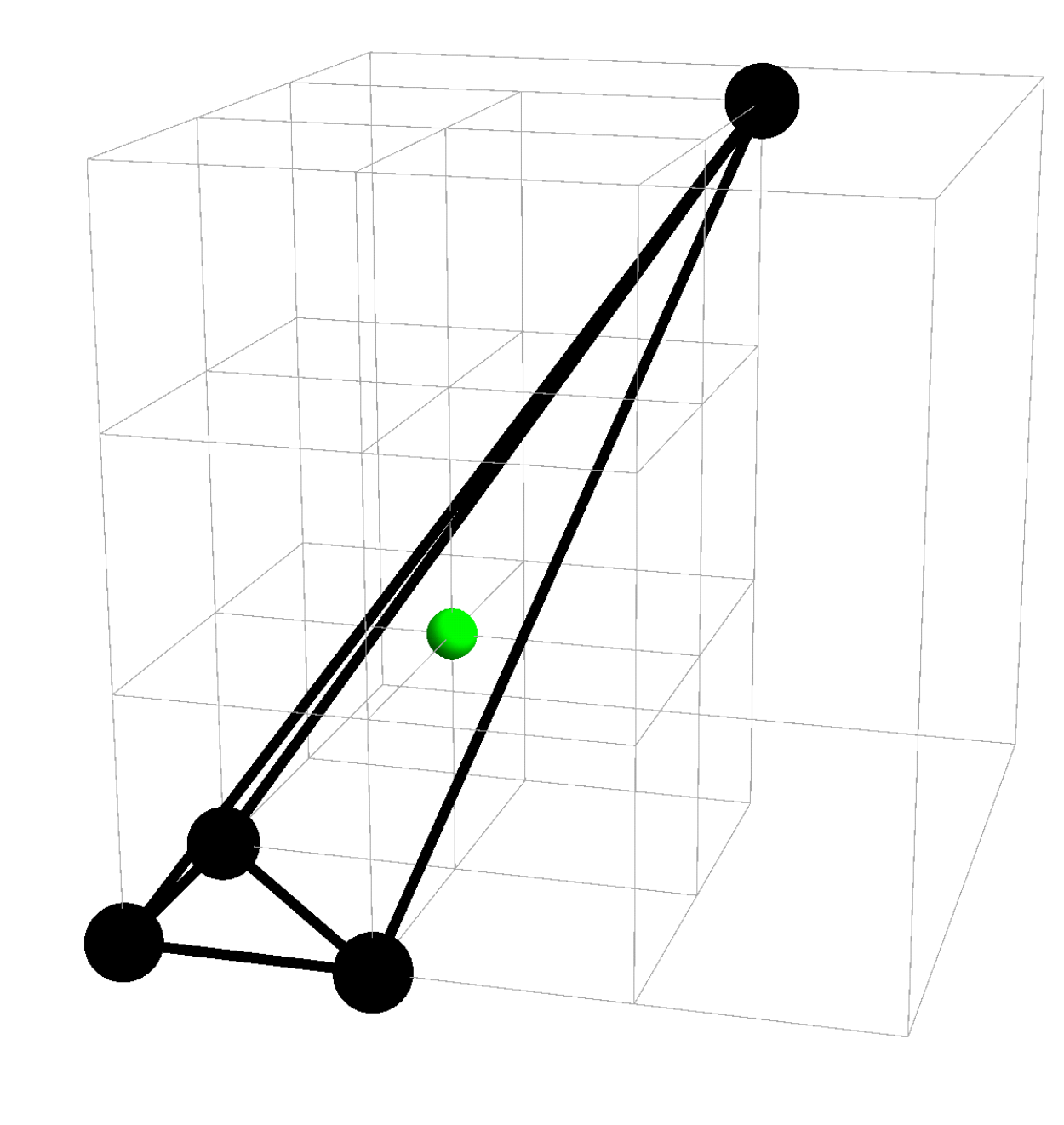}
\includegraphics[totalheight=4cm]{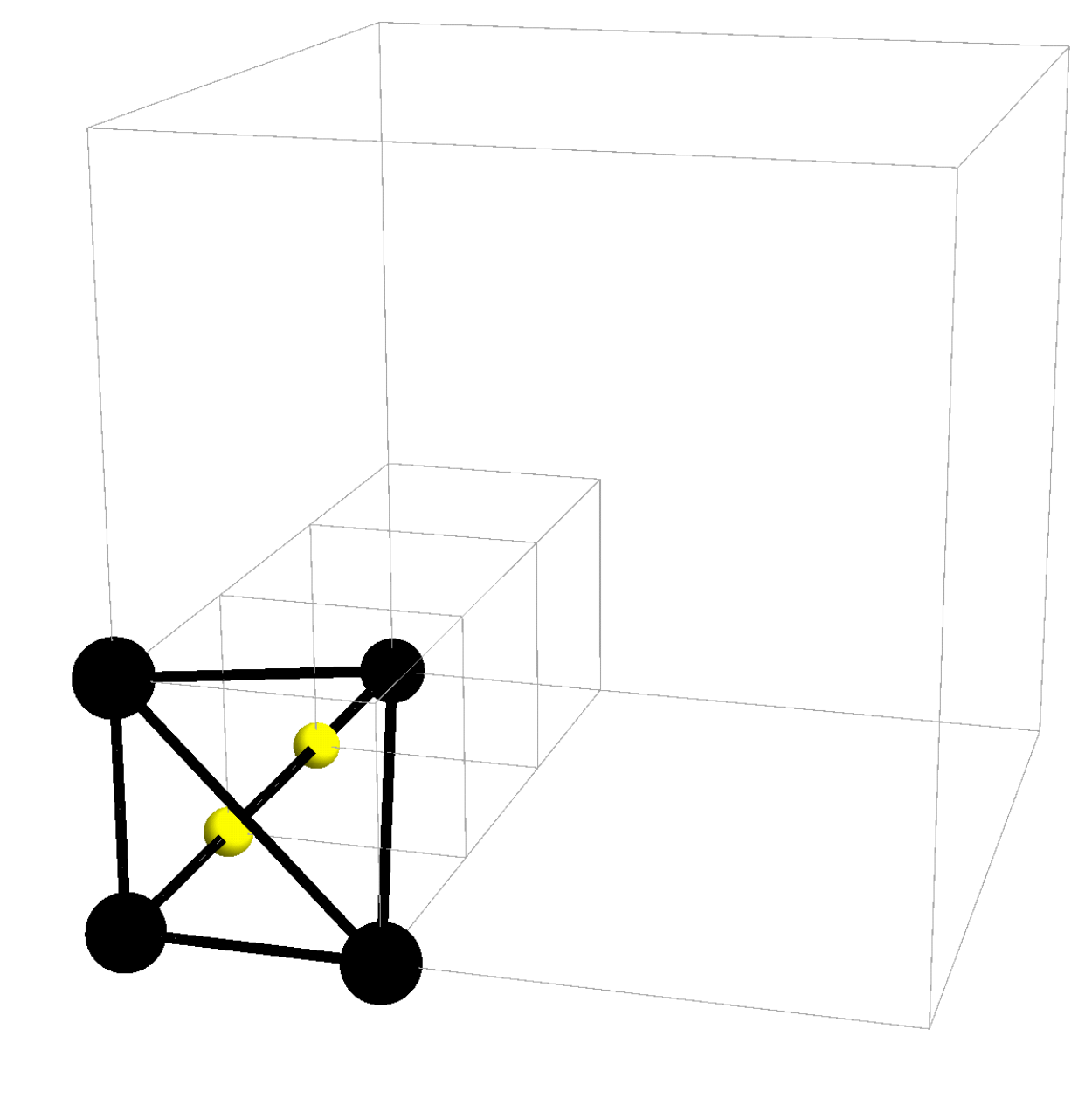}
\includegraphics[totalheight=4cm]{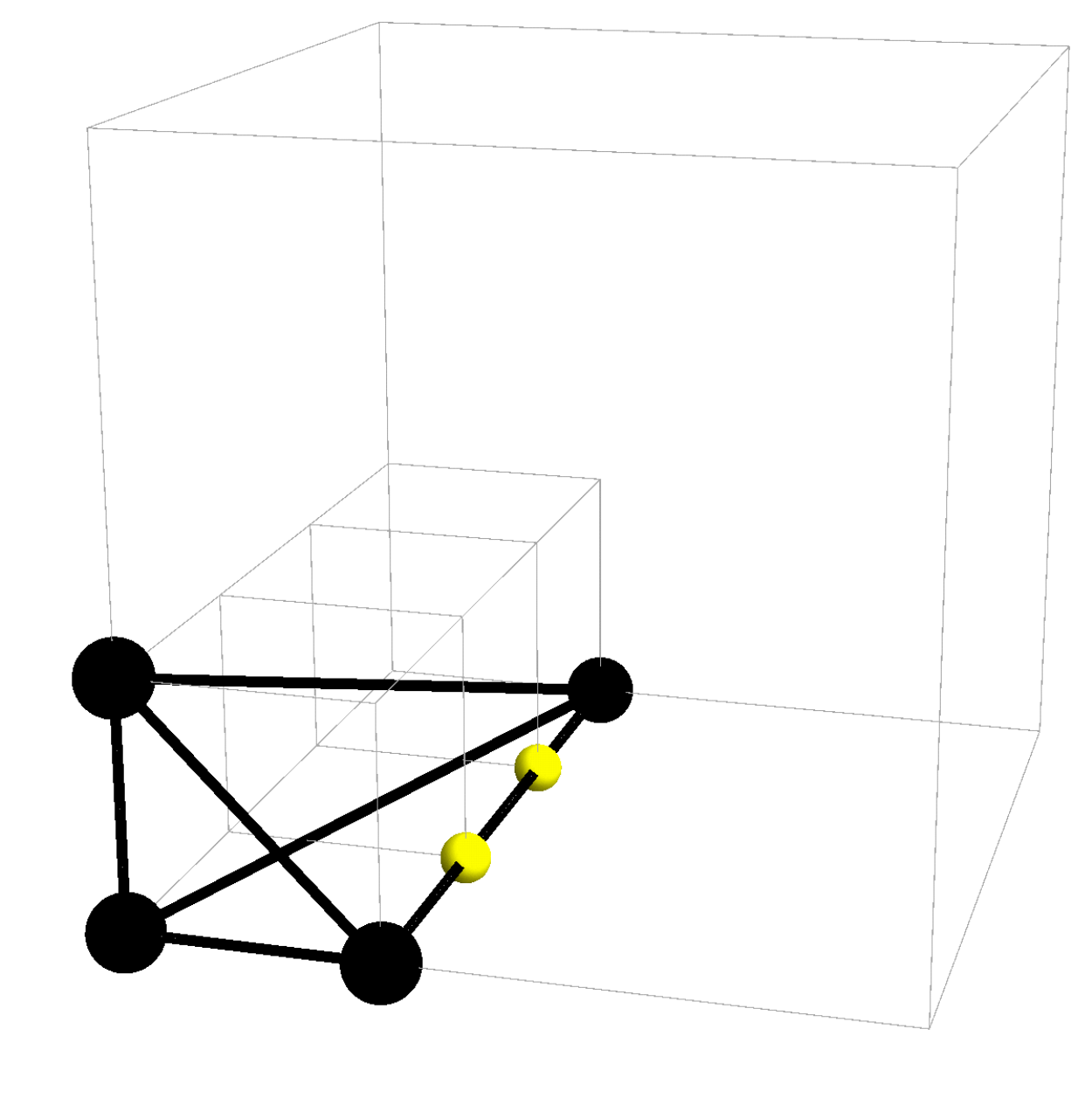}
\includegraphics[totalheight=4cm]{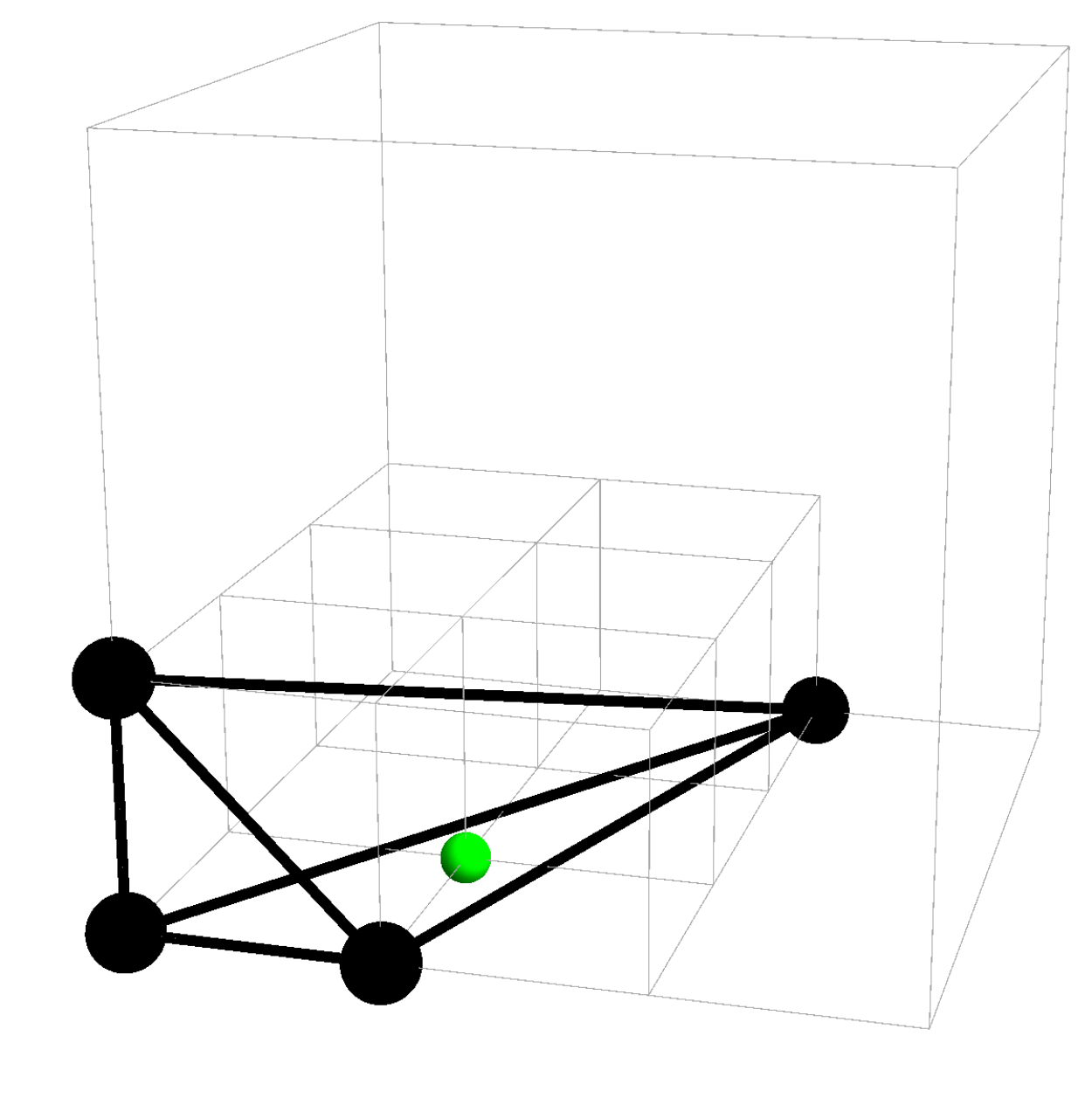}
\includegraphics[totalheight=4cm]{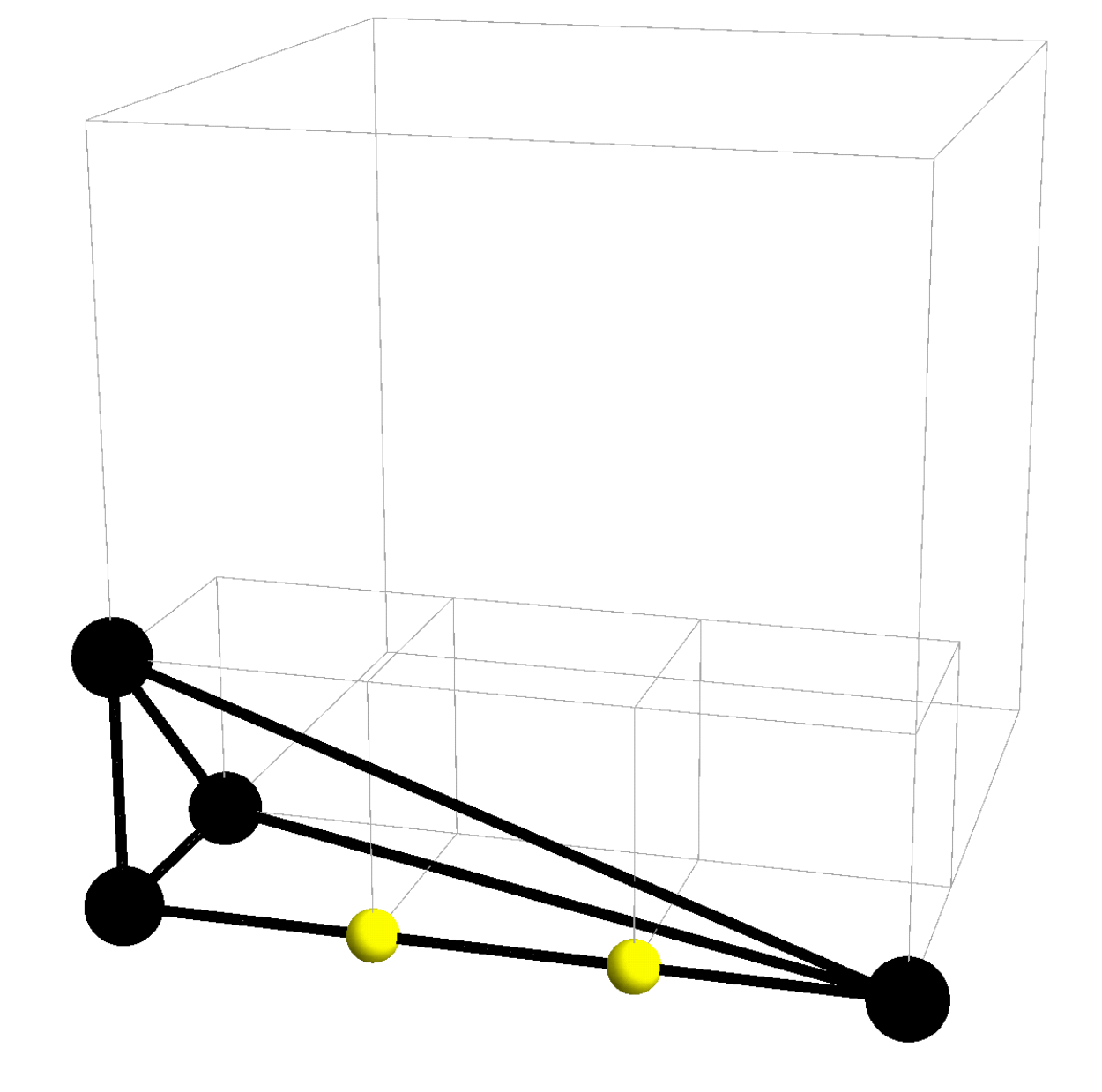}
\caption{The Hermite Normal Form toric tetrahedra of the orbifolds of the form $\mathbb{C}^{4}/\Gamma_{3}$. Lattice points on faces are colored green and lattice points on edges are colored yellow.}
  \label{c4n3hnf}
 \end{center}
 \end{figure}

%%%%%%%%%%%%% C4 %%%%%%%%%%%%%%%%%%%%%%%%%%%%
\begin{table}[ht!]
\begin{center}
\begin{tabular}{c|p{5mm}p{5mm}p{5mm}p{5mm}p{5mm}p{5mm}p{5mm}p{5mm}p{5mm}p{5mm}p{5mm}p{5mm}p{5mm}p{5mm}p{5mm}}
\hline
\hline
\multicolumn{11}{c}{$\mathbb{C}^{4}/\Gamma_{N}$}
\\
\hline
\hline
$N$ & 
1 & 2 & 3 & 4 & 5 & 6 & 7 & 8 & 9 & 10 
\\
\hline
$[1,1,1,1,1]$ & 
1& 1& 0& 1& 0& 0& 0& 1& 0& 0
\\
$[6,2,2,0,0]$ & 
0& 1& 1& 2& 1& 4& 1& 4& 2& 4
\\
$[4,2,0,1,0]$ & 
0& 0& 1& 1& 1& 1& 1& 1& 2& 1
\\
$[3,1,3,0,1]$ & 
0& 0& 1& 2& 1& 1& 1& 4& 2& 1
\\
$[2,1,2,0,0]$ & 
0& 0& 0& 1& 1& 2& 2& 8& 4& 5
\\
$[6,0,2,0,2]$ & 
0& 0& 0& 0& 1& 0& 0& 0& 0& 1
\\
$[24,0,0,0,0]$ & 
0& 0& 0& 0& 0& 1& 0& 0& 1& 4
\\
$[12,0,4,0,0]$ & 
0& 0& 0& 0& 0& 1& 1& 1& 2& 2
\\
$[8,0,0,2,0]$ & 
0& 0& 0& 0& 0& 0& 1& 0& 1& 0 
\\
$[6,0,6,0,0]$ & 
0& 0& 0& 0& 0& 0& 0& 1& 0& 0
\\
\hline
$g^{D=4}$ & 
1& 2& 3& 7& 5& 10& 7& 20& 14& 18
\\
\hline
\hline
$N$ & 
11 & 12 & 13 & 14 & 15 & 16 & 17 & 18 & 19 & 20 
\\
\hline
$[1,1,1,1,1]$ & 
0& 0& 0& 0& 0& 1& 0& 0& 0& 0
\\
$[6,2,2,0,0]$ & 
1& 9& 1& 4& 4& 9& 1& 8& 1& 9
\\
$[4,2,0,1,0]$ & 
1& 2& 1& 1& 1& 2& 1& 2& 1& 2
\\
$[3,1,3,0,1]$ & 
1& 3& 1& 1& 1& 6& 1& 2& 1& 3
\\
$[2,1,2,0,0]$ & 
4& 15& 5& 8& 12& 23& 7& 16& 8& 26
\\
$[6,0,2,0,2]$ & 
0& 0& 1& 0& 1& 0& 1& 0& 0& 3
\\
$[24,0,0,0,0]$ & 
2& 6& 3& 9& 7& 9& 7& 23& 9& 24
\\
$[12,0,4,0,0]$ & 
2& 4& 2& 4& 4& 5& 3& 8& 4& 8
\\
$[8,0,0,2,0]$ & 
0& 0& 1& 1& 0& 0& 0& 1& 1& 0
\\
$[6,0,6,0,0]$ & 
0& 2& 0& 0& 1& 3& 0& 0& 0& 2
\\
\hline
$g^{D=4}$ & 
11&41 &15 &28 &31 &58 &21 & 60& 25& 77
\\
\hline
\hline\end{tabular}
\caption{The number of inequivalent toric tetrahedra in $D(N)$ with symmetry characteristic $[g_{x_{1}^{4}},g_{x_{1}^{2}x_{2}},g_{x_{2}^{2}},g_{x_1 x_3},g_{x_4}]$.}
\label{tc4symdist}
\end{center}
\end{table}

\noindent\textbf{Example.} Consider $D(3)$ with orbifolds of the form $\mathbb{C}^{4}/\Gamma_{3}$ as shown in \fref{c4n3hnf}. There are $|D(3)|=13$ toric tetrahedra in the set. The size of $D(3)$ limits the choices for symmetry characteristics of the form $[\mathsf{g}_{x_{1}^{4}},\mathsf{g}_{x_{1}^{2}x_{2}},\mathsf{g}_{x_{2}^{2}},\mathsf{g}_{x_{1}x_{3}},\mathsf{g}_{x_{4}}]$ to 
\beal{ec1}
&&[1,1,1,1,1],[2,0,0,2,1],[2,0,2,2,0],[2,1,0,2,0],[3,0,1,0,3],[3,0,3,0,2],[3,1,1,0,2],\nn\\
&&[3,1,3,0,1],[3,2,1,0,1],[3,2,3,0,0],[3,3,1,0,0],[4,0,0,1,2],[4,0,2,1,1],[4,0,4,1,0],\nn\\
&&[4,1,0,1,1],[4,1,2,1,0],[4,2,0,1,0],[5,0,1,2,0],[6,0,0,0,3],[6,0,2,0,2],[6,0,4,0,1],\nn\\
&&[6,0,6,0,0],[6,1,0,0,2],[6,1,2,0,1],[6,1,4,0,0],[6,2,0,0,1],[6,2,2,0,0],[6,3,0,0,0],\nn\\
&&[7,0,1,1,1],[7,0,3,1,0],[7,1,1,1,0],[8,0,0,2,0],[9,0,1,0,2],[9,0,3,0,1],[9,0,5,0,0],\nn\\
&&[9,1,1,0,1],[9,1,3,0,0],[9,2,1,0,0],[10,0,0,1,1],[10,0,2,1,0],[10,1,0,1,0],\nn\\
&&[12,0,0,0,2],[12,0,2,0,1],[12,0,4,0,0],[12,1,0,0,1],[12,1,2,0,0],[12,2,0,0,0],\nn\\
&&[13,0,1,1,0]~~.\nn
\eea
This set further reduces if one identifies $\mathsf{g}_{x_{1}^{4}}$ to be a divisor of the order of the symmetric group $S_{4}$.\\

Under all $g\in S_4$, there are $3$ partitions of $D(3)$ which correspond to the distinct orbifolds of the form $\mathbb{C}^{4}/\Gamma_3$. This eliminates choices with $\mathsf{g}_{x_{1}^{4}}=12,13$. By explicit counting, it turns out that the characteristics which are assigned to $D(3)$ are
\beql{ec1}
[6,2,2,0,0],
[4,2,0,1,0],
[3,1,3,0,1]~~.\nn
\eeq
\\

%%%%%%%%%%%%% C5 %%%%%%%%%%%%%%%%%%%%%%%%%%%%
\begin{table}[ht!]
\begin{center}
\begin{tabular}{c|p{5mm}p{5mm}p{5mm}p{5mm}p{5mm}p{5mm}p{5mm}p{5mm}p{5mm}p{5mm}p{5mm}p{5mm}p{5mm}p{5mm}p{5mm}p{5mm}p{5mm}p{5mm}}
\hline
\hline
\multicolumn{14}{c}{$\mathbb{C}^{5}/\Gamma_{N}$}
\\
\hline
\hline
$N$ & 
1 & 2 & 3 & 4 & 5 & 6 & 7 & 8 & 9 & 10 & 11 & 12 & 13
\\
\hline
$[1,1,1,1,1,1,1]$ & 
1 &0 &0 &0 &1 &0 &0 &0 &0 & 0 & 0& 0 & 0
\\
$[10,4,2,1,1,0,0]$ & 
0 &1 &2 &4 &1 &2 &2 &6 &4 & 2 & 2& 8 & 2
\\
$[5,3,1,2,0,1,0]$ & 
0 &1 &1 &1 &0 &1 &1 &2 &1 & 1 & 1& 1 & 1
\\
$[15,3,3,0,0,1,0]$ & 
0 &0 &1 &2 &1 &1 &1 &5 &3 & 1 & 1& 5 & 1
\\
$[30,6,2,0,0,0,0]$ & 
0 &0 &0 &2 &2 &6 &2 &16 &6 & 9 & 4& 34 & 5
\\
$[20,6,0,2,0,0,0]$ & 
0 &0 &0 &1 &2 &4 &2 &4 &4 & 7 & 4& 11 & 5
\\
$[30,0,2,0,0,2,0]$ & 
0 &0 &0 &0 &1 &0 &0 &0 &0 & 1 & 0& 0 & 1
\\
$[60,6,0,0,0,0,0]$ & 
0 &0 &0 &0 &0 &4 &3 &9 &9 & 19 & 12& 47 & 18
\\
$[60,0,4,0,0,0,0]$ & 
0 &0 &0 &0 &0 &1 &1 &2 &2 & 2 & 2& 10 & 2
\\
$[20,2,0,2,2,0,0]$ & 
0 &0 &0 &0 &0 &0 &1 &0 &1 & 0 & 0& 0 & 1
\\
$[30,0,6,0,0,0,0]$ & 
0 &0 &0 &0 &0 &0 &0 &1 &0 & 0 & 0& 2 & 0
\\
$[40,0,0,4,0,0,0]$ & 
0 &0 &0 &0 &0 &0 &0 &0 &1 & 0 & 0& 0 & 0
\\
$[120,0,0,0,0,0,0]$ & 
0 &0 &0 &0 &0 &0 &0 &0 &1 & 5 & 3& 11 & 7
\\
$[20,0,4,2,0,0,0]$ & 
0 &0 &0 &0 &0 &0 &0 &0 &1 & 0 & 0& 0 & 0
\\
$[24,0,0,0,0,0,4]$ & 
0 &0 &0 &0 &0 &0 &0 &0 &0 & 0 & 1& 0 & 0
\\
\hline
$g^{D=4}$ & 
1& 2& 4& 10& 8& 19& 13& 45& 33& 47 & 30 & 129 & 43
\\
\hline
\hline
\end{tabular}
\caption{The number of inequivalent toric $4$-simplices in $D(N)$ with symmetry characteristic $[g_{x_{1}^{5}},g_{x_{1}^{3}x_{2}},g_{x_{1}x_{2}^{2}},g_{x_{1}^{2}x_{3}},g_{x_{2}x_{3}},g_{x_1 x_4},g_{x_5}]$.}
\label{tc5symdist}
\end{center}
\end{table}

%%%%%%%%%%%%%%%%%%%%%%%%%%%%%%%%%%%%%%%%%%%%%%%%%%%%%%%%%%%%
%%%%%%%%%%%%%%%%%%%%%%%%%%%%%%%%%%%%%%%%%%%%%%%%%%%%%%%%%%%%

\section{Discussions and Prospects}

By studying the world volume gauge theories of probe $D3$-branes and $M2$-branes, various toric singularities were identified and classified \cite{Davey:2009sr,Davey:2009qx,Davey:2009bp}. An open subset of the infinitely many probed toric singularities have been the abelian orbifolds of $\mathbb{C}^{3}$ and $\mathbb{C}^{4}$, and initial work on identifying associated quiver gauge theories \cite{Taki:2009wf} led to the work on counting distinct abelian orbifold theories and singularities \cite{HananyOrlando10,DaveyHananySeong10}.\\

In this work we have shown that it is possible to predict the number of distinct abelian orbifolds of the form $\mathbb{C}^{D}/\Gamma$ for any dimension $D$ where the order of the abelian group $\Gamma$ is a square-free product of primes. We have seen that an integral part of the computation are the discrete symmetries of the abelian orbifolds of $\mathbb{C}^{D}$ which are abelian subgroups of the permutation group $S_D$.\\

Such discrete symmetries appeared in previous work \cite{Beasley:2001zp,Feng:2002zw} as `nodal' quiver symmetries in the context of $3+1$ dimensional quiver gauge theories. We have shown in this work that such discrete symmetries can be identified directly from the toric diagram of the probed singularity for the abelian orbifolds of $\mathbb{C}^D$.\\

As an important and enlightening digression, in Section \sref{sp3}, we have used combinations of discrete symmetries constrained under the Cycle Index of the permutation group $S_D$ to give an upper bound on the number of distinct abelian orbifolds of the form $\mathbb{C}^D/\Gamma$ at a given order of $\Gamma$. Such observations underline the role of discrete symmetries in determining the number of distinct abelian orbifolds of $\mathbb{C}^D$.\\

There are several open questions which await us from here. Firstly, although we are able to predict the number of distinct abelian orbifolds of the form $\mathbb{C}^{D}/\Gamma$ where the order of $\Gamma$ is a square free product of primes, we are not able to do so for orders which are powers of prime. A solution to this problem would give us a truly complete picture of the infinite family of abelian orbifolds of $\mathbb{C}^D$.\\

Secondly, we have restricted ourselves to distinct abelian orbifolds of $\mathbb{C}^D$. In \cite{HananyOrlando10}, abelian orbifolds of the conifold $\mathcal{C}$ and $L_{aba}$ theories have been counted explicitly. In principle, we are not restricted to these toric singularities and are able to count distinct abelian orbifolds of any toric singularity using the techniques described in this work. From our observation that the number of distinct abelian orbifolds relies on the discrete symmetries of the toric singularity, we can reverse the relationship and ask whether two toric singularities have the same discrete symmetries if the number of distinct ways of orbifolding these singularities are the same.\\

We believe that further study of symmetries of abelian orbifolds of various toric singularities can give new valuable insights into underlying structures of the corresponding quiver gauge theories. \\

%%%%%%%%%%%%%%%%%%%%%%%%%%%%%%%%%%%%%%%%%%%%%%%%%%%%%%%%%%%%%%%%%%%%%%%%%%%%%%%%%%%%%%%%%%%%%%%%%%%%%%%%%%%%%%%%%%%%%%%%%%%%
\section*{Acknowledgements}

Discussions with Sebastian Franco, Yang-Hui He, Vishnu Jejjala, Sanjaye Ramgoolam, John Davey, Noppadol Mekareeya, Spyros Sypsas and Giuseppe Torri are gratefully acknowledged. We thank the 2010 Simons Workshop in Mathematics and Physics for hospitality during part of this work. 
R.-K.~S.~ would like to thank Jurgis Pasukonis for useful discussions and help in improving the programming code for the counting. He would also like to thank the Korea Institute for Advanced Studies in Seoul  for their kind hospitality during part of this work. He is grateful for the warm encouragement from his parents.
A.~H.~ would like to thank \'Ecole Polytechnique, Paris, and the KITPC, Beijing for their kind hospitality during various stages of this work.
This work is partially supported under ERC Advanced Grant 226371. 

\clearpage

%%%%%%%%%%%%%%%%%%%%%%%%%%%%%%%%%%%%%%%%%%%%%%%%%%%%%%%%%%%%%%%%%%%%%%%%%%%%%%%%%%%%%%%%%%%%%%%%%%%%%%%%%%%%%%%%%%%%%%%%%%%%

\appendix

%%%%%%%%%%%%%%%%%%%%%%%%%%%%%%%%%%%%%%%%%%%%%%%%%%%%%%%%%%%%%%%%%%%%%%
\section{Symmetries of Abelian Orbifolds of $\mathbb{C}^{3}$, their Orbifold Actions and Lattice $2$-Simplices}

The numbering of orbifolds (\#) is the same as in \cite{DaveyHananySeong10}, where the individual toric diagrams are illustrated.\\

%%%%%%%%%%%%%%%%%%%%%%%%%%%%%%%%%%%%%%%%%%%
%%%%%%%%%%%%%%%%%%% N=1..3 ================
\begin{table}[ht]

\begin{center}
\begin{tabular}{m{1cm}|m{2.2cm}|m{3cm}|m{4cm}|m{1cm}m{1cm}m{1cm}}
\hline \hline
\# & Orbifold & Orbifold Action & $I_{0}$ (Corners) & $\mathsf{g}_{x_{1}^{3}}$ & $\mathsf{g}_{x_{1}x_{1}^{2}}$ & $\mathsf{g}_{x_{3}}$  \\
\hline \hline

%%%%%%%%%%%%%%%%%%% N=1 %%%%%%%%%%%%%%%%%%%
(1.1)& $\mathbb{C}^{3}/\mathbb{Z}_{1}$ & $\bvec (0,0,0)\\(0,0,0) \evec$ & %\includegraphics*[height=3.5cm]{C3Tn1m1i1.pdf} 
$\{(0,0),(1,0),(0,1)\}$
& 1 & 1& 1
\\
\hline
\multicolumn{4}{r|}{Total} & 1 & 1 & 1
\\ 
\hline \hline

%%%%%%%%%%%%%%%%%%% N=2 %%%%%%%%%%%%%%%%%%%
(2.1)& $\mathbb{C}^{3}/\mathbb{Z}_{2}$ & $\bvec (0,1,1)\\(0,0,0) \evec$ &
$\{(0,0),(1,0),(0,2)\}$
& 3 & 1 & 0
\\
\hline
\multicolumn{4}{r|}{Total} & 3 & 1 & 1
\\
\hline \hline

%%%%%%%%%%%%%%%%%%% N=3 %%%%%%%%%%%%%%%%%%%
(3.1) & $\mathbb{C}^{3}/\mathbb{Z}_{3}$ & $\bvec (0,1,2)\\(0,0,0) \evec$ &
$\{(0,0),(1,0),(0,3)\}$
& 3 & 1 & 0
\\
\hline
(3.2) & $\mathbb{C}^{3}/\mathbb{Z}_{3}$ & $\bvec (1,1,1)\\(0,0,0) \evec$ &
$\{(0,0),(1,0),(2,3)\}$
& 1 & 1 & 1
\\
\hline
\multicolumn{4}{r|}{Total} & 4 & 2 & 1
\\
\hline \hline

%\end{tabular}
%\end{center}

%\caption{Orbifold Actions and corresponding Toric Diagrams for $\mathbb{C}^{3}/\Gamma_N$ orbifolds with order $N=1\dots 10$ %\textbf{(Part 1/7)}.}
%\label{t1}
%\end{table}

%\clearpage

%%%%%%%%%%%%%%%%%%%%%%%%%%%%%%%%%%%%%%%%%%%
%%%%%%%%%%%%%%%%%%% N=4..6 ================
%\begin{table}[ht]

%\begin{center}
%\begin{tabular}{m{1cm}|m{2.2cm}|m{3cm}|m{4.8cm}|m{1cm}m{1cm}m{1cm}}
%\hline \hline
%\# & Orbifold & Orbifold Action & Toric Diagram & $\mathsf{g}_{x_{1}^{3}}$ & $\mathsf{g}_{x_{1}x_{1}^{2}}$ & $\mathsf{g}_{x_{3}}$  \\
%\hline \hline
%%%%%%%%%%%%%%%%%%% N=4 %%%%%%%%%%%%%%%%%%%
(4.1) & $\mathbb{C}^{3}/\mathbb{Z}_{4}$ & $\bvec (0,1,3)\\(0,0,0) \evec$ & 
$\{(0,0),(1,0),(0,4)\}$
& 3 & 1 & 0
\\ 
\hline
(4.2) & $\mathbb{C}^{3}/\mathbb{Z}_{4}$ & $\bvec (1,1,2)\\(0,0,0) \evec$ & 
$\{(0,0),(1,0),(2,4)\}$
& 3 & 1 & 0
\\ 
\hline
(4.3) & $\mathbb{C}^{3}/\mathbb{Z}_{2}\times\mathbb{Z}_{2}$ & $\bvec (1,0,1)\\(0,1,1) \evec$ & 
$\{(0,0),(2,0),(0,2)\}$
& 1 & 1 & 1
\\
\hline
\multicolumn{4}{r|}{Total} & 7 & 3 & 1
\\ 
\hline \hline

%%%%%%%%%%%%%%%%%%% N=5 %%%%%%%%%%%%%%%%%%%
(5.1) & $\mathbb{C}^{3}/\mathbb{Z}_{5}$ & $\bvec (0,1,4)\\(0,0,0) \evec$ &
$\{(0,0),(1,0),(0,5)\}$
& 3 & 1 & 0
\\
\hline 

%\end{tabular}
%\end{center}

%\caption{Orbifold Actions and corresponding Toric Diagrams for $\mathbb{C}^{3}/\Gamma_N$ orbifolds with order $N=1\dots 10$ \textbf{(Part 2/7)}.}
%\label{t2}
%\end{table}

%\clearpage

%%%%%%%%%%%%%%%%%%%%%%%%%%%%%%%%%%%%%%%%%%%%%%%%
%%%%%%%%%%%%%%%%%%%%%%%%%%%%%%%%%%%%%%%%%%%%%%%%

%\begin{table}[ht]

%\begin{center}
%\begin{tabular}{m{1cm}|m{2.2cm}|m{3cm}|m{4.8cm}|m{1cm}m{1cm}m{1cm}}
%\hline \hline
%\# & Orbifold & Orbifold Action & Toric Diagram & $\mathsf{g}_{x_{1}^{3}}$ & $\mathsf{g}_{x_{1}x_{1}^{2}}$ & $\mathsf{g}_{x_{3}}$  \\
%\hline \hline

%%%%%%%%%%%%%%%%%%% N=5 %%%%%%%%%%%%%%%%%%%

(5.2) & $\mathbb{C}^{3}/\mathbb{Z}_{5}$ & $\bvec (1,1,3)\\(0,0,0) \evec$ &
$\{(0,0),(1,0),(2,5)\}$
& 3 & 1 & 0
\\
\hline
\multicolumn{4}{r|}{Total} & 6 & 2 & 0
\\
\hline \hline

\end{tabular}
\end{center}

\caption{The symmetry counting of distinct abelian orbifolds $\mathbb{C}^{3}$ with corresponding orbifold actions and toric triangles given in terms of $I_{0}$ (corner points in Cartesian coordinates) \textbf{(Part 1/3)}.}
\label{tc3allp1}
\end{table}

\clearpage

%%%%%%%%%%%%%%%%%%%%%%%%%%%%%%%%%%%%%%%%%%%%%%%%
%%%%%%%%%%%%%%%%%%%%%%%%%%%%%%%%%%%%%%%%%%%%%%%%

\begin{table}[ht]

\begin{center}
\begin{tabular}{m{1cm}|m{2.2cm}|m{3cm}|m{4cm}|m{1cm}m{1cm}m{1cm}}
\hline \hline
\# & Orbifold & Orbifold Action & $I_{0}$ (Corners) & $\mathsf{g}_{x_{1}^{3}}$ & $\mathsf{g}_{x_{1}x_{1}^{2}}$ & $\mathsf{g}_{x_{3}}$  \\
\hline \hline

%%%%%%%%%%%%%%%%%%% N=6 %%%%%%%%%%%%%%%%%%%
(6.1) & $\mathbb{C}^{3}/\mathbb{Z}_{6}$ & $\bvec (0,1,5)\\(0,0,0) \evec$ &
$\{(0,0),(1,0),(0,6)\}$
& 3 & 1 & 0
\\
\hline
(6.2) &  $\mathbb{C}^{3}/\mathbb{Z}_{6}$ & $\bvec (1,1,4)\\(0,0,0) \evec$ &
$\{(0,0),(1,0),(2,6)\}$
& 3 & 1 & 0
\\
\hline
(6.3) &  $\mathbb{C}^{3}/\mathbb{Z}_{6}$ & $\bvec (1,2,3)\\(0,0,0) \evec$ &
$\{(0,0),(1,0),(3,6)\}$
& 6 & 0 & 0
\\
\hline
\multicolumn{4}{r|}{Total} & 12 & 2 & 0
\\
\hline \hline
(7.1) & $\mathbb{C}^{3}/\mathbb{Z}_{7}$ & $\bvec (0,1,6)\\(0,0,0) \evec$ & 
$\{(0,0),(1,0),(0,7)\}$
& 3 & 1 & 0
\\ 
\hline
(7.2) & $\mathbb{C}^{3}/\mathbb{Z}_{7}$ & $\bvec (1,1,5)\\(0,0,0) \evec$ & 
$\{(0,0),(1,0),(2,7)\}$
& 3 & 1 & 0
\\ 
\hline
(7.3) & $\mathbb{C}^{3}/\mathbb{Z}_{7}$ & $\bvec (1,2,4)\\(0,0,0) \evec$ & 
$\{(0,0),(1,0),(3,7)\}$
& 2 & 0 & 2
\\
\hline
\multicolumn{4}{r|}{Total} & 8 & 2 & 2
\\ 
\hline \hline

%%%%%%%%%%%%%%%%%%% N=8 %%%%%%%%%%%%%%%%%%%
(8.1) & $\mathbb{C}^{3}/\mathbb{Z}_{8}$ & $\bvec (0,1,7)\\(0,0,0) \evec$ &
$\{(0,0),(1,0),(0,8)\}$
& 3 & 1 & 0
\\
\hline %\hline

%\end{tabular}
%\end{center}

%\caption{Orbifold Actions and corresponding Toric Diagrams for $\mathbb{C}^{3}/\Gamma_N$ orbifolds with order $N=1\dots 10$ \textbf{(Part 4/7)}.}
%\label{t3}
%\end{table}

%\clearpage

%%%%%%%%%%%%%%%%%%%%%%%%%%%%%%%%%%%%%%%%%%%%%%%%
%%%%%%%%%%%%%%%%%%%%%%%%%%%%%%%%%%%%%%%%%%%%%%%%

%\begin{table}[ht]

%\begin{center}
%\begin{tabular}{m{1cm}|m{2.2cm}|m{3cm}|m{4.8cm}|m{1cm}m{1cm}m{1cm}}
%\hline \hline
%\# & Orbifold & Orbifold Action & Toric Diagram & $\mathsf{g}_{x_{1}^{3}}$ & $\mathsf{g}_{x_{1}x_{1}^{2}}$ & $\mathsf{g}_{x_{3}}$  \\
%\hline \hline

%%%%%%%%%%%%%%%%%%% N=8 %%%%%%%%%%%%%%%%%%%
(8.2) & $\mathbb{C}^{3}/\mathbb{Z}_{8}$ & $\bvec (1,1,6)\\(0,0,0) \evec$ &
$\{(0,0),(1,0),(2,8)\}$
& 3 & 1& 0
\\
\hline 
(8.3) & $\mathbb{C}^{3}/\mathbb{Z}_{8}$ & $\bvec (1,2,5)\\(0,0,0) \evec$ &
$\{(0,0),(1,0),(3,8)\}$
& 3 & 1 & 0
\\
\hline
(8.4) & $\mathbb{C}^{3}/\mathbb{Z}_{8}$ & $\bvec (1,3,4)\\(0,0,0) \evec$ & 
$\{(0,0),(1,0),(4,8)\}$
& 3 & 1& 0
\\
\hline
(8.5) & $\mathbb{C}^{3}/\mathbb{Z}_{4}\times\mathbb{Z}_{2}$ & $\bvec (1,0,3)\\(0,1,1) \evec$ &
$\{(0,0),(2,0),(0,4)\}$
& 3 &1 & 0
\\
\hline
\multicolumn{4}{r|}{Total} & 15 & 5 & 0
\\
\hline \hline

\end{tabular}
\end{center}

\caption{The symmetry counting of distinct abelian orbifolds $\mathbb{C}^{3}$ with corresponding orbifold actions and toric triangles given in terms of $I_{0}$ (corner points in Cartesian coordinates) \textbf{(Part 2/3)}.}
\label{tc3allp2}
\end{table}

\clearpage

%%%%%%%%%%%%%%%%%%%%%%%%%%%%%%%%%%%%%%%%%%%%%%%%%%%%%%
%%%%%%%%%%%%%%%%%%%%%%%%%%%%%%%%%%%%%%%%%%%%%%%%%%%%%%

\begin{table}[ht]

\begin{center}
\begin{tabular}{m{1cm}|m{2.2cm}|m{3cm}|m{4cm}|m{1cm}m{1cm}m{1cm}}
\hline \hline
\# & Orbifold & Orbifold Action & $I_{0}$ (Corners) & $\mathsf{g}_{x_{1}^{3}}$ & $\mathsf{g}_{x_{1}x_{1}^{2}}$ & $\mathsf{g}_{x_{3}}$  \\
\hline \hline

%\end{tabular}
%\end{center}

%\caption{Orbifold Actions and corresponding Toric Diagrams for $\mathbb{C}^{3}/\Gamma_N$ orbifolds with order $N=1\dots 10$ \textbf{(Part 5/7)}.}
%\label{t3b}
%\end{table}

%\clearpage

%%%%%%%%%%%%%%%%%%%%%%%%%%%%%%%%%%%%%%%%%%%%%%%%
%%%%%%%%%%%%%%%%%%%%%%%%%%%%%%%%%%%%%%%%%%%%%%%%

%\begin{table}[ht]

%\begin{center}
%\begin{tabular}{m{1cm}|m{2.2cm}|m{3cm}|m{4.8cm}|m{1cm}m{1cm}m{1cm}}
%\hline \hline
%\# & Orbifold & Orbifold Action & Toric Diagram & $\mathsf{g}_{x_{1}^{3}}$ & $\mathsf{g}_{x_{1}x_{1}^{2}}$ & $\mathsf{g}_{x_{3}}$  \\
%\hline \hline

%%%%%%%%%%%%%%%%%%% N=9 %%%%%%%%%%%%%%%%%%%
(9.1) & $\mathbb{C}^{3}/\mathbb{Z}_{9}$ & $\bvec (0,1,8)\\(0,0,0) \evec$ & 
$\{(0,0),(1,0),(0,9)\}$
& 3 & 1& 0
\\ \hline
(9.2) & $\mathbb{C}^{3}/\mathbb{Z}_{9}$ & $\bvec (1,1,7)\\(0,0,0) \evec$ & 
$\{(0,0),(1,0),(2,9)\}$
& 3 & 1 & 0
\\ \hline
(9.3) & $\mathbb{C}^{3}/\mathbb{Z}_{9}$ & $\bvec (1,2,6)\\(0,0,0) \evec$ & 
$\{(0,0),(1,0),(3,9)\}$
& 6 & 0 & 0
\\ \hline
(9.4) & $\mathbb{C}^{3}/\mathbb{Z}_{3}\times\mathbb{Z}_{3}$ & $\bvec (0,1,2)\\(1,0,2) \evec$ & 
$\{(0,0),(3,0),(0,3)\}$
& 1 & 1 & 1
\\
\hline
\multicolumn{4}{r|}{Total} & 13 & 3 & 1
\\ 
\hline \hline

%\end{tabular}
%\end{center}

%\caption{Orbifold Actions and corresponding Toric Diagrams for $\mathbb{C}^{3}/\Gamma_N$ orbifolds with order $N=1\dots 10$ \textbf{(Part 6/7)}.}
%\label{t4}
%\end{table}

%\clearpage

%%%%%%%%%%%%%%%%%%%%%%%%%%%%%%%%%%%%%%%%%%%%%%%%
%%%%%%%%%%%%%%%%%%%%%%%%%%%%%%%%%%%%%%%%%%%%%%%%

%\begin{table}[ht]

%\begin{center}
%\begin{tabular}{m{1cm}|m{2.2cm}|m{3cm}|m{4.8cm}|m{1cm}m{1cm}m{1cm}}
%\hline \hline
%\# & Orbifold & Orbifold Action & Toric Diagram & $\mathsf{g}_{x_{1}^{3}}$ & $\mathsf{g}_{x_{1}x_{1}^{2}}$ & $\mathsf{g}_{x_{3}}$  \\
%\hline \hline

%%%%%%%%%%%%%%%%%%% N=10 %%%%%%%%%%%%%%%%%%%
(10.1) & $\mathbb{C}^{3}/\mathbb{Z}_{10}$ & $\bvec (0,1,9)\\(0,0,0) \evec$ &
$\{(0,0),(1,0),(0,10)\}$
& 3 & 1 & 0
\\ \hline
(10.2) & $\mathbb{C}^{3}/\mathbb{Z}_{10}$ & $\bvec (1,1,8)\\(0,0,0) \evec$ &
$\{(0,0),(1,0),(2,10)\}$
& 3 & 1 & 0
\\ \hline
(10.3) & $\mathbb{C}^{3}/\mathbb{Z}_{10}$ & $\bvec (1,2,7)\\(0,0,0) \evec$ &
$\{(0,0),(1,0),(3,10)\}$
& 6 & 0 & 0
\\ \hline
(10.4) & $\mathbb{C}^{3}/\mathbb{Z}_{10}$ & $\bvec (1,4,5)\\(0,0,0) \evec$ &
$\{(0,0),(1,0),(5,10)\}$
& 6 & 0 & 0
\\
\hline
\multicolumn{4}{r|}{Total} & 18 & 2 & 0
\\ 
\hline \hline

\end{tabular}
\end{center}

\caption{The symmetry counting of distinct abelian orbifolds of $\mathbb{C}^{3}$ with corresponding orbifold actions and toric triangles given in terms of $I_{0}$ (corner points in Cartesian coordinates) \textbf{(Part 3/3)}.}
\label{tc3allp2}
\end{table}

\clearpage

%%%%%%%%%%%%%%%%%%%%%%%%%%%%%%%%%%%%%%%%%%%%%%%%%%%%%%%%%%%%%%%%%%%%%%
\section{Symmetries of Abelian Orbifolds of $\mathbb{C}^{4}$, their Orbifold Actions and Lattice $3$-Simplices}

%%%%%%%%%%%%%%%%%%%%%%%%%%%%%%%%%%%%%%%%%%%%%%%
%%%%%%%%%%%%%%%%%%%%%%%%%%%%%%%%%%%%%%%%%%%%%%%
\begin{table}[ht]

\begin{center}
\begin{tabular}{m{1cm}|m{2cm}|m{2.8cm}|m{2.5cm}|m{0.6cm}m{0.6cm}m{0.6cm}m{0.6cm}m{0.6cm}}
\hline \hline
\# & Orbifold & Orbifold Action & $I_0$ (Corners) & $\mathsf{g}_{x_{1}^{4}}$ & $\mathsf{g}_{x_{1}^{2}x_{2}}$ & $\mathsf{g}_{x_{2}^{2}}$ & $\mathsf{g}_{x_{1}x_{3}}$ & $\mathsf{g}_{x_{4}}$ \\
\hline \hline

%%%%%%%%%%%%%%%%%%% N=1 %%%%%%%%%%%%%%%%%%%
(1.1) & $\mathbb{C}^{4}/\mathbb{Z}_{1}$ & $\bvec (0,0,0,0)\\(0,0,0,0)\\(0,0,0,0) \evec$ & 
$\bvset{c} (0,0,0) \\ (1,0,0)\\(0,1,0) \\ (0,0,1)\evset$
& 1 & 1& 1 & 1 & 1
\\
\hline
\multicolumn{4}{r|}{Total} &1 &1 &1 &1 &1
\\ 
\hline \hline

%%%%%%%%%%%%%%%%%%% N=2 %%%%%%%%%%%%%%%%%%%
(2.1) & $\mathbb{C}^{4}/\mathbb{Z}_{2}$ & $\bvec (0,0,1,1)\\(0,0,0,0)\\(0,0,0,0) \evec$ &
$\bvset{c} (0,0,0)\\(1,0,0)\\(0,1,0)\\(0,0,2) \evset$
& 6 &2 & 2& 0 & 0
\\
\hline
(2.2) & $\mathbb{C}^{4}/\mathbb{Z}_{2}$ & $\bvec (1,1,1,1)\\(0,0,0,0)\\(0,0,0,0) \evec$ &
$\bvset{c} (0,0,0)\\(1,0,0)\\(0,1,0)\\(1,1,2)\evset$
& 1 & 1& 1& 1& 1
\\
\hline
\multicolumn{4}{r|}{Total} &7 &3 &3 &1 &1
\\
\hline \hline
%\end{tabular}
%\end{center}

%\caption{Orbifold Actions and corresponding Toric Diagrams for $\mathbb{C}^{4}/\Gamma_N$ orbifolds with order $N=1\dots 6$ \textbf{(Part 1/10)}.}
%\label{t5}
%\end{table}

%%%%%%%%%%%%%%%%%%%%%%%%%%%%%%%%%%%%%%%%%%%%%%%
%%%%%%%%%%%%%%%%%%%%%%%%%%%%%%%%%%%%%%%%%%%%%%%
%\begin{table}[ht]

%\begin{center}
%\begin{tabular}{m{1cm}|m{0.8cm}|m{2.2cm}|m{3cm}|m{6cm}|m{2cm}}
%\hline \hline
%\# & $N$ & Orbifold & Orbifold Action & Toric Diagram & Multiplicity\\
%\hline \hline

%%%%%%%%%%%%%%%%%%% N=3 %%%%%%%%%%%%%%%%%%%
(3.1) &  $\mathbb{C}^{4}/\mathbb{Z}_{3}$ & $\bvec (0,0,1,2)\\(0,0,0,0)\\(0,0,0,0) \evec$ &
$\bvset{c}(0,0,0)\\(1,0,0)\\(0,1,0)\\(0,0,3)\evset$
& 6 & 2 & 2& 0 & 0
\\
\hline
(3.2) &  $\mathbb{C}^{4}/\mathbb{Z}_{3}$ & $\bvec (0,1,1,1)\\(0,0,0,0)\\(0,0,0,0) \evec$ &
$\bvset{c} (0,0,0)\\(1,0,0)\\(0,1,0)\\(0,2,3) \evset$
& 4 & 2 & 0 & 1 & 0
\\
\hline
(3.3) &  $\mathbb{C}^{4}/\mathbb{Z}_{3}$ & $\bvec (1,1,2,2)\\(0,0,0,0)\\(0,0,0,0) \evec$ &
$\bvset{c}(0,0,0)\\(1,0,0)\\(0,1,0)\\(1,1,3)\evset$
& 3 & 1 & 3 & 0 & 1
\\
\hline
\multicolumn{4}{r|}{Total} &13 &5 &5 &1 &1\\
\hline \hline
\end{tabular}
\end{center}

\caption{The symmetry counting of distinct abelian orbifolds of $\mathbb{C}^{4}$ with corresponding orbifold actions and toric tetrahedra given in terms of $I_{0}$ (corner points in Cartesian coordinates) \textbf{(Part 1/5)}.}
\label{tc4allp1}
\end{table}

\clearpage

%%%%%%%%%%%%%%%%%%%%%%%%%%%%%%%%%%%%%%%%%%%
%%%%%%%%%%%%%%%%%%% N=4..4 ================
\begin{table}[ht]

\begin{center}
\begin{tabular}{m{1cm}|m{2cm}|m{2.8cm}|m{2.5cm}|m{0.6cm}m{0.6cm}m{0.6cm}m{0.6cm}m{0.6cm}}
\hline \hline
\# & Orbifold & Orbifold Action & $I_0$ (Corners) & $\mathsf{g}_{x_{1}^{4}}$ & $\mathsf{g}_{x_{1}^{2}x_{2}}$ & $\mathsf{g}_{x_{2}^{2}}$ & $\mathsf{g}_{x_{1}x_{3}}$ & $\mathsf{g}_{x_{4}}$ \\
\hline \hline

%%%%%%%%%%%%%%%%%%% N=4 %%%%%%%%%%%%%%%%%%%
(4.1) &  $\mathbb{C}^{4}/\mathbb{Z}_{4}$ & $\bvec (0,0,1,3)\\(0,0,0,0)\\(0,0,0,0) \evec$ & 
$\bvset{c} (0,0,0)\\(1,0,0)\\(0,1,0)\\(0,0,4) \evset$
& 6 & 2& 2& 0 & 0
\\ 
\hline 
(4.2) &  $\mathbb{C}^{4}/\mathbb{Z}_{4}$ & $\bvec (0,1,1,2)\\(0,0,0,0)\\(0,0,0,0) \evec$ &
$\bvset{c}(0,0,0)\\(1,0,0)\\(0,1,0)\\(0,2,4)\evset$
& 12 & 2 & 0 & 0 & 0
\\
\hline
(4.3) &  $\mathbb{C}^{4}/\mathbb{Z}_{4}$ & $\bvec (1,1,3,3)\\(0,0,0,0)\\(0,0,0,0) \evec$ &
$\bvset{c}(0,0,0)\\(1,0,0)\\(0,1,0)\\(1,1,4)\evset$
& 3 & 1& 3 & 0 &1
\\
\hline %\hline
%\end{tabular}
%\end{center}

%\caption{Orbifold Actions and corresponding Toric Diagrams for $\mathbb{C}^{4}/\Gamma_N$ orbifolds with order $N=1\dots 6$ \textbf{(Part 3/10)}.}
%\label{t5c}
%\end{table}

%%%%%%%%%%%%%%%%%%%%%%%%%%%%%%%%%%%%%%%%%%%%%%%
%%%%%%%%%%%%%%%%%%%%%%%%%%%%%%%%%%%%%%%%%%%%%%%

%\begin{table}[ht]

%\begin{center}
%\begin{tabular}{m{1cm}|m{0.8cm}|m{2.2cm}|m{3cm}|m{6cm}|m{2cm}}
%\hline \hline
%\# & $N$ & Orbifold & Orbifold Action & Toric Diagram & Multiplicity\\
%\hline \hline
(4.4)  & $\mathbb{C}^{4}/\mathbb{Z}_{4}$ & $\bvec (1,2,2,3)\\(0,0,0,0)\\(0,0,0,0) \evec$ &
$\bvset{c}(0,0,0)\\(1,0,0)\\(0,1,0)\\(1,2,4)\evset$ 
& 6 & 2& 2& 0 & 0
\\
\hline
(4.5)  & $\mathbb{C}^{4}/\mathbb{Z}_{4}$ & $\bvec (1,1,1,1)\\(0,0,0,0)\\(0,0,0,0) \evec$ &
$\bvset{c}(0,0,0)\\(1,0,0)\\(0,1,0)\\(3,3,4) \evset$ 
& 1 & 1& 1& 1& 1
\\
\hline
(4.6)  & $\mathbb{C}^{4}/\mathbb{Z}_{2}\times\mathbb{Z}_{2}$ & $\bvec (0,1,0,1)\\(0,0,1,1)\\(0,0,0,0) \evec$ &
$\bvset{c}(0,0,0)\\(1,0,0)\\(0,2,0)\\(0,0,2)\evset$
& 4 & 2 & 0 & 1 & 0
\\
\hline
(4.7) & $\mathbb{C}^{4}/\mathbb{Z}_{2}\times\mathbb{Z}_{2}$ & $\bvec  (0,0,1,1)\\(1,1,1,1)\\(0,0,0,0) \evec$ &
$\bvset{c}(0,0,0)\\(1,0,0)\\(0,2,0)\\(1,0,2)\evset$
& 3 & 1& 3 &0 &1
\\
\hline
\multicolumn{4}{r|}{Total} &35 &11 &11 &2 &3
\\
\hline \hline 

\end{tabular}
\end{center}

\caption{The symmetry counting of distinct abelian orbifolds of $\mathbb{C}^{4}$ with corresponding orbifold actions and toric tetrahedra given in terms of $I_{0}$ (corner points in Cartesian coordinates) \textbf{(Part 2/5)}.}
\label{tc4allp2}
\end{table}

\clearpage

%%%%%%%%%%%%%%%%%%%%%%%%%%%%%%%%%%%%%%%%%%%
%%%%%%%%%%%%%%%%%%% N=5..5 ================
\begin{table}[ht]

\begin{center}
\begin{tabular}{m{1cm}|m{2cm}|m{2.8cm}|m{2.5cm}|m{0.6cm}m{0.6cm}m{0.6cm}m{0.6cm}m{0.6cm}}
\hline \hline
\# & Orbifold & Orbifold Action & $I_0$ (Corners) & $\mathsf{g}_{x_{1}^{4}}$ & $\mathsf{g}_{x_{1}^{2}x_{2}}$ & $\mathsf{g}_{x_{2}^{2}}$ & $\mathsf{g}_{x_{1}x_{3}}$ & $\mathsf{g}_{x_{4}}$ \\
\hline \hline

%%%%%%%%%%%%%%%%%%% N=4 %%%%%%%%%%%%%%%%%%%
%%%%%%%%%%%%%%%%%%% N=5 %%%%%%%%%%%%%%%%%%%
(5.1) & $\mathbb{C}^{4}/\mathbb{Z}_{5}$ & $\bvec (0,0,1,4)\\(0,0,0,0)\\(0,0,0,0) \evec$ & 
$\bvset{c} (0,0,0)\\(1,0,0)\\(0,1,0)\\(0,0,5) \evset$
& 6 & 2&2&0 &0
\\ 
\hline
(5.2) & $\mathbb{C}^{4}/\mathbb{Z}_{5}$ & $\bvec (0,1,1,3)\\(0,0,0,0)\\(0,0,0,0) \evec$ &
$\bvset{c} (0,0,0)\\(1,0,0)\\(0,1,0)\\(0,2,5) \evset$
& 12 & 2 & 0 & 0 & 0
\\
\hline% \hline

%\end{tabular}
%\end{center}

%\caption{Orbifold Actions and corresponding Toric Diagrams for $\mathbb{C}^{4}/\Gamma_N$ orbifolds with order $N=1\dots 6$ \textbf{(Part 5/10)}.}
%\label{t6}
%\end{table}

%%%%%%%%%%%%%%%%%%%%%%%%%%%%%%%%%%%%%%%%%%%%%%%
%%%%%%%%%%%%%%%%%%%%%%%%%%%%%%%%%%%%%%%%%%%%%%%

%\begin{table}[ht]

%\begin{center}
%\begin{tabular}{m{1cm}|m{0.8cm}|m{2.2cm}|m{3cm}|m{6cm}|m{2cm}}
%\hline \hline
%\# & $N$ & Orbifold & Orbifold Action & Toric Diagram & Multiplicity\\
%\hline \hline
(5.3) & $\mathbb{C}^{4}/\mathbb{Z}_{5}$ & $\bvec (1,1,4,4)\\(0,0,0,0)\\(0,0,0,0) \evec$ &
$\bvset{c} (0,0,0)\\(1,0,0)\\(0,1,0)\\(1,1,5) \evset$
& 3 & 1 & 3 & 0 & 1
\\
\hline
(5.4) & $\mathbb{C}^{4}/\mathbb{Z}_{5}$ & $\bvec (1,2,3,4)\\(0,0,0,0)\\(0,0,0,0) \evec$ &
$\bvset{c} (0,0,0)\\(1,0,0)\\(0,1,0)\\(1,2,5) \evset$
& 6 & 0 & 2 & 0 & 2
\\
\hline
(5.5) & $\mathbb{C}^{4}/\mathbb{Z}_{5}$ & $\bvec (1,1,1,2)\\(0,0,0,0)\\(0,0,0,0) \evec$ &
$\bvset{c} (0,0,0)\\(1,0,0)\\(0,1,0)\\(2,2,5) \evset$
& 4 & 2 & 0 & 1 & 0
\\
\hline
\multicolumn{4}{r|}{Total} &31 &7 &7 &1 &3
\\
\hline \hline

(6.1) & $\mathbb{C}^{4}/\mathbb{Z}_{6}$ & $\bvec (0,0,1,5)\\(0,0,0,0)\\(0,0,0,0) \evec$ & 
$\bvset{c} (0,0,0)\\(1,0,0)\\(0,1,0)\\(0,0,6) \evset$
 & 6 & 2 & 2 & 0 & 0
\\ 
\hline
(6.2) & $\mathbb{C}^{4}/\mathbb{Z}_{6}$ & $\bvec (0,1,1,4)\\(0,0,0,0)\\(0,0,0,0) \evec$ &
$\bvset{c}(0,0,0)\\(1,0,0)\\(0,1,0)\\(0,2,6) \evset$
& 12 & 2 & 0 & 0 & 0
\\
\hline \hline

\end{tabular}
\end{center}

\caption{The symmetry counting of distinct abelian orbifolds of $\mathbb{C}^{4}$ with corresponding orbifold actions and toric tetrahedra given in terms of $I_{0}$ (corner points in Cartesian coordinates) \textbf{(Part 3/5)}.}
\label{tc4allp3}
\end{table}

\clearpage

%%%%%%%%%%%%%%%%%%%%%%%%%%%%%%%%%%%%%%%%%%%
%%%%%%%%%%%%%%%%%%% N=6..6 PART I ================
\begin{table}[ht]

\begin{center}
\begin{tabular}{m{1cm}|m{2cm}|m{2.8cm}|m{2.5cm}|m{0.6cm}m{0.6cm}m{0.6cm}m{0.6cm}m{0.6cm}}
\hline \hline
\# & Orbifold & Orbifold Action & $I_0$ (Corners) & $\mathsf{g}_{x_{1}^{4}}$ & $\mathsf{g}_{x_{1}^{2}x_{2}}$ & $\mathsf{g}_{x_{2}^{2}}$ & $\mathsf{g}_{x_{1}x_{3}}$ & $\mathsf{g}_{x_{4}}$ \\
\hline \hline

%%%%%%%%%%%%%%%%%%% N=6 %%%%%%%%%%%%%%%%%%%
(6.3)& $\mathbb{C}^{4}/\mathbb{Z}_{6}$ & $\bvec (0,1,2,3)\\(0,0,0,0)\\(0,0,0,0) \evec$ &
$\bvset{c} (0,0,0)\\(1,0,0)\\(0,1,0)\\(0,3,6) \evset$
& 24 & 0 & 0 & 0 & 0
\\
\hline 

(6.4) & $\mathbb{C}^{4}/\mathbb{Z}_{6}$ & $\bvec (1,1,5,5)\\(0,0,0,0)\\(0,0,0,0) \evec$ &
$\bvset{c} (0,0,0)\\(1,0,0)\\(0,1,0)\\(1,1,6) \evset$
& 3 & 1 & 3 & 0 & 1
\\
\hline
(6.5) & $\mathbb{C}^{4}/\mathbb{Z}_{6}$ & $\bvec
(1,2,4,5)\\(0,0,0,0)\\(0,0,0,0) \evec$ & 
$\bvset{c} (0,0,0)\\(1,0,0)\\(0,1,0)\\(1,2,6) \evset$
& 12 & 0 & 4 & 0 & 0
\\
\hline
(6.6) & $\mathbb{C}^{4}/\mathbb{Z}_{6}$ & $\bvec (1,3,3,5)\\(0,0,0,0)\\(0,0,0,0) \evec$ &
$\bvset{c} (0,0,0)\\(1,0,0)\\(0,1,0)\\(1,3,6) \evset$
& 6 & 2 & 2 & 0 & 0
\\
\hline %\hline

%\end{tabular}
%\end{center}

%\caption{Orbifold Actions and corresponding Toric Diagrams for $\mathbb{C}^{4}/\Gamma_N$ orbifolds with order $N=1\dots 6$ \textbf{(Part 8/10)}.}
%\label{t7b}
%\end{table}

%\clearpage

%%%%%%%%%%%%%%%%%%%%%%%%%%%%%%%%%%%%%%%%%%%
%%%%%%%%%%%%%%%%%%% N=6..6 PART II ================
%\begin{table}[ht]

%\begin{center}
%\begin{tabular}{m{1cm}|m{0.8cm}|m{2.2cm}|m{3cm}|m{6cm}|m{2cm}}
%\hline \hline
%\# & $N$ & Orbifold & Orbifold Action & Toric Diagram & Multiplicity\\
%\hline \hline

%%%%%%%%%%%%%%%%%%% N=6 %%%%%%%%%%%%%%%%%%%
(6.7) & $\mathbb{C}^{4}/\mathbb{Z}_{6}$ & $\bvec (1,3,4,4)\\(0,0,0,0)\\(0,0,0,0) \evec$ &
$\bvset{c} (0,0,0)\\(1,0,0)\\(0,1,0)\\(2,2,6) \evset$
& 12 & 2 & 0 & 0 & 0
\\
\hline
(6.8) & $\mathbb{C}^{4}/\mathbb{Z}_{6}$ & $\bvec (1,1,1,3)\\(0,0,0,0)\\(0,0,0,0) \evec$ &
$\bvset{c} (0,0,0)\\(1,0,0)\\(0,1,0)\\(3,5,6) \evset$
& 4 & 2 & 0 & 1 & 0
\\
\hline
(6.9) & $\mathbb{C}^{4}/\mathbb{Z}_{6}$ & $\bvec 
(1,1,2,2)\\(0,0,0,0)\\(0,0,0,0) \evec$ &
$\bvset{c} (0,0,0)\\(1,0,0)\\(0,1,0)\\(4,4,6) \evset$
& 6 & 2 & 2 & 0 & 0
\\
\hline \hline

\end{tabular}
\end{center}

\caption{The symmetry counting of distinct abelian orbifolds of $\mathbb{C}^{4}$ with corresponding orbifold actions and toric tetrahedra given in terms of $I_{0}$ (corner points in Cartesian coordinates) \textbf{(Part 4/5)}.}
\label{tc4allp4}
\end{table}

\clearpage

%%%%%%%%%%%%%%%%%%% N=6..6 PART II ================
\begin{table}[ht]

\begin{center}
\begin{tabular}{m{1cm}|m{2cm}|m{2.8cm}|m{2.5cm}|m{0.6cm}m{0.6cm}m{0.6cm}m{0.6cm}m{0.6cm}}
\hline \hline
\# & Orbifold & Orbifold Action & $I_0$ (Corners) & $\mathsf{g}_{x_{1}^{4}}$ & $\mathsf{g}_{x_{1}^{2}x_{2}}$ & $\mathsf{g}_{x_{2}^{2}}$ & $\mathsf{g}_{x_{1}x_{3}}$ & $\mathsf{g}_{x_{4}}$ \\
\hline \hline

(6.10) & $\mathbb{C}^{4}/\mathbb{Z}_{6}$ & 
$\bvec (2,3,3,4)\\(0,0,0,0)\\(0,0,0,0) \evec$ &%\includegraphics*[height=6cm]{C4Tn6m10i1.pdf} 
$\bvset{c}(0,0,0)\\(1,0,0)\\(0,2,0)\\(1,0,3)\evset$
& 6 & 2 & 2 & 0 & 0
\\
\hline
\multicolumn{4}{r|}{Total} &91 &15 &15 &1 &1
\\
\hline\hline
\end{tabular}
\end{center}

\caption{The symmetry counting of distinct abelian orbifolds of $\mathbb{C}^{4}$ with corresponding orbifold actions and toric tetrahedra given in terms of $I_{0}$ (corner points in Cartesian coordinates) \textbf{(Part 5/5)}.}
\label{tc4allp5}
\end{table}

%%%%%%%%%%%%%%%%%%%%%%%%%%%%%%%%%%%%%%%%%%%%%%%%%%%%%%%%%%%%%%%%%%%
%%%%%%%%%%%%%%%%%%%%%%%%%%%%%%%%%%%%%%%%%%%%%%%%%%%%%%%%%%%%%%%%%%%

\section{Symmetries of Abelian Orbifolds of $\mathbb{C}^5$, their Orbifold Actions and Lattice $4$-Simplices}

\begin{table}[ht]

\begin{center}
\begin{tabular}{m{1cm}|m{2cm}|m{2.8cm}|m{2.5cm}|m{0.6cm}m{0.6cm}m{0.6cm}m{0.6cm}m{0.6cm}m{0.6cm}m{0.6cm}}
\hline \hline
\# & Orbifold & Orbifold Action & $I_0$ (Corners) & 
$\mathsf{g}_{x_{1}^{5}}$ & $\mathsf{g}_{x_{1}^{3}x_{2}}$ & $\mathsf{g}_{x_{1}^{2}x_{2}^{2}}$ & $\mathsf{g}_{x_{1}^{2}x_{3}}$ & $\mathsf{g}_{x_{2}x_{3}}$ & $\mathsf{g}_{x_{1}x_{4}}$ & $\mathsf{g}_{x_{5}}$\\
\hline \hline

(1.1) & $\mathbb{C}^{5}/\mathbb{Z}_{1}$ &
$\bvec (0,0,0,0,0)\\(0,0,0,0,0)\\(0,0,0,0,0)\\(0,0,0,0,0) \evec$ &
$\bvset{c} (0,0,0,0)\\(1,0,0,0)\\(0,1,0,0)\\(0,0,1,0)\\(0,0,0,1) \evset$ &
1& 1& 1& 1& 1& 1& 1
\\
\hline
\multicolumn{4}{r|}{Total} &1 &1 &1 &1 &1 &1 &1
\\
\hline \hline

(2.1) & $\mathbb{C}^{5}/\mathbb{Z}_{2}$ &
$\bvec (0,0,0,1,1)\\(0,0,0,0,0)\\(0,0,0,0,0)\\(0,0,0,0,0) \evec$ &
$\bvset{c} (0,0,0,0)\\(1,0,0,0)\\(0,1,0,0)\\(0,0,1,0)\\(0,0,0,2)\evset$ &
10&4 &2 &1 &1 &0 &0
\\
\hline \hline

\end{tabular}
\end{center}

\caption{The symmetry counting of distinct abelian orbifolds of $\mathbb{C}^{5}$ with corresponding orbifold actions and toric $4$-simplices given in terms of $I_{0}$ (corner points in Cartesian coordinates) \textbf{(Part 1/2)}.}
\label{tc5allp1}
\end{table}

\begin{table}[ht]

\begin{center}
\begin{tabular}{m{1cm}|m{2cm}|m{2.8cm}|m{2.5cm}|m{0.6cm}m{0.6cm}m{0.6cm}m{0.6cm}m{0.6cm}m{0.6cm}m{0.6cm}}
\hline \hline
\# & Orbifold & Orbifold Action & $I_0$ (Corners) & 
$\mathsf{g}_{x_{1}^{5}}$ & $\mathsf{g}_{x_{1}^{3}x_{2}}$ & $\mathsf{g}_{x_{1}^{2}x_{2}^{2}}$ & $\mathsf{g}_{x_{1}^{2}x_{3}}$ & $\mathsf{g}_{x_{2}x_{3}}$ & $\mathsf{g}_{x_{1}x_{4}}$ & $\mathsf{g}_{x_{5}}$\\
\hline \hline

(2.2) & $\mathbb{C}^{5}/\mathbb{Z}_{2}$ &
$\bvec (0,1,1,1,1)\\(0,0,0,0,0)\\(0,0,0,0,0)\\(0,0,0,0,0) \evec$ &
$\bvset{c} (0,0,0,0)\\(1,0,0,0)\\(0,1,0,0)\\(0,0,1,0)\\(0,1,1,2) \evset$ &
5&3 &1 &2 &0 &1 &0
\\
\hline
\multicolumn{4}{r|}{Total} &15 &7 &3 &3 &1 &1 &0
\\
\hline \hline

(3.1) & $\mathbb{C}^{5}/\mathbb{Z}_{3}$ &
$\bvec (0,0,0,1,2)\\(0,0,0,0,0)\\(0,0,0,0,0)\\(0,0,0,0,0) \evec$ &
$\bvset{c} (0,0,0,0)\\(1,0,0,0)\\(0,1,0,0)\\(0,0,1,0)\\(0,0,0,3) \evset$ &
10&4 &2 &1 &1 &0 &0
\\
\hline

(3.2) & $\mathbb{C}^{5}/\mathbb{Z}_{3}$ &
$\bvec (0,0,1,1,1)\\(0,0,0,0,0)\\(0,0,0,0,0)\\(0,0,0,0,0) \evec$ &
$\bvset{c} (0,0,0,0)\\(1,0,0,0)\\(0,1,0,0)\\(0,0,1,0)\\(0,0,2,3) \evset$ &
10&4 &2 &1 &1 &0 &0
\\
\hline

(3.3) & $\mathbb{C}^{5}/\mathbb{Z}_{3}$ &
$\bvec (0,1,1,2,2)\\(0,0,0,0,0)\\(0,0,0,0,0)\\(0,0,0,0,0) \evec$ &
$\bvset{c} (0,0,0,0)\\(1,0,0,0)\\(0,1,0,0)\\(0,0,1,0)\\(0,1,1,3) \evset$ &
15&3 &3 &0 &0 &1 &0
\\
\hline

(3.4) & $\mathbb{C}^{5}/\mathbb{Z}_{3}$ &
$\bvec (1,1,1,1,2)\\(0,0,0,0,0)\\(0,0,0,0,0)\\(0,0,0,0,0) \evec$ &
$\bvset{c} (0,0,0,0)\\(1,0,0,0)\\(0,1,0,0)\\(0,0,1,0)\\(1,1,1,3) \evset$ &
5&3 &1 &2 &0 &1 &0
\\
\hline

\multicolumn{4}{r|}{Total} &40 &14 &8 &4 &2 &2 &0
\\
\hline \hline

\end{tabular}
\end{center}

\caption{The symmetry counting of distinct abelian orbifolds of $\mathbb{C}^{5}$ with corresponding orbifold actions and toric $4$-simplices given in terms of $I_{0}$ (corner points in Cartesian coordinates) \textbf{(Part 2/2)}.}
\label{tc5allp2}
\end{table}

\clearpage

%%%%%%%%%%%%%%%%%%%%%%%%%%%%%%%%%%%%%%%%%%%%%%%%%%%%%%%%%%%
%%%%%%%%%%%%%%%%%%%%%%%%%%%%%%%%%%%%%%%%%%%%%%%%%%%%%%%%%%%
\section{Examples of Identifying Symmetries using Barycentric Coordinates \label{s3app}}

%%%%%%%%%%%%%%%%%%%%%%%%%%%%%%%%%%%%%%%%%%%%%%%%%%%%%%%%%%%%%%%%%%%%%%%%%%%%%%
\subsection{Example: Lattice Triangles corresponding to Abelian Orbifolds of $\mathbb{C}^{3}$ \label{s3a}}

Consider the orbifold of the form $\mathbb{C}^{3}/\mathbb{Z}_{7}$ with the orbifold actions
\beql{e16}
A_{1}~=~
\bvec
(1,1,5)\\
(0,0,0)
\evec
~~~~,~~~~
A_{2}~=~
\bvec
(1,2,4)\\
(0,0,0)
\evec~~.
\eeq\\

\begin{figure}[ht!]
\begin{center}
\includegraphics[totalheight=7cm]{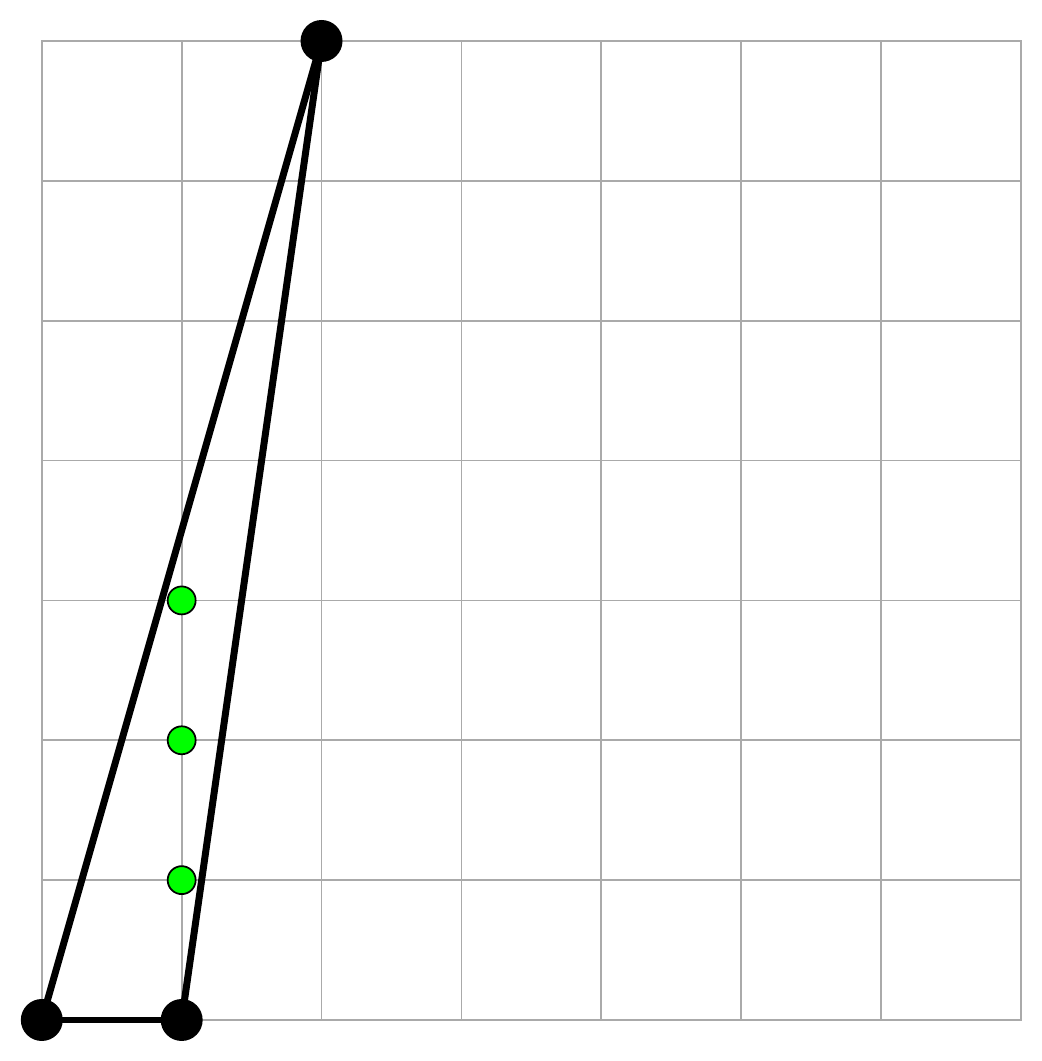}
\includegraphics[totalheight=7cm]{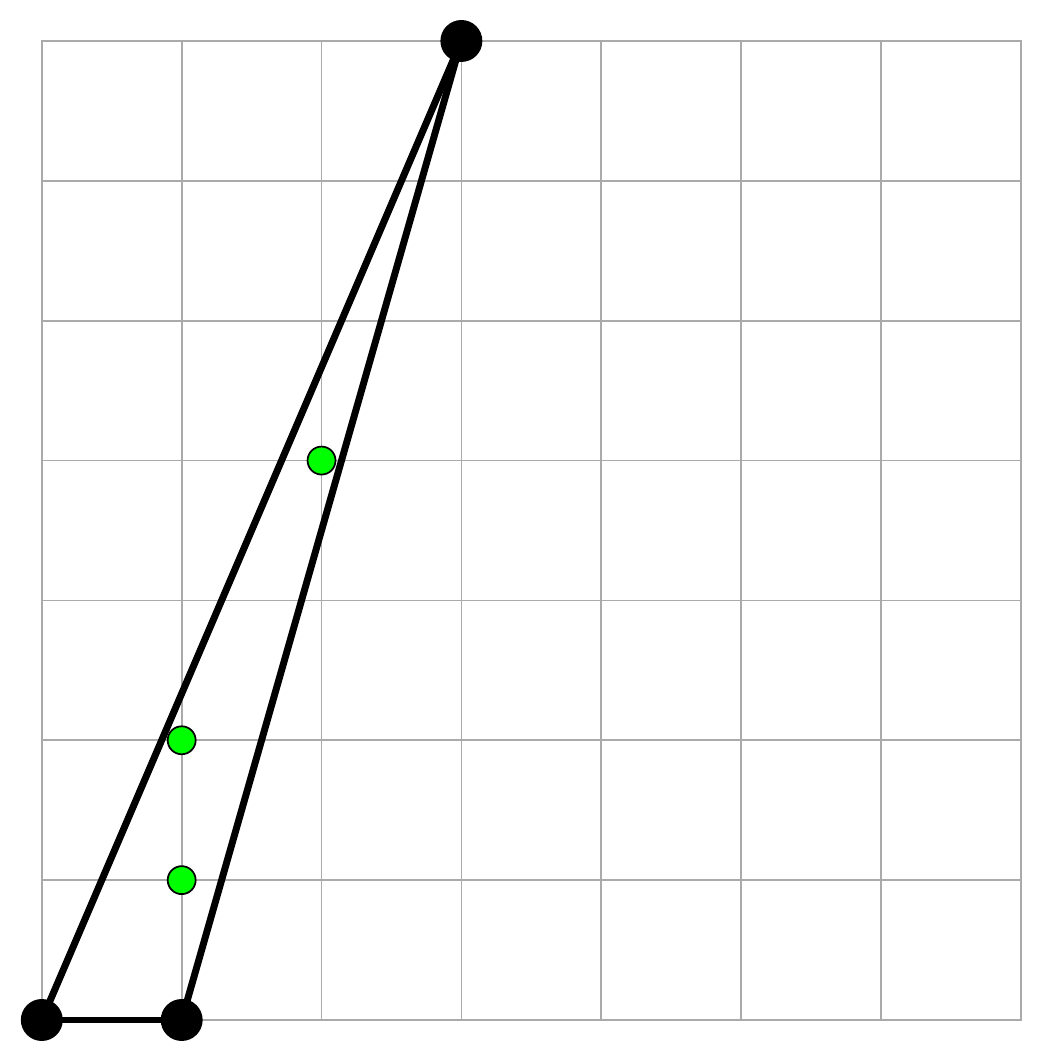}
\caption{Toric triangles of $\mathbb{C}^{3}/\mathbb{Z}_{7}$ with scaling $s_{2}=1$ corresponding to orbifold actions $A_1=((1,1,5),(0,0,0))$ and $A_2=((1,2,4),(0,0,0))$ respectively. Internal toric points $w_{k}\in I_{2}$ are colored green.}
  \label{f1}
 \end{center}
\end{figure}

\noindent\textbf{The scaled Toric Diagram.} The corresponding toric $2$-simplices are shown in \fref{f1} with each having $|I_2|=3$ internal lattice points colored green in the diagram.\\

\begin{figure}[ht!]
\begin{center}
\includegraphics[totalheight=7cm]{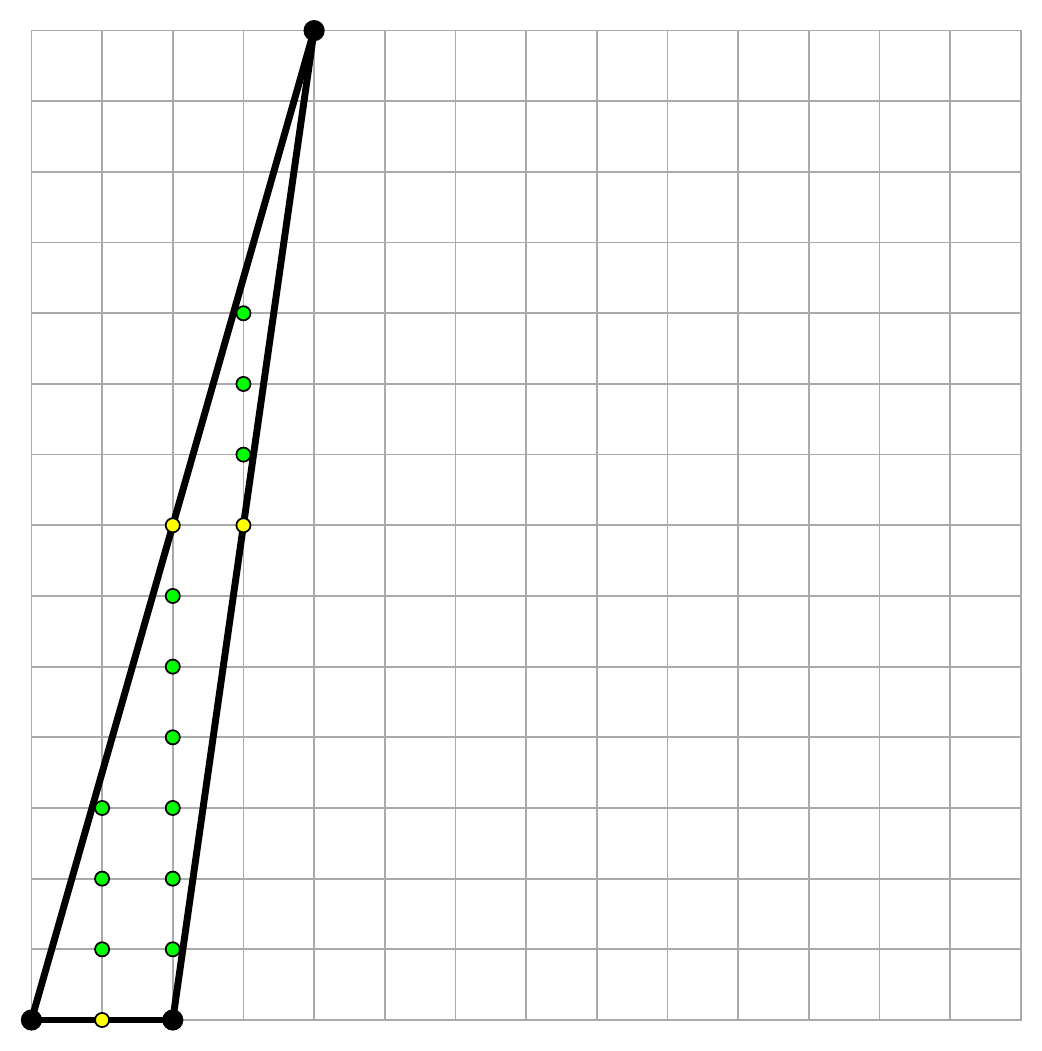}
\includegraphics[totalheight=7cm]{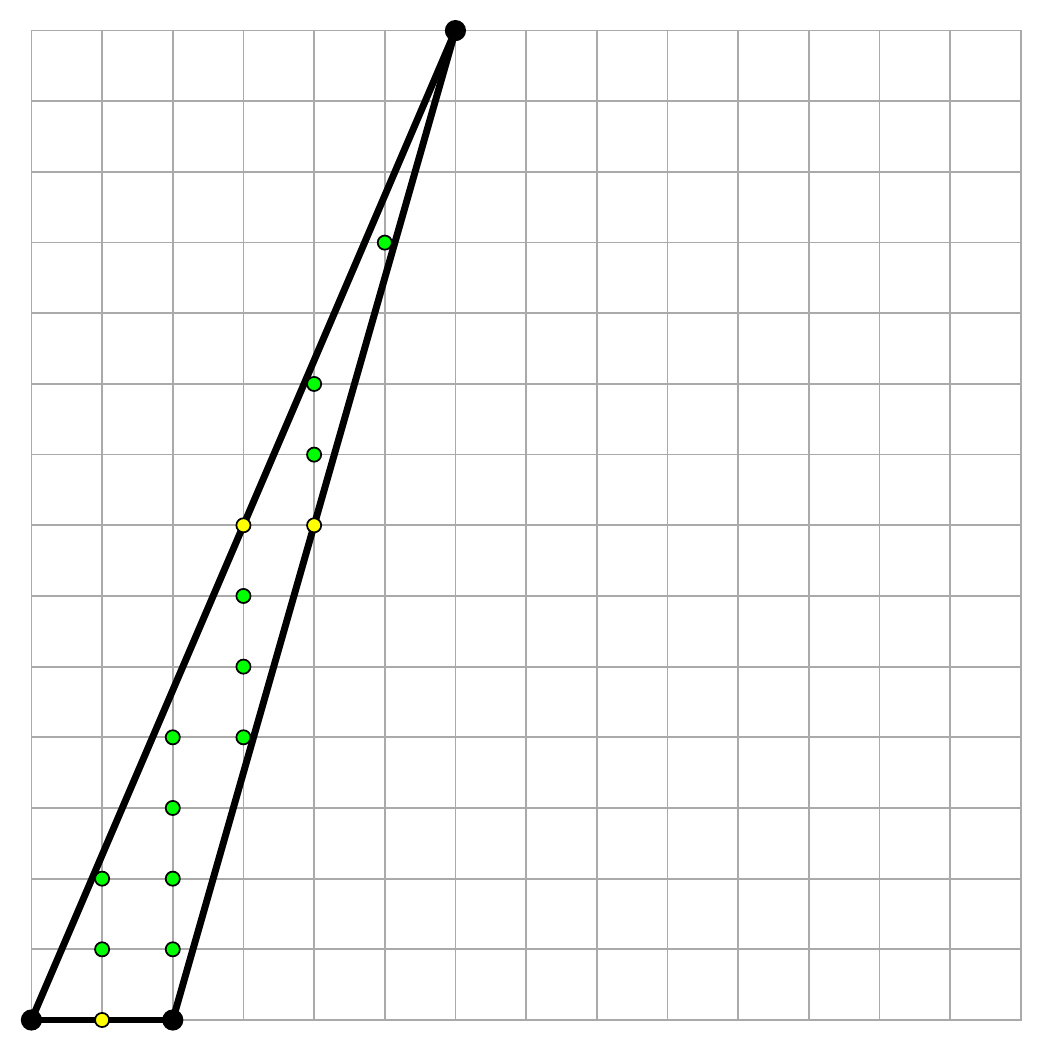}
\caption{Toric triangles of $\mathbb{C}^{3}/\mathbb{Z}_{7}$ with scaling $s_{1}=2$ corresponding to orbifold actions $A_1=((1,1,5),(0,0,0))$ and $A_2=((1,2,4),(0,0,0))$ respectively. Lattice points on edges are colored yellow ($I_1$) and internal toric points $(I_2)$ are colored green.}
  \label{f2}
 \end{center}
\end{figure}

There are no lattice points on the edges of the toric diagrams in \fref{f1}, $|I_1|=\emptyset$. To make them `visible' for the purpose of obtaining the topological character of the toric diagram, we increase the scaling to $s_{1}=2$. This results in the toric diagrams in \fref{f2}. Accordingly, the overall scaling coefficient required for the computation of the topological character is $s=\max{(s_1,s_2)}=\max{(2,1)}=2$.\\

\textbf{The Topological Character.} Let us call the toric triangles corresponding to the orbifold actions $A_1$ and $A_2$ as $\sigma_{1}^{2}$ and $\sigma_{2}^{2}$ respectively. The respective topological characters $\tau_1$ and $\tau_2$ are
\beal{e17}
\tau_1~~=~~
&\Bigg\{&
\underbrace{
(0,0,1),
(0,1,0),
(1,0,0)
}_{I_{0}},
\nn\\
&&
\underbrace{
\left(0,\frac{1}{2},\frac{1}{2}\right),
\left(\frac{1}{2},0,\frac{1}{2}\right),
\left(\frac{1}{2},\frac{1}{2},0\right)
}_{I_{1}},\nn\\
&&
\underbrace{
\left(\frac{1}{14},\frac{5}{14},\frac{4}{7}\right),
\left(\frac{4}{7},\frac{5}{14},\frac{1}{14}\right),
\left(\frac{1}{7},\frac{3}{14},\frac{9}{14}\right),
\left(\frac{9}{14},\frac{3}{14},\frac{1}{7}\right),
\left(\frac{3}{14},\frac{1}{14},\frac{5}{7}\right),
\left(\frac{5}{7},\frac{1}{14},\frac{3}{14}\right)
}_{I_{2}},
\nn\\
&&
\underbrace{
\left(\frac{1}{7},\frac{5}{7},\frac{1}{7}\right),
\left(\frac{2}{7},\frac{3}{7},\frac{2}{7}\right),
\left(\frac{3}{7},\frac{1}{7},\frac{3}{7}\right)
\left(\frac{1}{14},\frac{6}{7},\frac{1}{14}\right),
\left(\frac{3}{14},\frac{4}{7},\frac{3}{14}\right),
\left(\frac{5}{14},\frac{2}{7},\frac{5}{14}\right)
}_{I_{2}}
\Bigg\}~~,
\nn\\
\eea
and
\beal{e18}
\tau_2~~=~~
&\Bigg\{&
\underbrace{
(0,0,1),
(0,1,0),
(1,0,0)
}_{I_{0}},\nn\\
&&
\underbrace{
\left(0,\frac{1}{2},\frac{1}{2}\right),
\left(\frac{1}{2},0,\frac{1}{2}\right),
\left(\frac{1}{2},\frac{1}{2},0\right)
}_{I_{1}},\nn\\
&&
\underbrace{
\left(\frac{1}{14},\frac{2}{7},\frac{9}{14}\right),
\left(\frac{2}{7},\frac{9}{14},\frac{1}{14}\right),
\left(\frac{9}{14},\frac{1}{14},\frac{2}{7}\right),
\left(\frac{1}{14},\frac{11}{14},\frac{1}{7}\right),
\left(\frac{1}{7},\frac{1}{14},\frac{11}{14}\right),
\left(\frac{11}{14},\frac{1}{7},\frac{1}{14}\right)
}_{I_{2}},\nn\\
&&
\underbrace{
\left(\frac{1}{7},\frac{4}{7},\frac{2}{7}\right),
\left(\frac{4}{7},\frac{2}{7},\frac{1}{7}\right),
\left(\frac{2}{7},\frac{1}{7},\frac{4}{7}\right),
\left(\frac{3}{14},\frac{5}{14},\frac{3}{7}\right),
\left(\frac{5}{14},\frac{3}{7},\frac{3}{14}\right),
\left(\frac{3}{7},\frac{3}{14},\frac{5}{14}\right)
}_{I_{2}}
\Bigg\}~~.\nn\\
\eea
The elements of the characters above are barycentric coordinates of the topologically important points in $I_0$, $I_1$ and $I_2$ with an overall scaling $s=2$.\\

\noindent\textbf{The Symmetries.} The orbifold dimension is $D=3$. Accordingly, we consider cycles of $S_{3}$ corresponding to $C_{(1)(2)(3)}$, $C_{(1,2,3)}$, $C_{(1,3,2)}$, $C_{(2,3)(1)}$, $C_{(1,3)(2)}$ and $C_{(1,2)(3)}$.\\

\begin{figure}[ht!]
\begin{center}
\includegraphics[totalheight=7cm]{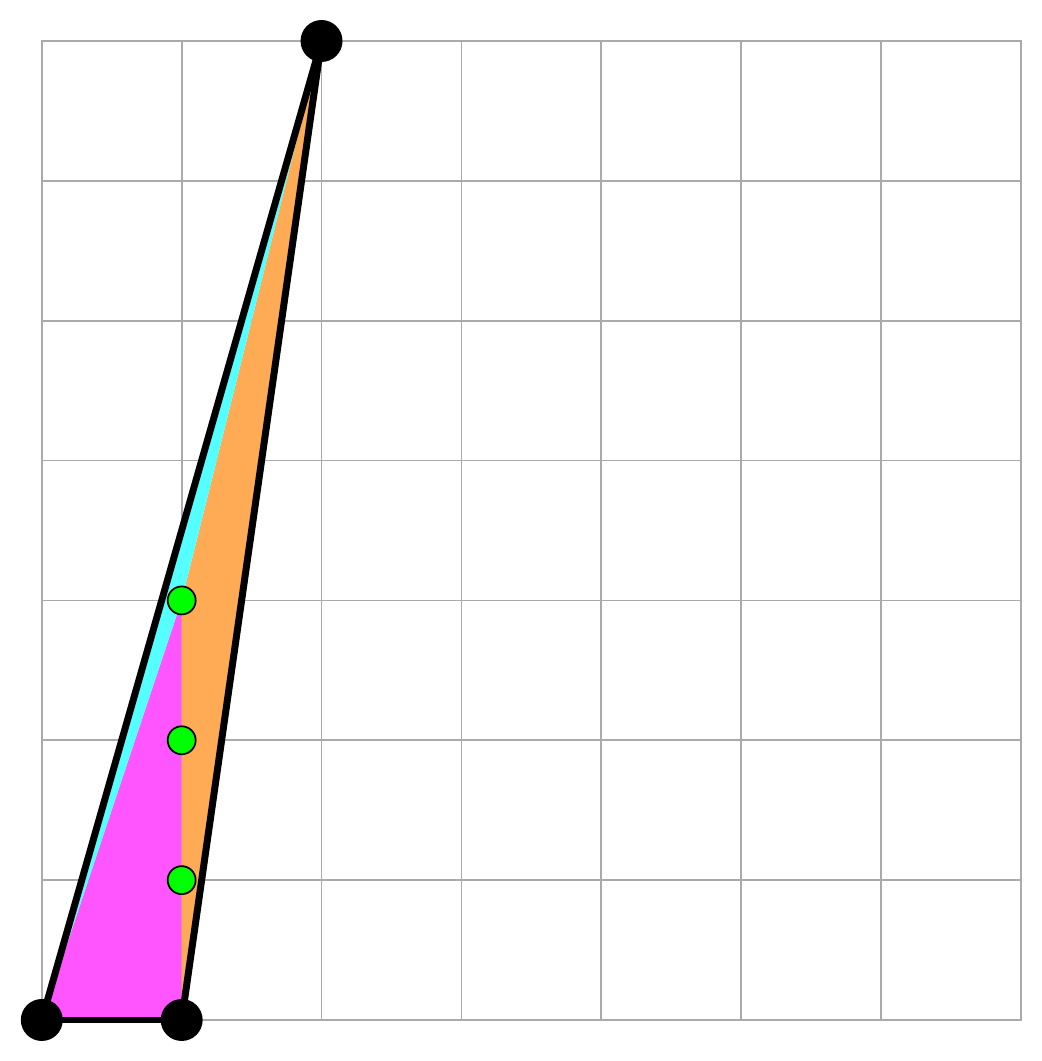}
\includegraphics[totalheight=7cm]{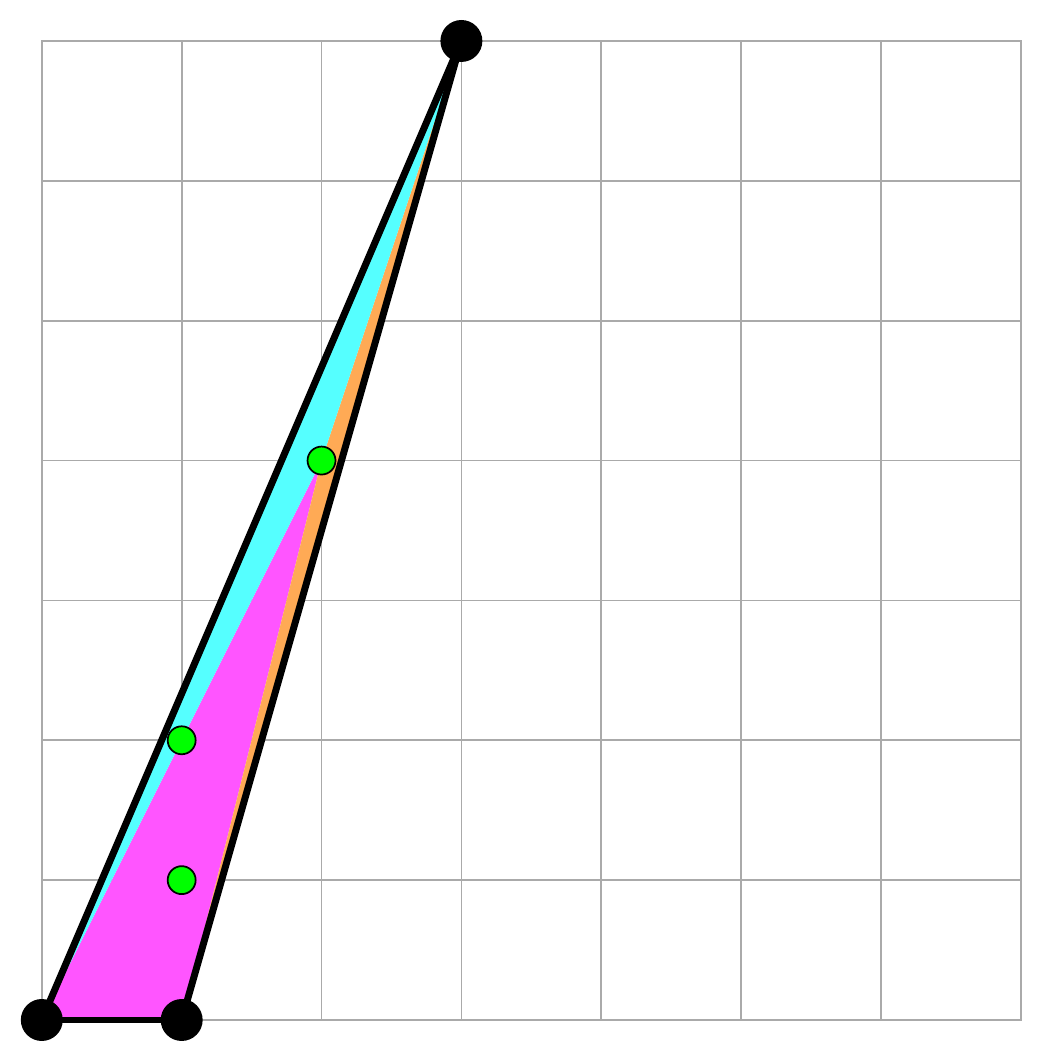}
\caption{Toric triangles of $\mathbb{C}^{3}/\mathbb{Z}_{7}$ with scaling $s_{2}=1$ corresponding to orbifold actions $A_1=((1,1,5),(0,0,0))$ and $A_2=((1,2,4),(0,0,0))$ respectively. For the diagram of $A_{1}$ on the left, the sub-triangles with areas proportional to the barycentric coordinates of the internal point $\left(\frac{3}{7},\frac{1}{7},\frac{3}{7}\right)\in I_{2}(f_{s_{2}=1}(\sigma^{2}_{1}))$ are colored magenta ($\frac{3}{7}$), cyan ($\frac{1}{7}$) and orange ($\frac{3}{7}$). For the diagram of $A_{2}$, the sub-triangles with areas proportional to the barycentric coordinates of the internal point $\left(\frac{4}{7},\frac{2}{7},\frac{1}{7}\right)\in I_{2}(f_{s_2=1}(\sigma_{2}^{2}))$ are colored magenta ($\frac{4}{7}$), cyan ($\frac{2}{7}$) and orange ($\frac{1}{7}$).}
  \label{f3}
 \end{center}
\end{figure}

Picking the transformation $C_{(1,3)(2)}$, we observe its action on the barycentric coordinates  $\left(\frac{3}{7},\frac{1}{7},\frac{3}{7}\right)\in I_{2}(f_{s_{2}=1}(\sigma^{2}_{1}))$ of an internal point from the first toric simplex $\sigma_{1}^{2}$ and the barycentric coordinates $\left(\frac{4}{7},\frac{2}{7},\frac{1}{7}\right)\in I_{2}(f_{s_2=1}(\sigma_{2}^{2}))$ of an internal point from the second toric simplex $\sigma_{2}^{2}$. As shown in \fref{f3}, the chosen internal points divide the toric triangles into three sub-triangles each corresponding to one component of the barycentric coordinates.\\

The transformation $C_{(1,3)(2)}$ swaps the barycentric coordinates axes $\hat{v}_{1}$ and $\hat{v}_{3}$ such that $C_{(1,3)(2)}:\left(\frac{3}{7},\frac{1}{7},\frac{3}{7}\right)\mapsto\left(\frac{3}{7},\frac{1}{7},\frac{3}{7}\right)$ and $C_{(1,3)(2)}:\left(\frac{4}{7},\frac{2}{7},\frac{1}{7}\right)\mapsto\left(\frac{1}{7},\frac{2}{7},\frac{4}{7}\right)$. This transformation corresponds to swapping the cyan and orange colored sub-triangles in \fref{f3}. $C_{(1,3)(2)}$ leaves the internal point $\left(\frac{3}{7},\frac{1}{7},\frac{3}{7}\right)$ of $\sigma_{1}^{2}$ invariant. In comparison, $C_{(1,3)(2)}$ maps the internal point $\left(\frac{4}{7},\frac{2}{7},\frac{1}{7}\right)$ to a different point $\left(\frac{1}{7},\frac{2}{7},\frac{4}{7}\right)$ which is in fact not an element of the original topological character of $\sigma_{2}^{2}$ in \eref{e18}. Accordingly, $C_{(1,3)(2)}$ is not a symmetry of $\sigma_{2}^{2}$ and the corresponding orbifold with action $A_2$. In contrast, it turns out that $\tau_2$ is invariant under $C_{(1,3)(2)}$. Accordingly, $C_{(1,3)(2)}$ is a symmetry of $\sigma^{2}_{1}$.\\

%%%%%%%%%%%%%%%%%%%%%%%%%%%%%%%%%%%%%%%%%%%%%%%%%%%%%%%%%%%%%%%%%%%%%%%%%%%%%%
\subsection{Example: Lattice Tetrahedra corresponding to Abelian Orbifolds of $\mathbb{C}^{4}$ \label{s3b}}

Let us proceed with the abelian orbifold of the form $\mathbb{C}^{4}/\mathbb{Z}_{6}$ and orbifold action
\beql{e19}
A~~=~~\bvec
(0,1,1,4)\\
(0,0,0,0)\\
(0,0,0,0)
\evec~~.
\eeq\\

\begin{figure}[ht!]
\begin{center}
\includegraphics[totalheight=7cm]{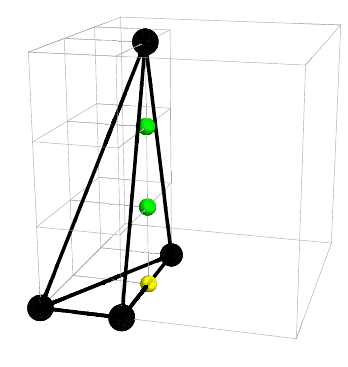}
\includegraphics[totalheight=7cm]{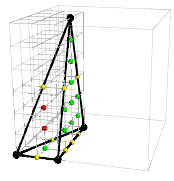}
\caption{Toric tetrahedra $\sigma^{3}=f_{1}(\sigma^{3})$ and $f_{2}(\sigma^{3})$ of $\mathbb{C}^{4}/\mathbb{Z}_{6}$ corresponding to orbifold action $A=((0,1,1,4),(0,0,0,0),(0,0,0,0))$ with optimal scaling $s_{1}=s_{2}=1$ for edge $I_{1}(\sigma^{3})$ and face $I_2(\sigma^{3})$ points, and optimal scaling $s_3=2$ for internal points $I_{3}(f_{2}(\sigma^{3}))$. Internal lattice points are colored red, while edge and face points are colored yellow and green respectively.}
  \label{f4}
 \end{center}
\end{figure}

\noindent\textbf{The scaled Toric Diagram.} The corresponding toric tetrahedron $\sigma^{3}$ for \eref{e19} is shown in \fref{f4}. With unit scaling $s_1=s_2=1$ there is $|I_1|=1$ lattice point on an edge and $|I_2|=2$ lattice points on the faces of the toric tetrahedron. For internal lattice points, we need to scale the tetrahedron with $s_{3}=2$ such that $|I_3|=2$. Accordingly, the optimal scaling coefficient for $\sigma^{3}$ is $s=\max{(s_1,s_2,s_3)}=\max{(1,1,2)}=2$.\\

\noindent\textbf{The Topological Character.} The topological character of $\sigma^3$ is
\beal{e20}
\tau(\sigma^{3})~~=~~
&\Bigg\{&
\underbrace{
(0,0,0,1),
(0,0,1,0),
(0,1,0,0),
(1,0,0,0)
}_{I_{0}},\nn\\
&&
\underbrace{
\left(0,0,\frac{1}{2},\frac{1}{2}\right),
\left(0,\frac{1}{2},\frac{1}{2},0\right),
\left(\frac{1}{2},\frac{1}{2},0,0\right),
\left(0,\frac{1}{2},0,\frac{1}{2}\right)
}_{I_{1}},\nn\\
&&
\underbrace{
\left(\frac{1}{2},0,\frac{1}{2},0\right),
\left(\frac{1}{2},0,0,\frac{1}{2}\right),
\left(0,\frac{1}{4},\frac{3}{4},0\right),
\left(0,\frac{3}{4},\frac{1}{4},0\right)
}_{I_{1}},\nn\\
&&
\underbrace{
\left(0,\frac{1}{4},\frac{1}{4},\frac{1}{2}\right),
\left(\frac{1}{2},\frac{1}{4},\frac{1}{4},0\right),
\left(\frac{1}{6},\frac{1}{6},\frac{2}{3},0\right),
\left(\frac{1}{6},\frac{2}{3},\frac{1}{6},0\right)
}_{I_{2}},\nn\\
&&
\underbrace{
\left(\frac{2}{3},\frac{1}{6},\frac{1}{6},0\right),
\left(\frac{1}{3},\frac{1}{3},\frac{1}{3},0\right),
\left(\frac{1}{3},\frac{1}{12},\frac{7}{12},0\right),
\left(\frac{1}{3},\frac{7}{12},\frac{1}{12},0\right)
}_{I_{2}},\nn\\
&&
\underbrace{
\left(\frac{1}{6},\frac{5}{12},\frac{5}{12},0\right),
\left(\frac{5}{6},\frac{1}{12},\frac{1}{12},0\right)
}_{I_{2}},
\underbrace{
\left(\frac{1}{6},\frac{1}{6},\frac{1}{6},\frac{1}{2}\right),
\left(\frac{1}{3},\frac{1}{12},\frac{1}{12},\frac{1}{2}\right)
}_{I_{3}}
\Bigg\}~~.\nn\\
\eea
\\

\begin{figure}[ht!]
\begin{center}
\includegraphics[totalheight=9cm]{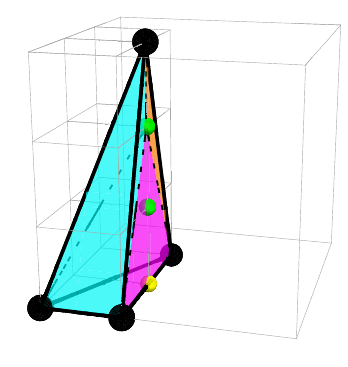}
\caption{Toric tetrahedra of $\mathbb{C}^{4}/\mathbb{Z}_{6}$ corresponding to orbifold action $A_3=((0,1,1,5),(0,0,0,0),(0,0,0,0))$ with optimal scaling $s_{2}=1$. The face point with barycentric coordinates $\left(\frac{2}{3},\frac{1}{6},\frac{1}{6},0\right)$ divides the tetrahedron into four sub-tetrahedra with one having a nil volume. The other three sub-tetrahedra have volumes $\frac{2}{3}$ (magenta), $\frac{1}{6}$ (cyan) and $\frac{1}{6}$ (orange).}
  \label{f5}
 \end{center}
\end{figure}

\noindent\textbf{The Symmetries.} Let us pick the lattice point on a face with barycentric coordinates $\left(\frac{2}{3},\frac{1}{6},\frac{1}{6},0\right)\in I_{2}$ as shown in \fref{f5}. The face point divides the tetrahedron into four sub-tetrahedra with volumes corresponding to  $\left(\frac{2}{3},\frac{1}{6},\frac{1}{6},0\right)$. One sub-tetrahedron has zero volume, the other three have normalized volumes $\frac{2}{3}$, $\frac{1}{6}$ and $\frac{1}{6}$ colored magenta, cyan and orange respectively in \fref{f5}.\\

Let us pick the $S_4$ transformation $C_{(1,3,4,2)}$ which acts on the barycentric coordinates axes $\{\hat{v}_{1},\hat{v}_{2},\hat{v}_{3},\hat{v}_{4}\}$ as $C_{(1,3,4,2)}:[\hat{v}_{1},\hat{v}_{2},\hat{v}_{3},\hat{v}_{4}]\mapsto[\hat{v}_{3},\hat{v}_{1},\hat{v}_{4},\hat{v}_{2}]$. The transformation corresponds to a cyclic permutation of the sub-tetrahedra in \fref{f5}. $C_{(1,3,4,2)}$ transforms the face point $\left(\frac{2}{3},\frac{1}{6},\frac{1}{6},0\right)$ into $\left(\frac{1}{6},\frac{2}{3},0,\frac{1}{6}\right)$ which is not an element of the topological character $\tau(\sigma^{3})$ in \eref{e20}. Accordingly, the lattice simplex $\sigma^{3}$ and its corresponding orbifold action $A$ are not symmetric under $C_{(3,1,2,4)}$.\\

Another transformation is $C_{(2,3)(1)(4)}$. It leaves the barycentric coordinates of the face point $\left(\frac{2}{3},\frac{1}{6},\frac{1}{6},0\right)$ invariant. In fact, the entire topological character $\tau(\sigma^{3})$ is invariant under $C_{(2,3)(1)(4)}$. Accordingly, $C_{(2,3)(1)(4)}$ is a symmetry of the lattice simplex $\sigma^{3}$ and its corresponding orbifold action $A$.\\

%%%%%%%%%%%%%%%%%%%%%%%%%%%%%%%%%%%%%%%%%%%%%%%%%%%%%%%%%%%%%%%%%%%%%%%%%

\section{Sequence Derivations for Abelian Orbifolds of $\mathbb{C}^D$ \label{appseq}}

Under observation \ref{prop2}, the convolution for $x_{7}$ is derived as
\beal{epre1}
\mathsf{g}_{x_{7}} 
&=& 
\chi_{7,1} * \chi_{7,2} * \chi_{7,3} * \chi_{7,4} * \chi_{7,5} * \chi_{7,6} * \left(\sum_{a=0}^{\infty}{t^{7^a}}\right) \nn\\
&=& 
\mathsf{u} * \chi_{7,2} * \chi_{7,3} * \chi_{7,4} * \chi_{7,5} * \chi_{7,6} \nn\\
\Rightarrow g_{x_{7}}(t) 
&=& 
t+t^7+2 t^8+6 t^{29}+6 t^{43}+t^{49}+2 t^{56}+3 t^{64}+6 t^{71}+\mathcal{O}(t^{101})~~.
\eea
\\

Under observation \ref{prop2}, the following sequences for abelian orbifolds of $\mathbb{C}^{7}$ are derived, 
\beal{epre2}
\mathsf{g}_{x_{1}^{5}x_{2}}
&=& 
\mathsf{u}*\mathsf{N}*\mathsf{N}^{2}*\mathsf{N}^{3}*\mathsf{N}^{4}*\chi_{2,1}*\left(t+32\sum_{a=0}^{\infty}{t^{2^{(a+2)}}}\right)\nn\\
&=& 
\mathsf{u}*\mathsf{u}*\mathsf{N}*\mathsf{N}^{2}*\mathsf{N}^{3}*\mathsf{N}^{4}*(t-t^2+32t^4)\nn\\
\Rightarrow g_{x_{1}^{5}x_{2}}(t) 
&=& 
t+32 t^2+122 t^3+683 t^4+782 t^5+3904 t^6+2802 t^7+12494 t^8+11133 t^9\nn\\
&&+25024 t^{10}+16106 t^{11}+83326 t^{12}+30942 t^{13}+89664 t^{14}+95404 t^{15}\nn\\
&&+213281 t^{16}+88742 t^{17}+356256 t^{18}+137562 t^{19}
+\mathcal{O}(t^{20})
\eea
\beal{epre3}
\mathsf{g}_{x_{1}^{4}x_{3}}
&=& 
\mathsf{u}*\mathsf{N}*\mathsf{N}^{2}*\mathsf{N}^{3}*\chi_{3,1}*\chi_{3,2}*\left(t+81\sum_{a=0}^{\infty}{t^{3^{(a+2)}}}\right)\nn\\
&=&
\mathsf{u}*\mathsf{u}*\mathsf{N}*\mathsf{N}^{2}*\mathsf{N}^{3}*\chi_{3,2}*(t-t^3+81t^9)
\nn\\
\Rightarrow g_{x_{1}^{4}x_{3}} 
&=&
t+15 t^2+40 t^3+156 t^4+156 t^5+600 t^6+402 t^7+1410 t^8+1291 t^9+2340 t^{10}\nn\\
&&+1464 t^{11}+6240 t^{12}+2382 t^{13}+6030 t^{14}+6240 t^{15}+11967 t^{16}+5220 t^{17}\nn\\
&&+19365 t^{18}+7242 t^{19}+24336 t^{20}+16080 t^{21}+21960 t^{22}
+\mathcal{O}(t^{23})
\eea
\beal{epre4}
\mathsf{g}_{x_{1}^{2}x_{5}}
&=&
\mathsf{u}*\mathsf{N}*\chi_{5,1}*\chi_{5,2}*\chi_{5,3}*\chi_{5,4}*\left(t+25\sum_{a=0}^{\infty}{t^{5^{(a+2)}}}\right)
\nn\\
&=&
\mathsf{u}*\mathsf{u}*\mathsf{N}*\chi_{5,2}*\chi_{5,3}*\chi_{5,4}*(t-t^5+25t^{25})
\nn\\
\Rightarrow g_{x_{1}^{2}x_{5}}
&=&
t+3 t^2+4 t^3+7 t^4+6 t^5+12 t^6+8 t^7+15 t^8+13 t^9+18 t^{10}+16 t^{11}+28 t^{12}\nn\\
&&+14 t^{13}+24 t^{14}+24 t^{15}+32 t^{16}+18 t^{17}+39 t^{18}+20 t^{19}+42 t^{20}+32 t^{21}+48 t^{22}\nn\\
&&+24 t^{23}+60 t^{24}+56 t^{25}+42 t^{26}+40 t^{27}+56 t^{28}+30 t^{29}
+\mathcal{O}(t^{30})
~~.
\eea
\\

For the abelian orbifolds of $\mathbb{C}^{8}$, we derive using observation \ref{prop2},
\beal{epre3}
\mathsf{g}_{x_{1}^{6}x_{2}}
&=& 
\mathsf{u}*\mathsf{N}*\mathsf{N}^{2}*\mathsf{N}^{3}*\mathsf{N}^{4}*\mathsf{N}^{5}*\chi_{2,1}*\left(t+64\sum_{a=0}^{\infty}{t^{2^{(a+2)}}}\right)\nn\\
&=& 
\mathsf{u}*\mathsf{u}*\mathsf{N}*\mathsf{N}^{2}*\mathsf{N}^{3}*\mathsf{N}^{4}*\mathsf{N}^{5}*(t-t^2+64t^4)\nn\\
\Rightarrow g_{x_{1}^{6}x_{2}}(t) 
&=& 
t+63 t^2+365 t^3+2731 t^4+3907 t^5+22995 t^6+19609 t^7+101251 t^8\nn\\
&&+99828 t^9+246141 t^{10}+177157 t^{11}+996815 t^{12}+402235 t^{13}+1235367 t^{14}\nn\\
&&+1426055 t^{15}+3484531 t^{16}+1508599 t^{17}+6289164 t^{18}+2613661 t^{19}
+\mathcal{O}(t^20)\nn\\
\eea
\beal{epre5}
\mathsf{g}_{x_{1}^{5}x_{3}}
&=& 
\mathsf{u}*\mathsf{N}*\mathsf{N}^{2}*\mathsf{N}^{3}*\mathsf{N}^{4}*\chi_{3,1}*\chi_{3,2}*\left(t+243\sum_{a=0}^{\infty}{t^{3^{(a+2)}}}\right)\nn\\
&=&
\mathsf{u}*\mathsf{u}*\mathsf{N}*\mathsf{N}^{2}*\mathsf{N}^{3}*\mathsf{N}^{4}*\chi_{3,2}*(t-t^3+243t^9)
\nn\\
\Rightarrow g_{x_{1}^{5}x_{3}}(t) 
&=&
t+31 t^2+121 t^3+652 t^4+781 t^5+3751 t^6+2803 t^7+11842 t^8+11254 t^9\nn\\
&&+24211 t^{10}+16105 t^{11}+78892 t^{12}+30943 t^{13}+86893 t^{14}+94501 t^{15}\nn\\
&&+201439 t^{16}+88741 t^{17}+348874 t^{18}+137563 t^{19}
+\mathcal{O}(t^{20})
\eea
\beal{epre6}
\mathsf{g}_{x_{1}^{3}x_{5}}
&=&
\mathsf{u}*\mathsf{N}*\mathsf{N}^{2}*\chi_{5,1}*\chi_{5,2}*\chi_{5,3}*\chi_{5,4}*\left(t+125\sum_{a=0}^{\infty}{t^{5^{(a+2)}}}\right)
\nn\\
&=&
\mathsf{u}*\mathsf{u}*\mathsf{N}*\mathsf{N}^{2}*\chi_{5,2}*\chi_{5,3}*\chi_{5,4}*(t-t^5+125t^{25})
\nn\\
\Rightarrow g_{x_{1}^{3}x_{5}}(t)
&=&
t+7 t^2+13 t^3+35 t^4+31 t^5+91 t^6+57 t^7+155 t^8+130 t^9+217 t^{10}+137 t^{11}\nn\\
&&+455 t^{12}+183 t^{13}+399 t^{14}+403 t^{15}+652 t^{16}+307 t^{17}+910 t^{18}+381 t^{19}\nn\\
&&+1085 t^{20}+741 t^{21}+959 t^{22}+553 t^{23}+2015 t^{24}+931 t^{25}
+\mathcal{O}(t^{26})
\eea
\beal{epre7}
\mathsf{g}_{x_{1}x_{7}}
&=&
\mathsf{u}*\chi_{7,1}*\chi_{7,2}*\chi_{7,3}*\chi_{7,4}*\chi_{7,5}*\chi_{7,6}*\left(t+7\sum_{a=0}^{\infty}{t^{7^{(a+2)}}}\right)
\nn\\
&=&
\mathsf{u}*\mathsf{u}*\chi_{7,2}*\chi_{7,3}*\chi_{7,4}*\chi_{7,5}*\chi_{7,6}*(t-t^7+7t^{7})
\nn\\
\Rightarrow g_{x_{1}x_{7}}(t)
&=&
t+t^2+t^3+t^4+t^5+t^6+t^7+3 t^8+t^9+t^{10}+t^{11}+t^{12}+t^{13}+t^{14}+t^{15}\nn\\
&&+3 t^{16}+t^{17}+t^{18}+t^{19}+t^{20}+t^{21}+t^{22}+t^{23}+3 t^{24}+t^{25}+t^{26}+t^{27}+t^{28}\nn\\
&&+7 t^{29}+t^{30}+t^{31}+3 t^{32}+t^{33}+t^{34}+t^{35}+t^{36}+t^{37}
+\mathcal{O}(t^{38})
~~.
\eea
\\

Finally, for the abelian orbifolds of $\mathbb{C}^{9}$ we derive using observation \ref{prop2}
\beal{epre8}
\mathsf{g}_{x_{1}^{7}x_{2}}
&=& 
\mathsf{u}*\mathsf{N}*\mathsf{N}^{2}*\mathsf{N}^{3}*\mathsf{N}^{4}*\mathsf{N}^{5}*\mathsf{N}^{6}*\chi_{2,1}*\left(t+128\sum_{a=0}^{\infty}{t^{2^{(a+2)}}}\right)\nn\\
&=& 
\mathsf{u}*\mathsf{u}*\mathsf{N}*\mathsf{N}^{2}*\mathsf{N}^{3}*\mathsf{N}^{4}*\mathsf{N}^{5}*\mathsf{N}^{6}*(t-t^2+128t^4)\nn\\
\Rightarrow g_{x_{1}^{7}x_{2}}(t) 
&=& t + 127 t^2 + 1094 t^3 + 10923 t^4 + 19532 t^5 + 138938 t^6 + 
 137258 t^7 + 804419 t^8  \nn\\
 &&+ 897354 t^9 + 2480564 t^10 + 1948718 t^{11} + 
 11949762 t^{12} + 5229044 t^{13} \nn\\
 && + 17431766 t^{14} + 21368008 t^{15} + 
 55142131 t^{16} + 25646168 t^{17}
 +\mathcal{O}(t^{18})
\eea
\beal{epre9}
\mathsf{g}_{x_{1}^{6}x_{3}}
&=& 
\mathsf{u}*\mathsf{N}*\mathsf{N}^{2}*\mathsf{N}^{3}*\mathsf{N}^{4}*\mathsf{N}^{5}*\chi_{3,1}*\chi_{3,2}*\left(t+729\sum_{a=0}^{\infty}{t^{3^{(a+2)}}}\right)\nn\\
&=&
\mathsf{u}*\mathsf{u}*\mathsf{N}*\mathsf{N}^{2}*\mathsf{N}^{3}*\mathsf{N}^{4}*\mathsf{N}^{5}*\chi_{3,2}*(t-t^3+729t^9)
\nn\\
\Rightarrow g_{x_{1}^{6}x_{3}}(t)
&=&
t+63 t^2+364 t^3+2668 t^4+3906 t^5+22932 t^6+19610 t^7+97218 t^8\nn\\
&&+100192 t^9+246078 t^{10}+177156 t^{11}+971152 t^{12}+402236 t^{13}+1235430 t^{14}\nn\\
&&+1421784 t^{15}+3312415 t^{16}+1508598 t^{17}+6312096 t^{18}
+\mathcal{O}(t^{19})
\eea
\beal{epre10}
\mathsf{g}_{x_{1}^{4}x_{5}}
&=&
\mathsf{u}*\mathsf{N}*\mathsf{N}^{2}*\mathsf{N}^{3}*\chi_{5,1}*\chi_{5,2}*\chi_{5,3}*\chi_{5,4}*\left(t+625\sum_{a=0}^{\infty}{t^{5^{(a+2)}}}\right)
\nn\\
&=&
\mathsf{u}*\mathsf{u}*\mathsf{N}*\mathsf{N}^{2}*\mathsf{N}^{3}*\chi_{5,2}*\chi_{5,3}*\chi_{5,4}*(t-t^5+625t^{25})
\nn\\
\Rightarrow g_{x_{1}^{4}x_{5}}(t)
&=&
t+15 t^2+40 t^3+155 t^4+156 t^5+600 t^6+400 t^7+1395 t^8+1210 t^9+2340 t^{10}\nn\\
&&+1468 t^{11}+6200 t^{12}+2380 t^{13}+6000 t^{14}+6240 t^{15}+11812 t^{16}+5220 t^{17}\nn\\
&&+18150 t^{18}+7240 t^{19}+24180 t^{20}+16000 t^{21}+22020 t^{22}
+\mathcal{O}(t^{23})
\eea
\beal{epre11}
\mathsf{g}_{x_{1}^{2}x_{7}}
&=&
\mathsf{u}*\mathsf{N}*\chi_{7,1}*\chi_{7,2}*\chi_{7,3}*\chi_{7,4}*\chi_{7,5}*\chi_{7,6}*\left(t+49\sum_{a=0}^{\infty}{t^{7^{(a+2)}}}\right)
\nn\\
&=&
\mathsf{u}*\mathsf{u}*\mathsf{N}*\chi_{7,2}*\chi_{7,3}*\chi_{7,4}*\chi_{7,5}*\chi_{7,6}*(t-t^7+49t^{49})
\nn\\
\Rightarrow g_{x_{1}^{2}x_{7}}(t)
&=&
t+3 t^2+4 t^3+7 t^4+6 t^5+12 t^6+8 t^7+17 t^8+13 t^9+18 t^{10}+12 t^{11}+28 t^{12}\nn\\
&&+14 t^{13}+24 t^{14}+24 t^{15}+37 t^{16}+18 t^{17}+39 t^{18}+20 t^{19}+42 t^{20}+32 t^{21}+36 t^{22}\nn\\
&&+24 t^{23}+68 t^{24}+31 t^{25}+42 t^{26}+40 t^{27}+56 t^{28}+36 t^{29}
+\mathcal{O}(t^{30})
~~.
\eea
\\

Under observation \ref{prop3}, the cycles $x_{2}x_{5}$ and $x_{3}x_{5}$ for $\mathbb{C}^{7}/\Gamma_N$ and $\mathbb{C}^{8}/\Gamma_N$ have
\beal{epre12}
\mathsf{g}_{x_{2}x_{5}}
&=&
\mathsf{u}*\chi_{2,1}*\chi_{5,1}*\chi_{5,2}*\chi_{5,3}*\chi_{5,4}*\left(t+2\sum_{a=0}^{\infty}{t^{2^{(a+2)}}}\right)*\left(t+5\sum_{a=0}^{\infty}{t^{5^{(a+2)}}}\right)\nn\\
&=&
\mathsf{u}*\mathsf{u}*\mathsf{u}*\chi_{5,2}*\chi_{5,3}*\chi_{5,4}*(t-t^2+2t^4)*(t-t^5+5t^{25})
\nn\\
\Rightarrow g_{x_{2}x_{5}}(t)
&=&
t+t^2+2 t^3+3 t^4+2 t^5+2 t^6+2 t^7+5 t^8+3 t^9+2 t^{10}+6 t^{11}+6 t^{12}+2 t^{13}\nn\\
&&+2 t^{14}+4 t^{15}+8 t^{16}+2 t^{17}+3 t^{18}+2 t^{19}+6 t^{20}+4 t^{21}+6 t^{22}+2 t^{23}+10 t^{24}\nn\\
&&+8 t^{25}+2 t^{26}+4 t^{27}+6 t^{28}+2 t^{29}+4 t^{30}+6 t^{31}+10 t^{32}
+\mathcal{O}(t^{33})
\eea
\beal{epre13}
\mathsf{g}_{x_{3}x_{5}}
&=&
\mathsf{u}*\chi_{3,1}*\chi_{3,2}*\chi_{5,1}*\chi_{5,2}*\chi_{5,3}*\chi_{5,4}*\left(t+3\sum_{a=0}^{\infty}{t^{3^{(a+2)}}}\right)*\left(t+5\sum_{a=0}^{\infty}{t^{5^{(a+2)}}}\right)\nn\\
&=&
\mathsf{u}*\mathsf{u}*\mathsf{u}*\chi_{3,2}*\chi_{5,2}*\chi_{5,3}*\chi_{5,4}*(t-t^3+3t^9)*(t-t^5+5t^{25})
\nn\\
\Rightarrow g_{x_{3}x_{5}}(t)
&=&
t+t^2+t^3+2 t^4+t^5+t^6+3 t^7+2 t^8+4 t^9+t^{10}+5 t^{11}+2 t^{12}+3 t^{13}+3 t^{14}\nn\\
&&+t^{15}+4 t^{16}+t^{17}+4 t^{18}+3 t^{19}+2 t^{20}+3 t^{21}+5 t^{22}+t^{23}+2 t^{24}+7 t^{25}+3 t^{26}\nn\\
&&+7 t^{27}+6 t^{28}+t^{29}+t^{30}+7 t^{31}+4 t^{32}+5 t^{33}+t^{34}+3 t^{35}
+\mathcal{O}(t^{36})
~~.
\eea
\\

Using observation \ref{prop4}, results in \eref{epre12} and \eref{epre13}, and the sequence $\mathsf{g}_{x_{2}x_{3}}$, are extended for $\mathbb{C}^{7}/\Gamma_N$, $\mathbb{C}^{8}/\Gamma_N$ and $\mathbb{C}^{9}/\Gamma_N$, as follows
\beal{epre13}
\mathsf{g}_{x_{1}^{2}x_{2}x_{3}}
&=& 
\mathsf{u}*\mathsf{N}*\mathsf{N}^{2}*\mathsf{N}^{3}*\chi_{2,1}*\chi_{3,1}
*\chi_{3,2}*\left(t+8\sum_{a=0}^{\infty}{t^{2^{(a+2)}}}\right)*\left(t+27\sum_{a=0}^{\infty}{t^{3^{(a+2)}}}\right)
\nn\\
&=&
\mathsf{u}*\mathsf{u}*\mathsf{u}*\mathsf{N}*\mathsf{N}^{2}*\mathsf{N}^{3}*\chi_{3,2}*(t-t^2+8t^4)*(t-t^3+27t^9)
\nn\\
\Rightarrow g_{x_{1}^{2}x_{2}x_{3}}(t)
&=&
t+7 t^2+14 t^3+44 t^4+32 t^5+98 t^6+60 t^7+226 t^8+171 t^9+224 t^{10}+134 t^{11}\nn\\
&&+616 t^{12}+186 t^{13}+420 t^{14}+448 t^{15}+1039 t^{16}+308 t^{17}+1197 t^{18}+384 t^{19}\nn\\
&&+1408 t^{20}+840 t^{21}+938 t^{22}+554 t^{23}+3164 t^{24}+839 t^{25}
+\mathcal{O}(t^{26})
\eea

\beal{epre14}
\mathsf{g}_{x_{1}^{3}x_{2}x_{3}}
&=&
\mathsf{u}*\mathsf{N}*\mathsf{N}^{2}*\mathsf{N}^{3}*\mathsf{N}^{4}*\chi_{2,1}*\chi_{3,1}
*\chi_{3,2}*\left(t+16\sum_{a=0}^{\infty}{t^{2^{(a+2)}}}\right)*\left(t+81\sum_{a=0}^{\infty}{t^{3^{(a+2)}}}\right)
\nn\\
&=&
\mathsf{u}*\mathsf{u}*\mathsf{u}*\mathsf{N}*\mathsf{N}^{2}*\mathsf{N}^{3}*\mathsf{N}^{4}*\chi_{3,2}*(t-t^2+16t^4)*(t-t^3+81t^9)
\nn\\
\Rightarrow g_{x_{1}^{3}x_{2}x_{3}}(t)
&=&
t+15 t^2+41 t^3+172 t^4+157 t^5+615 t^6+403 t^7+1666 t^8+1332 t^9+2355 t^{10}\nn\\
&&+1465 t^{11}+7052 t^{12}+2383 t^{13}+6045 t^{14}+6437 t^{15}+14719 t^{16}+5221 t^{17}\nn\\
&&+19980 t^{18}+7243 t^{19}+27004 t^{20}+16523 t^{21}+21975 t^{22}
+\mathcal{O}(t^{23})
\eea

\beal{epre15}
\mathsf{g}_{x_{1}^{4}x_{2}x_{3}}
&=&
\mathsf{u}*\mathsf{N}*\mathsf{N}^{2}*\mathsf{N}^{3}*\mathsf{N}^{4}*\mathsf{N}^{5}*\chi_{2,1}*\chi_{3,1}
*\chi_{3,2}*\left(t+32\sum_{a=0}^{\infty}{t^{2^{(a+2)}}}\right)*\left(t+243\sum_{a=0}^{\infty}{t^{3^{(a+2)}}}\right)
\nn\\
&=&
\mathsf{u}*\mathsf{u}*\mathsf{u}*\mathsf{N}*\mathsf{N}^{2}*\mathsf{N}^{3}*\mathsf{N}^{4}*\mathsf{N}^{5}*\chi_{3,2}*(t-t^2+32t^4)*(t-t^3+243t^9)
\nn\\
\Rightarrow g_{x_{1}^{4}x_{2}x_{3}}(t)
&=&
t+31 t^2+122 t^3+684 t^4+782 t^5+3782 t^6+2804 t^7+12866 t^8+11376 t^9+24242 t^{10}\nn\\
&&+16106 t^{11}+83448 t^{12}+30944 t^{13}+86924 t^{14}+95404 t^{15}+223327 t^{16}+88742 t^{17}\nn\\
&&+352656 t^{18}+137564 t^{19}+534888 t^{20}+342088 t^{21}
+\mathcal{O}(t^{22})
\eea

\beal{epre16}
\mathsf{g}_{x_{1}x_{2}x_{5}}
&=&
\mathsf{u}*\mathsf{N}\chi_{2,1}*\chi_{5,1}*\chi_{5,2}*\chi_{5,3}*\chi_{5,4}*
\left(t+4\sum_{a=0}^{\infty}{t^{2^{(a+2)}}}\right)*
\left(t+25\sum_{a=0}^{\infty}{t^{5^{(a+2)}}}\right)
\nn\\
&=&
\mathsf{u}*\mathsf{u}*\mathsf{u}*\mathsf{N}*\chi_{5,2}*\chi_{5,3}*\chi_{5,4}*
(t-t^2+4t^4)*(t-t^5+25t^{25})
\nn\\
\Rightarrow g_{x_{1}x_{2}x_{5}}(t)
&=&
t+3 t^2+5 t^3+11 t^4+7 t^5+15 t^6+9 t^7+31 t^8+18 t^9+21 t^{10}+17 t^{11}+55 t^{12}\nn\\
&&+15 t^{13}+27 t^{14}+35 t^{15}+76 t^{16}+19 t^{17}+54 t^{18}+21 t^{19}+77 t^{20}+45 t^{21}+51 t^{22}\nn\\
&&+25 t^{23}+155 t^{24} + 63 t^{25} + 45 t^{26} + 58 t^{27} + 99 t^{28} + 31 t^{29}
+\mathcal{O}(t^{30})
\eea

\beal{epre17}
\mathsf{g}_{x_{1}^{2}x_{2}x_{5}}
&=&
\mathsf{u}*\mathsf{N}*\mathsf{N}^{2}*\chi_{2,1}*\chi_{5,1}*\chi_{5,2}*\chi_{5,3}*\chi_{5,4}*
\left(t+8\sum_{a=0}^{\infty}{t^{2^{(a+2)}}}\right)*
\left(t+125\sum_{a=0}^{\infty}{t^{5^{(a+2)}}}\right)
\nn\\
&=&
\mathsf{u}*\mathsf{u}*\mathsf{u}*\mathsf{N}*\mathsf{N}^{2}*\chi_{5,2}*\chi_{5,3}*\chi_{5,4}*
(t-t^2+8t^4)*(t-t^5+125t^{25})
\nn\\
\Rightarrow g_{x_{1}^{2}x_{2}x_{5}}(t)
&=&
t+7 t^2+14 t^3+43 t^4+32 t^5+98 t^6+58 t^7+219 t^8+144 t^9+224 t^{10}+138 t^{11}\nn\\
&&+602 t^{12}+184 t^{13}+406 t^{14}+448 t^{15}+996 t^{16}+308 t^{17}+1008 t^{18}+382 t^{19}\nn\\
&&+1376 t^{20}+812 t^{21}+966 t^{22}+554 t^{23}+3066 t^{24}+963 t^{25}+1288 t^{26}
+\mathcal{O}(t^{27})
\eea

\beal{epre18}
\mathsf{g}_{x_{1}x_{3}x_{5}}
&=&
\mathsf{u}*\chi_{3,1}*\chi_{3,2}*\chi_{5,1}*\chi_{5,2}*\chi_{5,3}*\chi_{5,4}*
\left(t+9\sum_{a=0}^{\infty}{t^{3^{(a+2)}}}\right)*
\left(t+25\sum_{a=0}^{\infty}{t^{5^{(a+2)}}}\right)
\nn\\
&=&
\mathsf{u}*\mathsf{u}*\mathsf{u}*\chi_{3,2}*\chi_{5,2}*\chi_{5,3}*\chi_{5,4}*(t-t^3+9t^9)*(t-t^5+25t^{25})
\nn\\
\Rightarrow g_{x_{1}x_{3}x_{5}}(t)
&=&
t+3 t^2+4 t^3+8 t^4+6 t^5+12 t^6+10 t^7+18 t^8+22 t^9+18 t^{10}+16 t^{11}+32 t^{12}\nn\\
&&+16 t^{13}+30 t^{14}+24 t^{15}+40 t^{16}+18 t^{17}+66 t^{18}+22 t^{19}+48 t^{20}+40 t^{21}+48 t^{22}\nn\\
&&+24 t^{23}+72 t^{24}+57 t^{25}+48 t^{26}+85 t^{27}+80 t^{28}+30 t^{29}+72 t^{30}
+\mathcal{O}(t^{31})
~~.
\eea
\\

%%%%%%%%%%%%%%%%%%%%%%%%%%%%%%%%%%%%%%%%%%%%%%%%%%%%%%%%%%%%%%%%%%%%%%%%
%%%%%%%%%%%%%%%%%%%%%%%%%%%%%%%%%%%%%%%%%%%%%%%%%%%%%%%%%%%%%%%%%%%%%%%%

\end{document}